\documentclass[review]{elsarticle}
\usepackage[margin=2.0cm]{geometry}
\usepackage{lineno,hyperref}
\usepackage{subfig}
\usepackage{longtable}
\usepackage{caption}
\usepackage{hyperref}
\usepackage{enumitem}
\usepackage[table]{xcolor}
\usepackage{multirow}
\usepackage{subfloat}
\usepackage[linesnumbered,ruled,vlined]{algorithm2e}

\usepackage{float}
\usepackage{amsmath}
\usepackage{setspace}
\usepackage{pdflscape} 
\usepackage{geometry} 
\usepackage{graphicx}

\journal{Journal of \LaTeX\ Templates}
\makeatletter \def\ps@pprintTitle{  \let\@oddhead\@empty  \let\@evenhead\@empty  \def\@oddfoot{\hfill\thepage}  \def\@evenfoot{\thepage\hfill}} \makeatother
\begin{document}

\begin{frontmatter}

\title{Hierarchical Classification for Intrusion Detection System: Effective Design and Empirical Analysis}


\author[1]{Md. Ashraf Uddin}\ead{ashraf.uddin@deakin.edu.au}
\author[1]{Sunil Aryal}\ead{sunil.aryal@deakin.edu.au}  
\author[1]{Mohamed Reda Bouadjenek}
\author[1]{Muna Al-Hawawreh}
\author[2]{Md. Alamin Talukder} \ead{alamin.cse@iubat.edu}

\address[1]{School of Information Technology, Deakin University, Geelong, VIC 3125, Australia}
\address[2]{Department of Computer Science and Engineering, International University of Business Agriculture and Technology, Dhaka, Bangladesh}

\cortext[mycorrespondingauthor]{Corresponding authors: Md Ashraf Uddin}

\begin{abstract} With the increased use of network technologies like Internet of Things (IoT) in many real-world applications, new types of cyberattacks have been emerging. To safeguard critical infrastructures from these emerging threats, it is crucial to deploy an Intrusion Detection System (IDS) that can detect different types of attacks accurately while minimizing false alarms. Machine learning approaches have been used extensively in IDS and they are mainly using flat multi-class classification to differentiate normal traffic and different types of attacks. Though cyberattack types exhibit a hierarchical structure where similar granular attack subtypes can be grouped into more high-level attack types, hierarchical classification approach has not been explored well. In this paper, we investigate the effectiveness of hierarchical classification approach in IDS. We use a three-level hierarchical classification model to classify various network attacks, where the first level classifies benign or attack, the second level classifies coarse high-level attack types, and the third level classifies a granular level attack types. Our empirical results of using 10 different classification algorithms in 10 different datasets show that there is no significant difference in terms of overall classification performance (i.e., detecting normal and different types of attack correctly) of hierarchical and flat classification approaches. However, flat classification approach misclassify attacks as normal whereas hierarchical approach misclassify one type of attack as another attack type. In other words, the hierarchical classification approach significantly minimises attacks from misclassified as normal traffic, which is more important in critical systems. 

\end{abstract}

   
     
        



\end{frontmatter}

\section{Introduction}
\label{Introduction}

With the increasing dependence on network systems in our daily lives, the need for secure network systems has become more critical than ever. However, the increasing sophistication of cyber attacks poses a significant threat to network security, and traditional security measures alone are no longer sufficient\cite{injadat2020multi, talukder2023dependable}. The Intrusion Detection System (IDS) is considered an effective defensive tool that closely monitors network traffic and system logs and blocks any suspected traffic in the network or system\cite{tieppo2022hierarchical}. 

In IDS literature, different classification approaches have been used to classify normal traffic and different type of cyberattacks \cite{gupta2022cse, zhou2020building, talukder2022machine, talukder2023efficient, pradeep2021intrusion}. Though they are using different classification algorithms, most of them learn one flat multi-class classifier to differentiate normal and various attack categories at the same time. Because the training data in these cases are heavily unbalanced as most of the network traffic will be normal and there will be significantly less attacks, a flat classifier trained on such unbalanced data is biased towards to the normal class and more attacks will be misclassified as normal, i.e., they have high false negative rate. In network security, false negative prediction where attack is classified as a normal request can be catastrophic. Therefore, the main goal here is to minimise false negative results as much as possible while ensuring that genuine requests get access to the network. Also, some attack types are more similar to each other than others making them difficult to separate from each other well with a global flat classifier. For example, different types of Denial-of-Service (DoS) attacks such as DrDoS DNS, DrDoS SNMP, and
DrDoS LDAP can be difficult to differentiate between them while also separating them from attack types in other families (e.g., Malware).

\begin{figure}[!htbp]
                \centering
                \includegraphics[scale = .70]{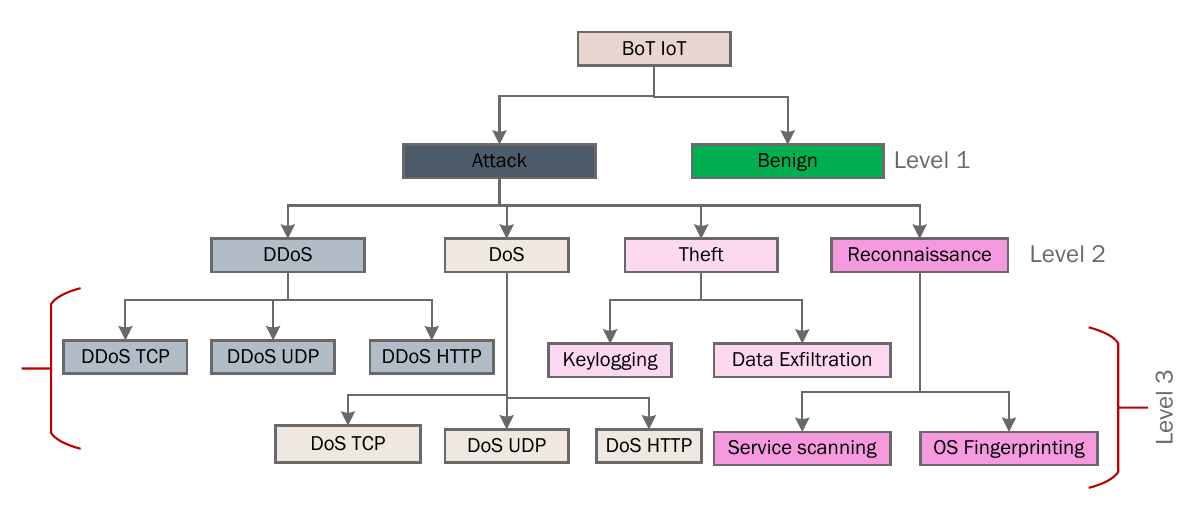}
                \caption{Hierarchical level of attacks in the Bot IoT dataset}
                \label{fig:botiot}
        \end{figure}

Because cyberattacks naturally exhibit a hierarchical structure where similar attack types can be grouped in more broader families of attacks, it would make sense to learn classification in a hierarchical fashion, where first normal traffic is separated from attacks and then attack families are separated followed by separating specific attack types within each family. For example, Figure \ref{fig:botiot} shows an example scenario of the BoT IoT dataset where attacks are grouped in a three level hierarchy: normal vs attack at first level, attacks are classified into four high-level attack families (DDoS, DoS, Theft, Reconnaissance) at the second level, and 10 specific attack types at the third/bottom level (e.g., Theft can be of two types Keylogging and Data Exfiltration). 

In hierarchical classification \cite{silla2011survey}, classes are arranged in a tree-like structure such that each node represents a subset of classes that are similar to each other in some way. The root node covers all the classes and each leaf nodes represents one specific class. The number of leaf nodes is the same as the number of classes. A separate classifier is trained at each internal node using training data belonging to the node only. In the testing phase, the class label prediction for a new data instance happens top-down following a specific branch of the tree-structure and the leaf node it lands is the predicted class label. Hierarchical classification approach involves a series of prediction decisions made at each level of the tree until the final classification is reached at the leaf node \cite{binkhonain2023machine}. In the example shown Figure \ref{fig:botiot}, network traffic is classified at three stages: (i) attack or normal at Level 1 (root); (ii) if predicted as attack at the root, then classified into one of the four attack families at Level 2; and (iii) finally into a specific attack type within the predicted attack family at Level 3. Because the classification happens at different granular levels, algorithms can use features/cues that are important to differentiate classes at a given level. 




In the context of cybersecurity, misclassifying an attack category poses a greater risk than blocking benign traffic as an attack can cause significant disruption or even halt system operations. Because the root level classification in the case of hierarchical approach is solely focused on differentiating attacks from normal traffic, it  is  is well-suited to minimize misclassifications of attacks as benign requests. Even if an attack is misidentified as a different attack family/type at the lower level, it still poses no threat as long as it is identified as an attack at the root. However, in the case of flat multiclass classification approach, the classifier attempts to separate different attack types (some of which have very subtle differences) and normal at the same time the chances of attacks being misclassified as normal is higher. It assumes that all classes (various attack types and normal) are independent and completely ignores and inherent relationship/similarity of some attach types. In real-wolrd situations, identifying and analyzing similar attacks as a group is more meaningful and informative for users than detecting each attack individually.For instance, DoS is a more comprehensible term for non-technical users compared to its subcategories such as neptune, smurf, back, teardrop, pod, and land. Moreover, machine learning algorithms can effectively learn and recognize DoS attacks due to the availability of a larger number of instances representing DoS as a whole, compared to its specific subcategories. 

There are some studies in the literature that attempt to use hierarchical classification approach in some way \cite{sarikaya2020class, eldahshan2022meta, alin2020towards, ahmim2019novel, wang2017hast, sarnovsky2020hierarchical, mohd2021intrusion}. These studies defined the hierarchical studies in their own way without relying on ground truth information nor using any well-established taxonomy. For example, some of them use stacked multiple one-vs-rest classification models \cite{alzaqebah2023hierarchical, sarnovsky2020hierarchical, mohd2021intrusion} and others defined hierarchical structure by combining some attack families/types together \cite{sarikaya2020class, eldahshan2022meta} arbitrarily or using two level hierarchy with normal vs attack at Level 1 and all attack types in Level 2 \cite{alin2020towards}. Also, these prior studies evaluated their models with a few classification algorithms in one or two datasets. As far as we known, there is no prior study that evaluated whether hierarchical approach is beneficial in IDS more systematically using multiple classification algorithms and datasets.


In this study, we conduct a comprehensive comparison of traditional multiclass flat and three-level hierarchical (attack vs normal, attack family, and attack types) approaches using 10 classification algorithms in 10 widely used IDS datasets in the literature. We aim to answer to the research question ``\textit{Is hierarchical classification approach  more effective in detecting network intrusions compared to a flat model?}''. We used the three-level hierarchy as most of these IDS datasets have ground truth label for these levels. In the case of datasets where attack family label is not available, we grouped attack types into families following the Cyber Kill Chain framework \cite{mihai2014cyber}. Our empirical results shows that there is no significant difference in terms of overall classification performance (i.e., classification of normal and different types of attack correctly) of hierarchical and flat classification approaches. However, the flat approach missclassify attacks as normal whereas hierarchical approach missclassify one type of attack as another attack type. In other words, the hierarchical classification approach significantly minimises the number of attacks from misclassified as normal traffic, which is more important in practice.

The structure of this paper is as follows. Section \ref{Literature Review} presents a review of the related literature. Section \ref{Proposed} details the hierarchical classification architecture and materials used in this study. In Section \ref{RESULTS AND DISCUSSION}, we present the results of our experiments conducted to compare them with their flat counterparts. Finally, Section \ref{Conclusion} summarizes the paper and outlines potential future research directions.

\section{Related works} 
\label{Literature Review}

In this section, we begin by analyzing the most recent hierarchical classification methods that are relevant to our study. Following that, we present several most cited literature in this domain to illustrate the current state-of-the-art intrusion detection systems.

\subsection{Hierarchical Classification in IDS}

Sarikaya et al.\cite{sarikaya2020class} proposed a hierarchical classification approach to detect network intrusion. They divided the UNSW-NB15 dataset into a 60\% training set and a 40\% testing set. In the first stage, they trained a Random Forest model on a dataset consisting of two targets: 0 (benign) and 1 (attacks). In the second stage, they trained a Random Forest (RF) model on two groups of attack samples: group 1 (DoS and Exploit attacks) and group 2 (other attacks). This grouping was done based on confusion matrix of a flat classification. For example, since most reconnaissance attacks are being identified as exploits, both of them are kept in the same group. Finally, in the third stage, they used a multiclassifier to identify all types of attack in group 2. The model was trained and tested using Weka tools. In order to improve the accuracy rates of various attack categories in IDS datasets, several researchers have directed their attention towards the development of the One-vs-rest approach. For instance, Mohd et al.\cite{mohd2021intrusion} introduced a hierarchical classification approach based on One-vs-rest to identify intrusion. They utilized the KDD CUP99 dataset to train five classifiers - Support Vector Machine(SVM), Probabilistic Neural Network (PNN), Decision Tree ( DT), Neuro-fuzzy Classifier(NFC), and Smooth Support Vector Machine (SSVM) - where each classifier was used four times to detect four different types of attacks at four levels, as well as normal data. The best classifiers were then placed at each level to construct the hierarchical model. Following the training and testing of a uniform One- vs-rest model, the most effective classifier at each level was chosen. Eldahshan et al.\cite{eldahshan2022meta} proposed a One-vs-rest based hierarchical classification model to detect network intrusion. The model has a pipeline of three stage: 1) Binary GWO (Grey Wolf Optimizer) feature selection 2) Extreme Leaning Machine (ELM) 3) Physical meta-heuristic-based parameter tuning using AOA (Archimedes Optimization Algorithm) and HBA (Honey Badger Algorithm).  The model was validated using UNSW NB 15 and CIC-IDS 2017 benchmark datasets. Notably, the study incorporated various optimization techniques for feature selection and hyperparameter tuning. While the implementation of the One-vs-rest approach was part of the methodology, this remains unclear whether One-vs-rest directly contributed to the improvement of attack category rates. To explore the efficacy of the One-vs-rest approach in intrusion detection systems, Alzaqebah et al.\cite{alzaqebah2023hierarchical} followed the similar approach to Eldahshan et al.\cite{eldahshan2022meta} and  presented a hierarchical intrusion detection system based on bio-inspired feature selection. They utilized a One-vs-rest classification strategy to detect both benign and various attack categories, converting a multi-classification problem to multiple binary classification problems. The model implemented an Extreme Learning Machine (ELM) to identify attacks at different levels, with each level detecting a specific type of attack. To determine the optimized set of features, weights, and biases for ELM, an enhanced Harris Hawks optimizer called IHHO was employed. The performance of the model was evaluated using the UNSW-NB15 datasets, which was used for both training and testing.

Several studies in this field have explored the use of a two-level hierarchical approach to improve the accuracy rates for detecting attack instances. For instances, Alin et al.\cite{alin2020towards} proposed a two-level hierarchical intrusion detection model. The first level consists of a binary classifier that identifies whether network traffic samples are benign or attack. The second level consists of a multi-classifier that identifies the specific type of attack if the sample is classified as an attack in the first level. The authors compared their hierarchical model with traditional multi-class flat models and demonstrated that hierarchical classification performs better using datasets such as KDD-CUP99, NSL-KDD, UNSW-NB15, and CIC-IDS2017. The experiment was carried out using basic models including Naïve Bayes (NB), DT, K Nearest Neighbors (KNN), Logistic Regression (LR), SVM, AdaBoost, RF, and Multi-Layer Perceptron (MLP).  Following Alin et al.\cite{alin2020towards}, Sarnovsky et al.\cite{sarnovsky2020hierarchical} also proposed a hierarchical classification method for detecting intrusions. Initially, they created a targeted binary datasets by labeling all attack categories as "attack" and combining them with normal samples. In the first level of the model, they trained a binary classifier on 70\% of the KDD-CUP99 data to distinguish between normal and attack samples. The remaining 30\% of the dataset was used for testing. In the second level, they developed an ensemble classifier using 70\% of the attack samples and consisting of five classifiers: Decision Tree (C4.5), Random Forests, ForestPA, and One-vs-rest. The remaining 30\% of the data was used for evaluating the performance of the model. Likewise, Ahmim et al.\cite{ahmim2019novel} presented a two-level intrusion detection model. The first level includes two classifiers trained on the same normalized datasets to classify network traffic as either benign or attack. The second level utilizes the outcome from the two models at the first level and the datasets with original labels to train a model for further classification of attacks. 
In most literature, One-vs-One and One-vs- rest were considered hierarchical classification approaches. Nevertheless, the hierarchical approach of category and subcategory of similar kinds of network attacks and their identification at multiple levels is less prevalent in the existing IDS works. After conducting a thorough search and review of the existing literature, we have identified several authors who have attempted to develop hierarchical classification models at the classifier levels \cite{sarikaya2020class, eldahshan2022meta, alin2020towards, ahmim2019novel, wang2017hast, sarnovsky2020hierarchical, mohd2021intrusion}. For example, Sarikaya et al. \cite{sarikaya2020class} and Alin et al. \cite{alin2020towards} have investigated two-level hierarchical models, with the latter examining only the UNSW-NB15 dataset. In contrast, Alin et al. \cite{alin2020towards} have trained and tested multiple machine learning algorithms for KDD-CUP99, NSL-KDD, UNSW-NB15, and CIC-IDS2017, where NSL-KDD is a modified version of KDD-CUP99. However, a limitation of these works \cite{sarikaya2020class, alin2020towards} is that they trained the first- and second-level classifiers using a subset of the original dataset. This training and testing mechanism can result in overfitting or underfitting of the model, leading to reduced generalization performance on new data. In addition, the model's efficacy can heavily depend on the random partitioning of data into training and testing sets, resulting in an unreliable estimation of its actual performance on unseen data. Most of the existing work ignored cross-validation techniques which can accurately evaluate the model's efficacy using multiple data splits. Further, the state-of-the-art works explored limited number of datasets, and machine learning algorithms and did not follow proper hierarchical structure of IDS datasets while investigating the performance of hierarchical classification in this field.

\subsection{ Influential Studies in IDS}

In this review, we have examined a selection of highly cited state-of-the-art studies in order to provide readers valuable insights into this field. In the field of IDS, a significant focus of researchers has been on exploring various deep learning and machine learning models. These studies have involved the application of diverse optimization techniques to enhance feature selection, hyperparameter tuning, and address data balancing challenges through data-level and class-level approaches. To start with, Jie et al.\cite{gu2021effective} applied Naïve Bayes algorithm to transform original features to a set of new features with high quality. The dataset with new converted feature was fed to a SVM to classify malicious or non malicious traffic. The feature embedding technique and model were implemented using R and LIBSVM. Performance was analyzed with respect to accuracy, precison, recall.  Four different types of dataset: UNSW-NB15, CIC-IDS2017, NSL-KDD, and Kyoto 2006+ were used to train the model. They attempted to address the issue of data quality. However, the dimension of the dataset is double the original one. Still, training time and complexity of the model was shown less than that of single SVM.

Injadat et al.\cite{injadat2020multi} suggested a machine-learning model to detect intrusion detection, including multi-stage optimization techniques. They performed preprocessing, selected the most correlated features with the target, and applied different hyperparameter tuning processes before training two classification algorithms. In the preprocessing, Z-Score normalization was applied to transform categorical data into numerical data. SMOTE was utilized to balance the number of minority and majority classes in the dataset. The authors examined two feature selection methods called Information Gain and correlation coefficient. Different hyperparameter optimization techniques, including random search, meta-heuristic, and Bayesian, are applied while training K-nearest neighbor and Random Forest classifier algorithms. Two different datasets: CICIDS 2017 and UNSW-NB 2015 were fed to the model for training it. The model was implemented using Python programing and analyzed the performance with respect to accuracy, precision, recall and false positive recognition. The authors’ finding shows that the combination of Tree Parzen Estimator (TPE) based Bayesian optimization and Random Forest outperforms other examined approaches for CICIDS 2017 and UNSW-NB 2015. The articles analyzed the complexities and effectiveness of the model with respect to feature set size, training sample size, and model performance. Additionally, the authors utilized big $O$ notation to analyze complexity of the model. However, the authors did not compare the performances of their suggested model with existing models. They addressed high dimensionality and imbalanced issue of the two IDS datasets. 

IDS datasets typically have a limited number of attack categories at level 2. Training models with such datasets often results in misclassification of many attacks in real-life scenarios. To address this issue, Kilincer et al.\cite{kilincer2021machine} combined the UNSW-NB15, CSE-CIC-IDS2018, ISCX2012, CIDDS-001, and NSL-KDD datasets to form a big dataset and scaled the features of the combined datasets using min-max normalisation. KNN, DT, and SVM classifiers were fed the merged dataset to evaluate their accuracy, precision, recall, and F-score. Matlab programming was used to implement the model. The study examined retrieval methods and insights into common datasets used to identify intrusion in numerous sectors, such as computer networks, the industrial internet of things, electrical networks, and android settings. Although the authors intend to address the problem of training IDS with datasets containing a limited variety of attack types and form a big dataset using the existing four datasets, the combined data suffer from imbalanced data issues because there are few common attacks among the merged dataset. Additionally, individual datasets vary in terms of network characteristics, environment, and time.  

Roy et al.\cite{roy2022lightweight} target to design a simplified machine learning model to detect intrusion into the Internet of Things. The author suggested an optimized machine learning-based intrusion detection system. First, they developed the pre-processing stage by applying variance inflation factor(VIF) to remove the multicollinearity of the feature sets, z-score normalization, SMOTE, and Tomek Link to balance the dataset. Next, they designed a stacking ensemble classifier where class level 0 includes K Neighbors, Random Forest, and XGBoost, and level 1(meta-classifier) finally classifies the network traffic using XG-Boost. The model was tested using two popular datasets: CIC-IDS2017 and NSL-KDD and evaluated concerning the accuracy, recall, precision, and F1 score. The model was implemented using TensorFlow, Keras of python programming. According to the author, the model is computationally efficient. However, no analysis was provided to back up the theory regarding training and testing time. In addition, training numerous base models is necessary for stacking models, which can take longer, especially when working with huge datasets. Stacking ensemble approach suffers from overfitting, where the final model grows overly complicated and begins to memorise the training data rather than generalise it.

To find out the optimal features and  investigate new datasets, Kilincer et al.\cite{kilincer2022comprehensive} applied Extra Tree to discover the most optimal features, and the dataset with optimal features is fed into Light Gradient Boosting Machine(LGBM) and XGBoost (Extreme Gradient Boosting) algorithms for classifying intrusion. To train their model, the authors formed two new datasets called CCiDD\_A and CCiDD\_B. Further, three public datasets: NSL-KDD, UNSW-NB15, and CSE-CIC-IDS2018, were used to analyze the performance of the proposed model. They implemented the model and analyzed performance concerning different performance metrics, including accuracy, precision, recall, and F1-score. Using solely Extra Trees to choose important features may not provide the intended results, especially if the data contains noisy features, as it may select irrelevant or duplicated features.

To reduce data dimensionality, Naseri et al.\cite{naseri2022feature} suggested Farmland Fertility Algorithm (FFA) for selecting the most relevant feature. The dataset with the chosen feature set is normalized before feeding them into an ensemble machine learning model. The ensemble model includes several classifiers, such as KNN, SVM, RF, and ADA\_BOOST. The model was trained using two datasets: NSL-KDD and UNSW-NB15. They measured the performance with respect to accuracy, precision, recall, and F1-score. The author investigated a new meta-heuristic optimization technique called Farmland Fertility Algorithm (FFA) to reduce the data dimensionality and showed higher accuracy. However, they used two old datasets: NSL-KDD and UNSW-NB15, which does not justify the impact of the FFA because our studies showed that the most classifier algorithms trained with these datasets even without eliminating features showed higher accuracy. 

Tttota et al. \cite{attota2021ensemble} intended to address the security and privacy issue of IoT devices. They suggested an ensemble-based machine learning approach for intrusion detection in IoT federated learning environments. The authors partitioned lightweight MQTT (Message Queuing Telemetry Transfer protocol) data into three datasets, each with a particular number of features: Bi-Flow, Packet, and Uni-Flow. The Grey Wolves Optimization approach was utilized on each dataset to perform feature engineering. In their works, first, a central server builds a neural network model and shares it with the IoT Gateway. Second, IoT devices train the base model using three views of the dataset and submit learning parameters to the server. Third, The server aggregates the learning parameters collected from each IoT and updates the base model to generate a global model. Finally, every IoT device is returned to the global model. Later, the server constructs a Random Forest-based ensemble mechanism for models trained with three datasets. Lightwight MQTT dataset. The model was implemented using Pysyft deep learning framework and compared to a traditional centralized trained model concerning the accuracy, precision, recall, and f-score. The authors achieved higher accuracy by incorporating an ensemble mechanism in a federated learning setting. However, other pressing issues of federated learning, such as efficient and secure communication between server and IoT devices, poison attacks, and device heterogeneity, are yet to be addressed.

Recently, many researchers have investigated different deep learning model to detect network intrusion. For instances, Imrana et al.\cite{imrana2021bidirectional} recommended bidirectional LSTM to identify the intrusion from the network traffic. They performed basic processing, such as scaling and transforming the NSL-KDD dataset, before feeding it into LSTM. NSL-KDD was to conduct the performance analysis.Authors implemented the model using Tensorflow and Keras library of the python programming language. The shortcoming of the model is that this was not trained and tested using a recent dataset.The research aim to reduce the rates of false alarm and increase the accuracy of U2R and R2L attack. The biLSTM might produce higher accuracy for other network traffic-related dataset having sequential properties as this model is suitable for sequential data. However, most network traffic datasets have both sequential patterns and discrete characteristics. Therefore, a hybrid model combining DL and ML might detect intrusion more correctly. Usually, deep learning involves higher complexity and deep learning training and testing times are higher than traditional machine learning algorithms. 

Jiang et al.\cite{jiang2020network} emphasized on balancing datasets and utilising deep learning approach to extract spatial and temporal features. To detect intrusions in the IDS dataset, the authors merged CNN and BiLSTM to create CNN-BiLSTM, which can extract temporal and geographical information from the dataset. They used Tomek Links to minimize the number of samples from the majority class and SMOTE to increase the number of samples from the minority class to create a balanced IDS dataset. They analysed the performance of their model with the NSL-KDD and UNSW-NB15 datasets and compared it to RF, AlexNet, LeNet-5, CNN, and BiLSTM in terms of accuracy, precision, recall, and F1-score. The model was implemented using TensorFlow, Keras of python programming. Their findings indicate that the model was unable to detect U2R and R2L attacks with greater precision due to the flaws of the SMOTE data balancing technique. SMOTE can increase the number of noisy samples since it generates new, inconsistent samples. To investigate deep learning algorithms for IoT data for higher accuracy. Saba et al.\cite{saba2022anomaly} proposed a CNN model for the Internet of Things network to detect intrusions. Two IoT datasets: NID Dataset and BoT-IoT, were utilized for training a CNN model. The model's performance was demonstrated only concerning accuracy. Although the authors underlined the need for deploying IDS in IoT networks, data pre-processing, including feature selection and the handling of imbalanced datasets, is overlooked, and performance measurement is inadequately conducted.


\section{Hierarchical Classification Model}
\label{Proposed}

Our analysis of the current state-of-the-art research reveals that most studies have focused on conventional flat multi-class classification approaches and have evaluated their models using a limited number of IDS datasets. This raises concerns about the effectiveness of these models in accurately detecting network intrusions in real-life scenarios. In addition, these models tend to produce a higher number of false negative cases (i.e., attacks are classified as normal), which is particularly concerning in practice. To address this gap, our study aims to investigate hierarchical classification using a large number of benchmark IDS datasets. There is still limited literature exploring hierarchical classification in IDS on a large scale, making our research valuable and unique in this context.

In this section, we present the design and architecture of our hierarchical classification approach. Our model utilizes a hierarchical structure and is trained using 10 different benchmark IDS datasets. Most IDS datasets' creators typically include three target/label columns: (i) class - binary attack or normal (Level 1); (ii) attack family (Level 2)- attack categories such as DDoS, DoS, Trojan Horse, and Spyware; and (iii) attack type  -  attack subcategories (Level 3) such as rootkit, worms, buffer overflow, tear drop, and more. Some datasets will have only class (Level 1) and attack category or type (Level 2 or 3) labels provided. Considering this taxonomy or hierarchical structure of the dataset, hierarchical classifications with multiple levels can be developed. This allows for a more granular and structured classification of the data, taking into account different levels of attack categories.

\begin{figure}[!htbp]
    \centering
    \subfloat[Hierarchical model]{\includegraphics[scale=.55]{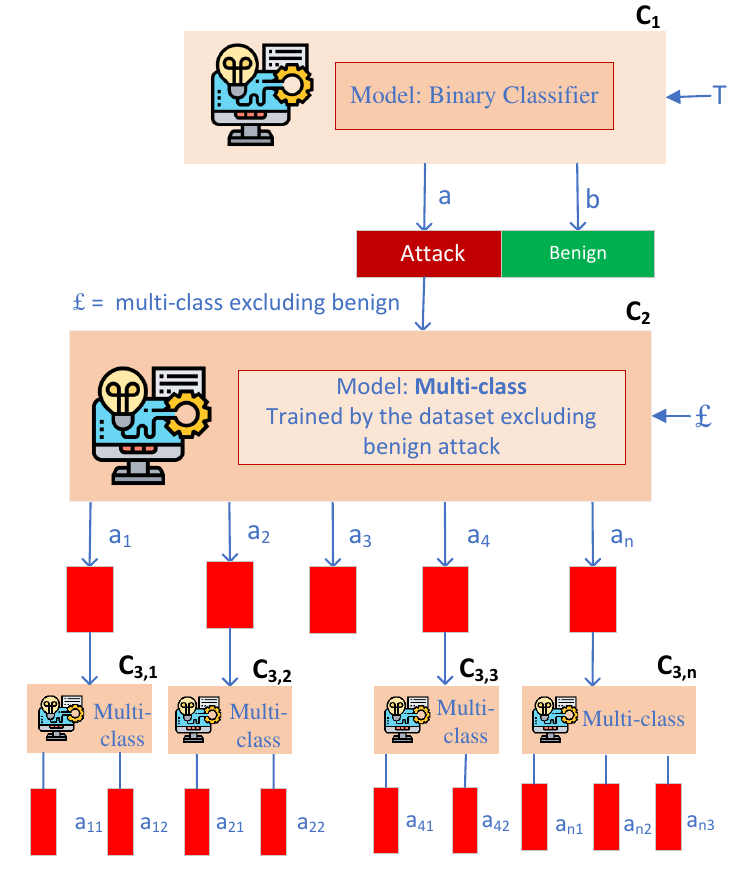}}
    \subfloat[Flat model]{\includegraphics[scale=.55]{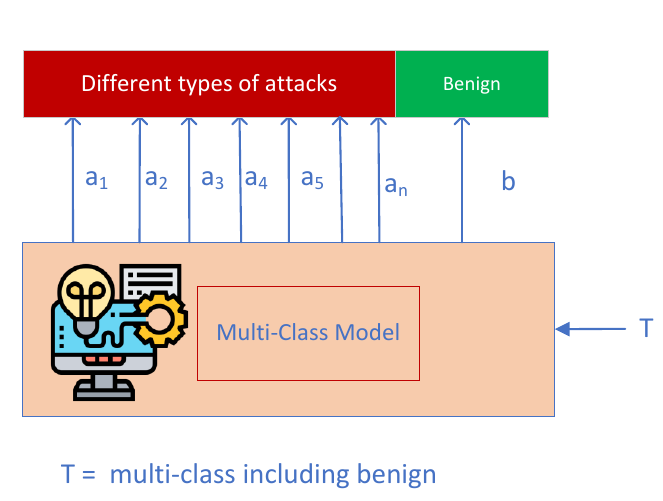}}    
    \caption{High level view of Hierarchical and Flat model}
    \label{fig:model}
\end{figure}

Figure \ref{fig:model} (a) and (b) show a model of hierarchical and flat classification approach respectively. In hierarchical classification, we train a binary classifier $C_1$ using the training dataset $T$ (which includes benign and attack traffic) at Level 1 to predict whether a traffic request is attack or normal. At Level 2, a multi-class classifier $C_2$ is trained using the training dataset of attacks only (excludes benign or normal traffic) to predict categories for attacks. Similarly, at Level 3, a separate multi-class classifier $C_{3,i}$ is trained for each attack category $i={1,2,\cdots,n}$ ($n$ is the number of attack categories/families in the given datasets) to predict specific attack type in the category. The number of classifiers required at third level depends on the number of attack categories (Level 2 labels)in the given IDS dataset. In the testing phase, a test data is classified at different levels using classifiers in a branch from the root to the leaf in the hierarchical structure: (i) attack vs normal using classifier $C_1$ at Level 1; (ii) attack family at Level 2 using classifier $C_2$ if it is predicted as attack by $C_1$; and (iii) specific attack type at Level 3 using classifier $C_{3,i}$ if it is predicted as of attack family $i$ by $C_2$. while flat model depicted in Figure \ref{fig:model} (b) identifies whether the test data is normal or of a particular type of attack directly using the single multi-class classifier trained to differentiate them all at the same time.

\begin{figure}[!htbp]
    \centering
    \includegraphics[scale = .55]{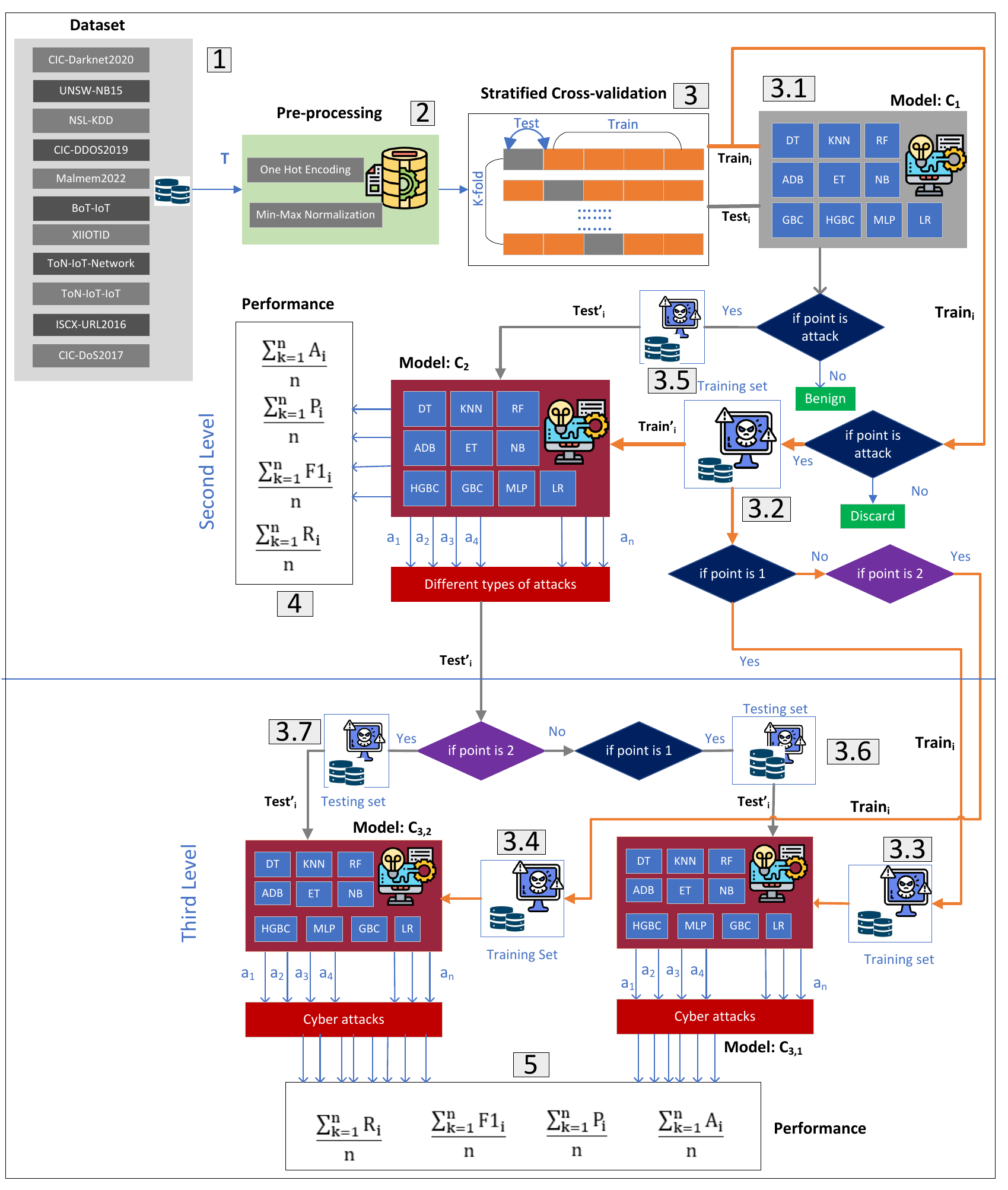}
    \caption{Training and testing phase of the two hierarchical model}
    \label{fig:1}
\end{figure}

Figure \ref{fig:1} presents a three-level hierarchical model that we implement in this article where we apply ten different machine learning algorithms in $10$ different datasets. The hierarchical model implemented in this study is described briefly below. Numbers in the square bracket at the start of each item refer to the relevant block in Figure \ref{fig:1}.   

\begin{itemize}
    \item {\bf [1]}Ten different IDS datasets to conduct the experiment are NSL-KDD, UNSW-NB15, CIC-DoS2017, CIC-DDoS2019, CIC-Darknet2020, CIC-MalMem2022, XIIOTID, ToN-IoT-Network, ToN-IoT-IoTs, BoT-IoT and ISCX-URL2016.
    
    \item {\bf [2]}Every dataset is passed through minimal pre-processing, such as converting categorical values to numerical values using one-hot encoding and normalising the features' values using min-max techniques.
    
    \item {\bf [3]}We apply $10$-fold stratified cross-validation for training and testing the model for every dataset. In every fold of the stratified cross-validation, the following steps are performed.
    \begin{itemize}
    
        \item[$\circ$] {\bf [3.1]} Firstly, the classifier $(C_1)$ which is a binary classifier at the first level is trained with all data instances labeled as either benign or attack. We used 10 different classifiers - Gaussian Naive Bayes (GNB), Random Forest (RF), Logistic Regression (LR), Gradient Boosting Classification (GBC), Extreme Tree (ET), Histogram Gradient Boosting Classifier (HGBC), K-Nearest Neighbour( KNN), Multilayer Perceptron( MLP), and Decision Tree (DT).

        \item[$\circ$] {\bf [3.2]} Secondly, we create a training set by removing the normal or benign points from the original training data. This dataset excluding benign instances is used to train a multi-class classifier called $C_2$ at the second level. At this level attack family used as the class label.
        
        \item[$\circ$] {\bf [3.3, 3.4]} Thirdly, we form training sets for the third level classifiers $(C_{3,1})$, $(C_{3,2}), \cdots, (C_{3,n})$ for $n$ families of attacks at this level. 
        
        \item[$\circ$] {\bf [3.5]} Fourth, the testing phase begins with testing the classifier $C_1$ at the first level. If a data instance is identified as an attack by $C_1$, the instance is predicted by the classifier $C_2$ at the second level for recognizing the attack family. 
        
        \item[$\circ$] {\bf [3.6, 3.7]} Next, if a data point is identified as attack family $i$, the data point is passed to the classifier $C_{3,i}$ at the third level to predict its specific attack type. 

    \end{itemize}
    
    \item {\bf [4, 5]} Finally, performance is measured using various performance metrics such as accuracy, precision, recall, and f1score at level 1, 2 and 3. Moreover, a comparison analysis is conducted between flat and hierarchical binary (Level 1), category (Level 2) and subcategory (Level 3) classes.    
 
\end{itemize}

\subsection{Taxonomy of IDS Dataset}

\begin{figure}[!htbp]
                \centering
                \includegraphics[scale = .65]{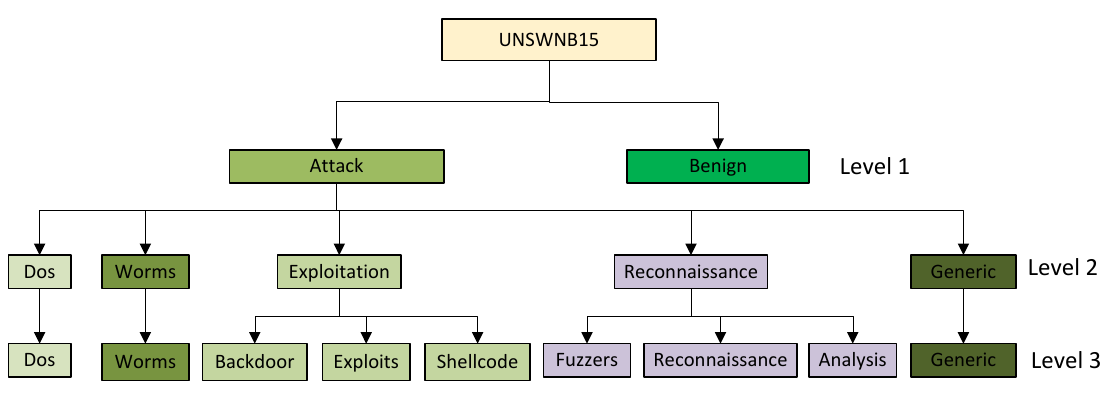}
                \caption{Hierarchical level of UNSW-NB15 dataset}
                \label{fig:unswn-b15}
        \end{figure}

Hierarchical classification involves grouping similar classes together in a hierarchical structure to enable efficient classification of new instances. In the context of cyber-attacks, similar attacks can be grouped together based on their operation, behaviour, or purpose, and then further sub-grouped into more specific categories. This allows for a more efficient and accurate classification of new instances based on their similarities to known attack patterns. In this work, we follow "Cyber Kill Chain" framework to group attacks to form the taxonomy of attacks for the datasets that do not contain level 3 attack classification. Martin et al.\cite{mihai2014cyber} designed a common framework called “Kill Chain” to comprehend and analyse the different phases of a cyberattack. For example, in UNSW-NB15, subcategories or level 3 categories are not provided. We considered given categories as Level 3 or subcategories and made level 2 by following "Kill Chain" framework from given categories in the dataset. We group Exploits, Backdoor and Shellcode as Exploitation attack. Backdoor, Exploits and Shellcode are malware-related attacks used to exploit vulnerabilities in computer systems and gain unauthorized access to sensitive data. Exploits are a type of attack that exploits vulnerabilities in a system or application and can be classified as exploit attacks. A generic attack is a common method of exploiting weaknesses in computer systems. Analysis and Fuzzers attacks refer to gathering information about a target and are typically classified as reconnaissance attacks used to test and evaluate a system’s security. Figure \ref{fig:unswn-b15} shows the level 2 and level 3 categories of this dataset. 

In this section, we present level 2 (attack family) and level 3 (specific attack types) categories of 10 different popular IDS dataset to train and test our hierarchical model. The attack family and the attack types of each dataset used in this study is presented in Table \ref{tab:datasetsubcategory}. The last column presents the ratio of attack instances count and total instances count in a datasets. 
\begin{table}[!htbp]
\centering
\caption{Dataset category, subcategory and attack class count}
\label{tab:datasetsubcategory}
\begin{tabular}{|l|l|l|}
\hline
\multicolumn{1}{|c|}{Dataset} &
  \multicolumn{1}{c|}{Attack family and attack types} &
  \multicolumn{1}{c|}{\begin{tabular}[c]{@{}c@{}}Attack Ratio   \\ (Attack/Total)\end{tabular}} \\ \hline
NSL-KDD &
  \begin{tabular}[c]{@{}l@{}}DoS   (neptune,  smurf, back, teardrop, pod,   land  ), Probe(satan,ipsweep\\ ,portsweep ,nmap ,mscan  ,apache2, processtable, snmpguess , saint ,\\ mailbomb, snmpgetattack, httptunnel, named ,ps, sendmail , xterm, xlock \\ ,xsnoop , sqlattack ,udpstorm, worm ), R2L ( guess\_passwd, warezmaster\\ ,warezclient, multihop, imap, ftp\_write, phf, spy), U2R ( buffer\_overflow ,\\ rootkit, loadmodule , perl )\end{tabular} &
  71463/148517 \\ \hline
UNSW-NB15 &
  \begin{tabular}[c]{@{}l@{}}DoS, Exploitation (Exploits, Backdoor, Shellcode ), Generic,\\  Reconnaissance (Fuzzers, Reconnaissance, Analysis), Worms\end{tabular} &
  164673/257673 \\ \hline
CIC-DoS2017 &
  \begin{tabular}[c]{@{}l@{}}High volume attack (ddossim, hulk, goldeneye), low volume (slowread, \\ slowloris,  rudy  slowheaders,  slowbody)\end{tabular} &
  14114/222914 \\ \hline
CIC-DDoS2019 &
  \begin{tabular}[c]{@{}l@{}}Syn Flood, E\_UDP(UDP, UDP-lag , UDPLag),  R\_TCP (MSSQL, \\ DrDoS\_MSSQL), R\_TCP/UDP(DrDoS\_DNS, DrDoS\_SNMP, LDAP,  \\ DrDoS\_LDAP, Portmap, NetBIOS, DrDoS\_NetBIOS, WebDDoS) , \\ R\_UDP (DrDoS\_NTP,  TFTP,    DrDoS\_UDP )\end{tabular} &
  333540/431371 \\ \hline
ISCX-URL2016 &
  Phishing,   Spam, Malware, Defacement &
  22228/30009 \\ \hline
CIC-Darkent2020 &
  \begin{tabular}[c]{@{}l@{}}VPN(V\_VOIP,  V\_Browsing,  V\_Audio-Streaming ,  V\_P2P ,   \\ V\_Video-Streaming,  V\_File-Transfer,   V\_Chat , V\_Email), \\ Tor(T\_VOIP,    T\_Browsing,  T\_Audio-Streaming   ,  T\_P2P ,\\  T\_Video-Streaming,  T\_File-Transfer, T\_Chat , T\_Email)\end{tabular} &
  24311/141530 \\ \hline
Malmem2022 &
  \begin{tabular}[c]{@{}l@{}}Spyware(Transponder,   Gator,  180solutions, CWS , TIBS) , Ransomware\\ (Shade, Ako,  Conti,    Maze,  Pysa),  Trojan( Refroso, Scar,  Emotet ,\\  Zeus, Reconyc)\end{tabular} &
  29298/58596 \\ \hline
ToN-IoT-Network &
  \begin{tabular}[c]{@{}l@{}}DDoS, DoS,   Ransomware, Scanning, Password, Exploitation (Backdoor, \\ Injection, XSS,  MITM)\end{tabular} &
  161043/461043 \\ \hline
ToN-IoT-IoTs &
  \begin{tabular}[c]{@{}l@{}}DDoS, DoS,   Ransomware, Scanning, Password, Exploitation (Backdoor,\\  Injection, XSS,  MITM\end{tabular} &
  2836524/5351760 \\ \hline
XIIOTID &
  \begin{tabular}[c]{@{}l@{}}Reconnaissance(Scanning,   Generic , Discovering, Fuzzing), RDOS,\\ Weaponization(BruteForce, Insider Malcious,  Dictionary), Lateral \\ (MQTT,  Modbus, TCP), Lateral(MQTT,  Modbus, TCP),  Exfiltration,\\ Tampering, C\&C,  Exploitation (Rshell,  MITM)\end{tabular} &
  259418/595090 \\ \hline
BoT-IoT &
  \begin{tabular}[c]{@{}l@{}}DDoS(TCP,   UDP,  HTTP),  DoS(UDP, TCP, HTTP) , Reconnaissance\\ (Servic   Scan, O Fingerprint),  Theft ( Keylogging, Data Exfiltration)\end{tabular} &
  3668045/3668522 \\ \hline
\end{tabular}
\end{table}

\subsection{Data Pre-Processing}

The data processing steps are briefly described below. 

\begin{itemize}
    \item Encoding: We usually use various encoding systems such as label encoding, one hot encoding on the dataset before feeding data into machine learning algorithms. In this work, we apply one hot encoding to convert categorical features into numerical features. One-hot encoding is chosen over other encoding systems because this can capture nonlinear relationships between categorical variables and the target variable. However, one-hot encoding might not be optimal for datasets with numerous categories, since this can result in high dimensionality and memory problems.

    \item Data Scaling: We apply Min-Max normalization that eliminates the impact of  different feature's value in the datasets on the performance of machine learning algorithms. Min-Max approach scales each feature's values to a range between 0 and 1. The min-max normalisation formula is follows:
    $X_{norm} = \frac{X - X_{min}}{X_{max} - X_{min}}$
    Where $X_norm$ is the normalized value of $X$, $X_min$ is the minimum value of $X$, and $X_max$ is the maximum value of $X$.
   
   \item  Dataset Partition: Stratified cross-validation splits dataset into folds while preserving the balanced proportion of each class in each fold. The stratified cross-validation provides a more accurate estimate of model performance, particularly when working with imbalanced datasets in which one class is more samples than the other. In this study, we trained and evaluated both the flat and hierarchical models using stratified 10-fold cross-validation. We divided the dataset into ten folds of equal size, with each fold containing a proportional representation of the different classes. This process is repeated for all ten folds and reported the average classification performances. The both models acquire more reliable and accurate estimates of the performance by employing this cross-validation method.    
\end{itemize}

\section{Results and Discussion}
\label{RESULTS AND DISCUSSION}

In this section, we evaluate the effectiveness of a flat multi-classification model and a hierarchical multi-classification model. Since there are 10 distinct IDS datasets and presenting the performance metrics for all 10 classifiers for both models, including confusion matrices, precision, recall, and f1 scores for each dataset, is not feasible. Therefore, we provide the performance of the top performing classification algorithms for both hierarchical and flat model. Detailed results of all classifiers in all datasets are provided in the Supplementary document. 

\subsection{Experimental Setup and Implementation}

Our experimental study was performed on an Intel Xeon E5-2670 CPU (8 cores, 16 threads), 128GB DDR3 RAM, 2x Nvidia GTX 1080 Ti. Python 3.9 was used to execute our code. The study utilized ten different machine learning models and primarily relied on Pandas and NumPy libraries for data pre-processing. Since the framework was developed using Python, the widely recognized Scikit-learn toolkit was utilized to leverage its wide range of algorithms and resources for data scientists, including effective accuracy and precision estimation metrics. In this work, we employed machine-learning algorithms from Scikit-learn~\cite{pedregosa2011scikit}.

\subsection{Performance Metrics}

In this study, we used accuracy, precision, recall, and F1-score that are essential for assessing the performance of an IDS model. However, their significance can vary depending on the system's specific objectives and requirements. Accuracy quantifies the proportion of accurate classifications made by the IDS. However, relying solely on accuracy is not the most suitable performance metric for IDS, as this might not accurately reflect the system's capability to identify attacks, which are a minority class within the dataset. Precision refers to the proportion of genuine positive detection out of all positive detection. High precision is essential in IDS in order to minimise false negatives (normal incorrectly detected as attack), which can result in false alarms. Recall measures the system's ability to reliably identify all instances of a particular class of attack. Low recall suggests that the system is missing many attacks, which can pose a significant security risk.
The F1 score is a combination of precision and recall that quantifies the proportion of true positive identification relative to the total number of positive instances in the dataset. F1-score is a valuable metric for IDS because this considers both false positives and false negatives and provides a balanced score between precision and recall.







\subsection{Results of Hierarchical Model and Flat Model}

In this section, we compare the Hierarchical (HR) model and Flat (FL) model in terms of their weighted average accuracy, precision, recall, and F1 score for the top performing classification algorithms. Further, we assess the efficiency of the hierarchical model by examining its false negative rate, which represents the number of attacks incorrectly classified as benign. As a final step in our analysis, we compare the time complexity of both models before concluding the paper.

\subsubsection{Top Hierarchical Classifier Performance (Binary Target)}

\begin{figure}[!htbp]
   \centering
    \includegraphics[scale = .55]{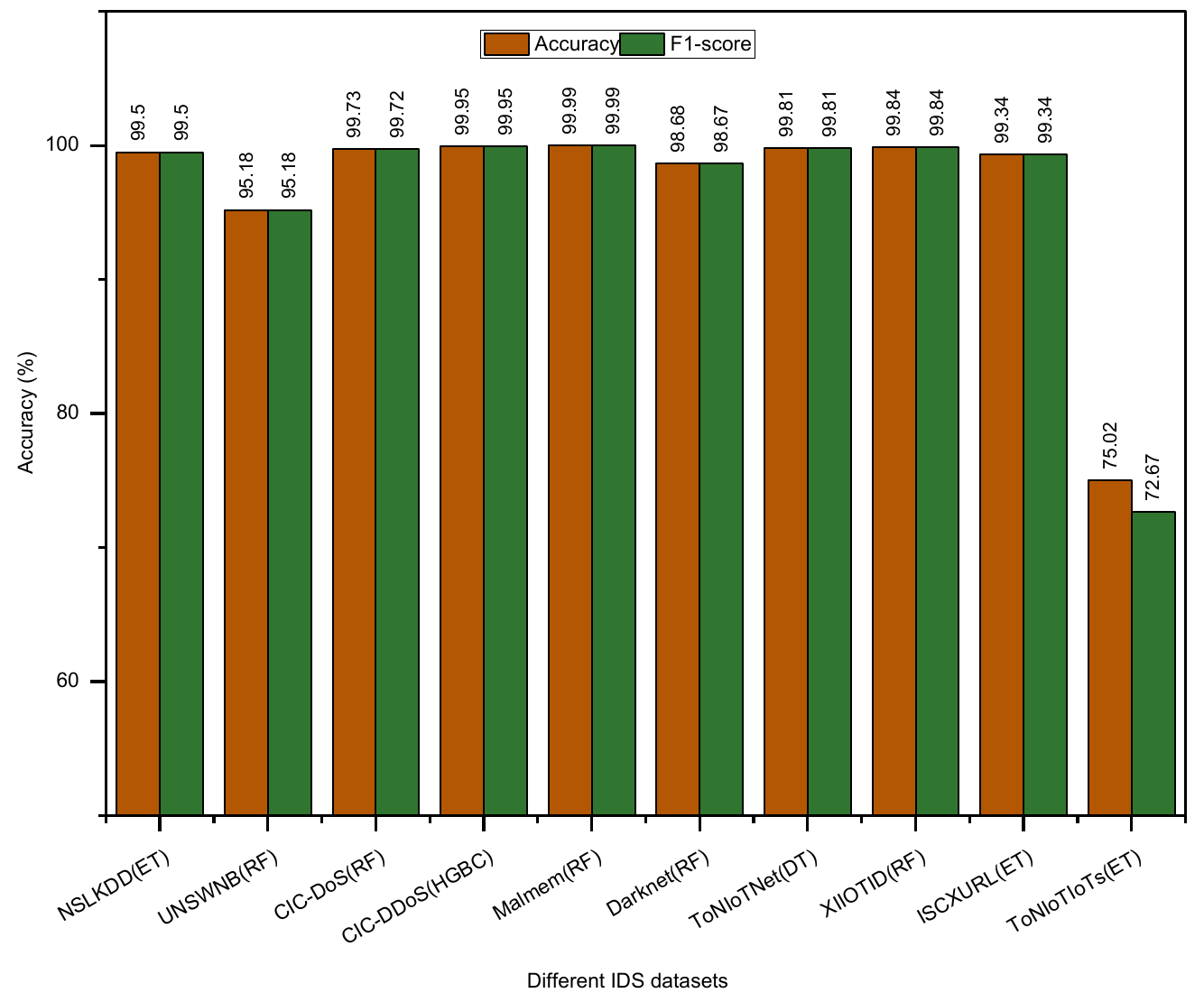}
    \caption{The accuracy and F1-score of top classifiers at level 1 on 10 datasets}
    \label{fig:bhraccuracyf1score}
\end{figure}

In HR model, we detect binary target (benign or attack) at its root level or level 1. Figure \ref{fig:bhraccuracyf1score} depicts the accuracy and F1-score of the best classifiers among 10 classifiers for each dataset using the HR model at level 1 for the binary target labelled as benign or attack. Classifiers are selected as the top performers based on their accuracy and F1-score in distinguishing between benign and attack instances in the respective datasets. The majority of datasets show excellent performance with high accuracy and F1-score, ranging from 95\% to near-perfect scores.

Graph in Figure \ref{fig:bhraccuracyf1score}  shows that  RF classifier commonly performs well on multiple datasets, indicating its effectiveness in classifying IDS dataset. The CIC\-DDoS dataset stands out with the HGBC classifier achieving higher accuracy and F1\-score of 99.95\%. The ToN-IoT- IoTs dataset shows relatively lower performance compared to others, suggesting the need for further improvements in detecting attacks in IoT scenarios.

\subsubsection{Top Classifier Performance: HR vs. FL Model (Level 2 and 3)}

 Table \ref{tab:hrflperformance} provides the performance results of the HR and FL models for the top-performing classifiers on 10 different IDS datasets. The analysis focuses on accuracy, precision, recall, and F1-score. RF classifier performs well for most of the datasets including UNSW-NB15, CIC-DoS2017, CIC-Darknet2020, and XIIOTID datasets. Table \ref{tab:hrflperformance} shows that the HR model performs well on datasets such as NSL-KDD, CIC-DoS2017, CIC-DDoS2019, CIC-Darknet2020, and XIIOTID datasets. On the other hand, the FL model shows better performance on datasets like UNSW- NB15, Malmem2022, ISCXURL2016, ToN-IoT-Network, and ToN-IoT-IoTs datasets.

\begin{table}[!htbp]
\centering
\caption{Top Performing Classifiers in HR and FL Models at Level 2 and Level 3}
\label{tab:hrflperformance}
\begin{tabular}{|c|c|c|cc|cc|cc|cc|}
\hline
\multirow{2}{*}{Dataset} & \multirow{2}{*}{Level} & \multirow{2}{*}{Classifier} & \multicolumn{2}{c|}{Accuracy} & \multicolumn{2}{c|}{Precision} & \multicolumn{2}{c|}{Recall} & \multicolumn{2}{c|}{F1-score} \\ \cline{4-11} 
 &  &  & \multicolumn{1}{c|}{HR} & FL & \multicolumn{1}{c|}{HR} & FL & \multicolumn{1}{c|}{HR} & FL & \multicolumn{1}{c|}{HR} & FL \\ \hline
\multirow{2}{*}{NSL-KDD} & 2 & ET & \multicolumn{1}{c|}{\textbf{99.47}} & 99.45 & \multicolumn{1}{c|}{\textbf{99.46}} & 99.45 & \multicolumn{1}{c|}{\textbf{99.47}} & 99.45 & \multicolumn{1}{c|}{\textbf{99.46}} & 99.45 \\ \cline{2-11} 
 & 3 & ET & \multicolumn{1}{c|}{\textbf{99.29}} & 99.27 & \multicolumn{1}{c|}{\textbf{99.25}} & 99.21 & \multicolumn{1}{c|}{\textbf{99.29}} & 99.27 & \multicolumn{1}{c|}{\textbf{99.26}} & 99.23 \\ \hline
\multirow{2}{*}{UNSW-NB15} & 2 & RF & \multicolumn{1}{c|}{83.44} & \textbf{83.59} & \multicolumn{1}{c|}{\textbf{82.52}} & \textbf{82.52} & \multicolumn{1}{c|}{83.44} & \textbf{83.59} & \multicolumn{1}{c|}{82.36} & \textbf{82.39} \\ \cline{2-11} 
 & 3 & RF & \multicolumn{1}{c|}{82.72} & \textbf{82.76} & \multicolumn{1}{c|}{82.41} & \textbf{82.75} & \multicolumn{1}{c|}{82.72} & \textbf{82.76} & \multicolumn{1}{c|}{81.29} & \textbf{81.74} \\ \hline
\multirow{2}{*}{CIC-DoS2017} & 2 & RF & \multicolumn{1}{c|}{\textbf{99.71}} & 99.70 & \multicolumn{1}{c|}{\textbf{99.71}} & 99.7 & \multicolumn{1}{c|}{\textbf{99.71}} & 99.70 & \multicolumn{1}{c|}{\textbf{99.7}} & \textbf{99.70} \\ \cline{2-11} 
 & 3 & RF & \multicolumn{1}{c|}{\textbf{99.66}} & 99.65 & \multicolumn{1}{c|}{\textbf{99.66}} & 99.65 & \multicolumn{1}{c|}{\textbf{99.66}} & 99.65 & \multicolumn{1}{c|}{\textbf{99.66}} & 99.64 \\ \hline
\multirow{2}{*}{CIC-DDoS2019} & 2 & RF & \multicolumn{1}{c|}{\textbf{96.81}} & \textbf{96.81} & \multicolumn{1}{c|}{\textbf{97.39}} & \textbf{97.39} & \multicolumn{1}{c|}{\textbf{96.81}} & \textbf{96.81} & \multicolumn{1}{c|}{\textbf{96.98}} & \textbf{96.98} \\ \cline{2-11} 
 & 3 & HGBC & \multicolumn{1}{c|}{\textbf{94.4}} & 93.31 & \multicolumn{1}{c|}{\textbf{94.42}} & 93.59 & \multicolumn{1}{c|}{\textbf{94.40}} & 93.31 & \multicolumn{1}{c|}{\textbf{94.03}} & 93.27 \\ \hline
\multirow{2}{*}{Malmem2022} & 2 & RF & \multicolumn{1}{c|}{87.90} & \textbf{87.95} & \multicolumn{1}{c|}{87.89} & \textbf{87.95} & \multicolumn{1}{c|}{87.9} & \textbf{87.95} & \multicolumn{1}{c|}{87.89} & \textbf{87.95} \\ \cline{2-11} 
 & 3 & RF & \multicolumn{1}{c|}{75.77} & \textbf{76.32} & \multicolumn{1}{c|}{75.62} & \textbf{76.2} & \multicolumn{1}{c|}{75.77} & \textbf{76.32} & \multicolumn{1}{c|}{75.57} & \textbf{76.17} \\ \hline
\multirow{2}{*}{CIC-Darknet2020} & 2 & RF & \multicolumn{1}{c|}{\textbf{98.66}} & 98.64 & \multicolumn{1}{c|}{\textbf{98.65}} & 98.63 & \multicolumn{1}{c|}{\textbf{98.66}} & 98.64 & \multicolumn{1}{c|}{\textbf{98.65}} & 98.63 \\ \cline{2-11} 
 & 3 & RF & \multicolumn{1}{c|}{\textbf{98.27}} & 98.24 & \multicolumn{1}{c|}{\textbf{98.23}} & 98.20 & \multicolumn{1}{c|}{\textbf{98.27}} & 98.24 & \multicolumn{1}{c|}{\textbf{98.14}} & 98.11 \\ \hline
\multirow{2}{*}{ISCXURL2016} & 2 & HGBC & \multicolumn{1}{c|}{98.55} & \textbf{98.64} & \multicolumn{1}{c|}{98.55} & \textbf{98.64} & \multicolumn{1}{c|}{98.55} & \textbf{98.64} & \multicolumn{1}{c|}{98.55} & \textbf{98.64} \\ \cline{2-11} 
 & 3 & HGBC & \multicolumn{1}{c|}{98.23} & \textbf{98.46} & \multicolumn{1}{c|}{98.24} & \textbf{98.46} & \multicolumn{1}{c|}{98.23} & \textbf{98.46} & \multicolumn{1}{c|}{98.23} & \textbf{98.46} \\ \hline
\multirow{2}{*}{ToN-IoT-Network} & 2 & ET & \multicolumn{1}{c|}{\textbf{99.53}} & \textbf{99.53} & \multicolumn{1}{c|}{\textbf{99.53}} & \textbf{99.53} & \multicolumn{1}{c|}{\textbf{99.53}} & \textbf{99.53} & \multicolumn{1}{c|}{\textbf{99.53}} & \textbf{99.53} \\ \cline{2-11} 
 & 3 & ET & \multicolumn{1}{c|}{\textbf{99.36}} & \textbf{99.36} & \multicolumn{1}{c|}{\textbf{99.38}} & 99.37 & \multicolumn{1}{c|}{\textbf{99.36}} & \textbf{99.36} & \multicolumn{1}{c|}{\textbf{99.37}} & \textbf{99.37} \\ \hline
\multirow{2}{*}{ToN-IoT-IoTs} & 2 & ET & \multicolumn{1}{c|}{73.53} & \textbf{73.56} & \multicolumn{1}{c|}{74.71} & \textbf{74.83} & \multicolumn{1}{c|}{73.53} & \textbf{73.56} & \multicolumn{1}{c|}{\textbf{70.67}} & \textbf{70.67} \\ \cline{2-11} 
 & 3 & ET & \multicolumn{1}{c|}{73.16} & \textbf{73.19} & \multicolumn{1}{c|}{74.02} & \textbf{74.15} & \multicolumn{1}{c|}{73.16} & \textbf{73.19} & \multicolumn{1}{c|}{\textbf{70.16}} & \textbf{70.16} \\ \hline
\multirow{2}{*}{XIIOTID} & 2 & RF & \multicolumn{1}{c|}{99.83} & \textbf{99.84} & \multicolumn{1}{c|}{99.83} & \textbf{99.84} & \multicolumn{1}{c|}{99.83} & \textbf{99.84} & \multicolumn{1}{c|}{99.83} & \textbf{99.84} \\ \cline{2-11} 
 & 3 & RF & \multicolumn{1}{c|}{99.83} & \textbf{99.84} & \multicolumn{1}{c|}{\textbf{99.83}} & \textbf{99.83} & \multicolumn{1}{c|}{99.83} & \textbf{99.84} & \multicolumn{1}{c|}{99.82} & \textbf{99.83} \\ \hline
\end{tabular}
\end{table}

Overall, RF and ET classifiers show promising performance in various IDS datasets across different levels and models. HR and FL models generally exhibit comparable results, with slight variations in accuracy, precision, recall, and F1-score.

\subsubsection{HR vs FL Model Performance in Attack Detection}

\begin{figure}[!htbp]
   \centering
    \includegraphics[scale = .60]{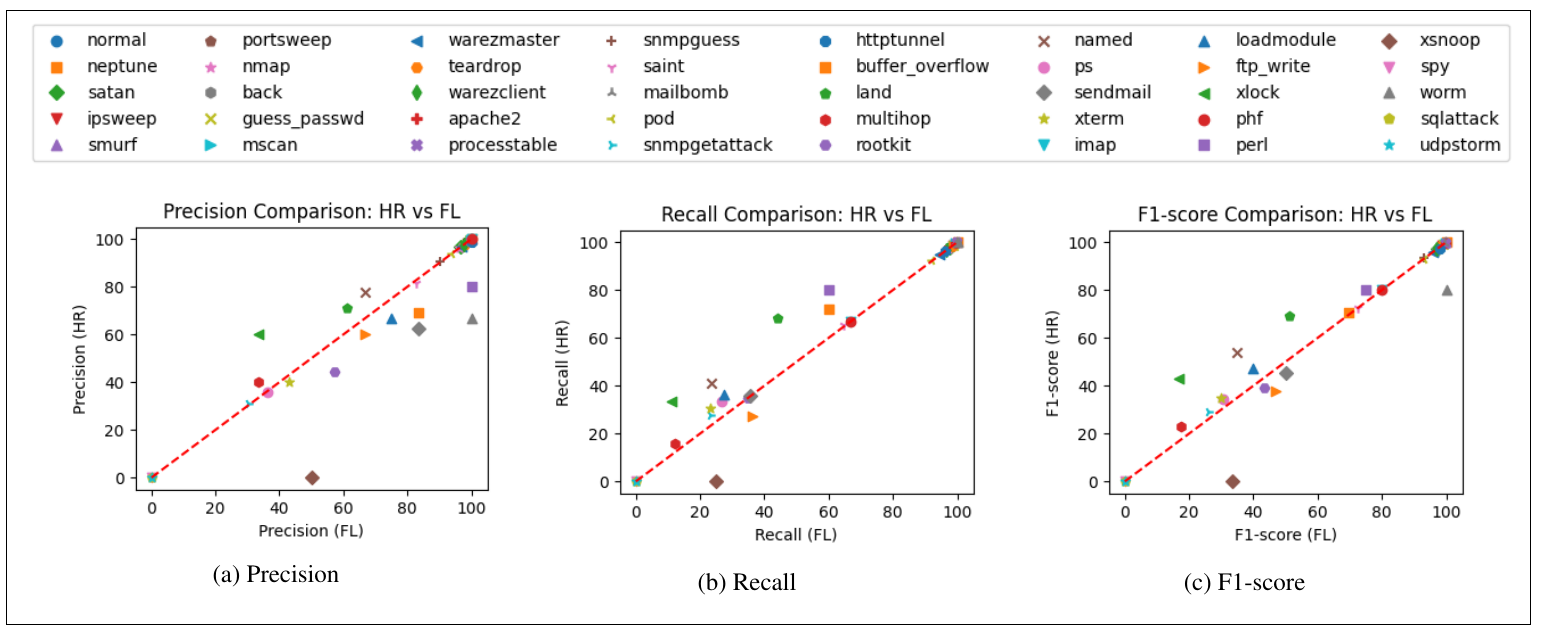}
    \caption{The difference of performance between HR and FL model on NSL-KDD's attacks}
    \label{fig:nslkddattack}
\end{figure}

\begin{figure}[!htbp]
   \centering
    \includegraphics[scale = .60]{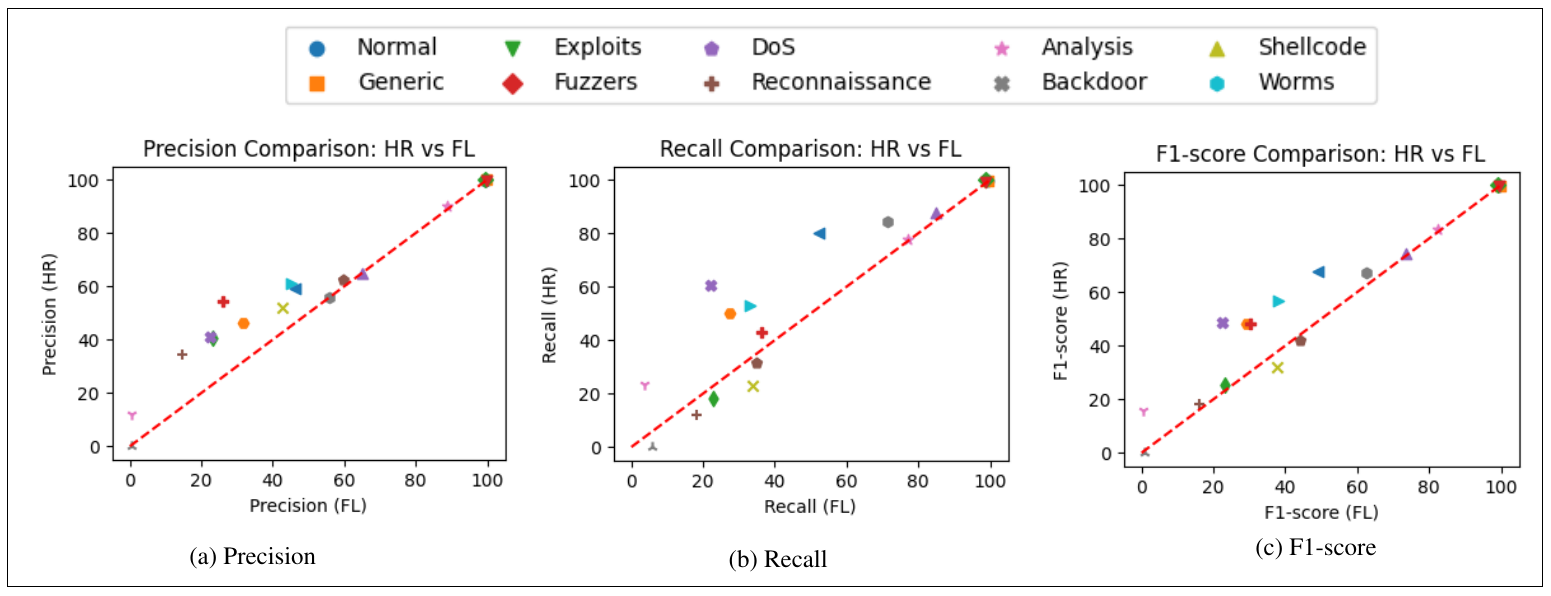}
    \caption{The difference of performance between HR and FL model on UNSW-NB15's attacks}
    \label{fig:unswnb15attack}
\end{figure}

In this section, we analyze the precision, recall and F1-score for each attack category at Level 3 for the top performing classifiers of the HR and FL models. We present results of the NSL-KDD, UNSW-NB15, CIC-DDOS2019 datasets in Figures \ref{fig:nslkddattack}, \ref{fig:unswnb15attack}, and \ref{fig:cicddosattack}, respectively. Results on other datasets are provided in the supplementary files. There are more points above the diagonal lines in all plots with an exception of precision in NSL-KDD. Consistently better recall and F1-score results in all datasets indicate the HR model detects more attack types correctly.

 \begin{figure}[!htbp]
   \centering
    \includegraphics[scale = .550]{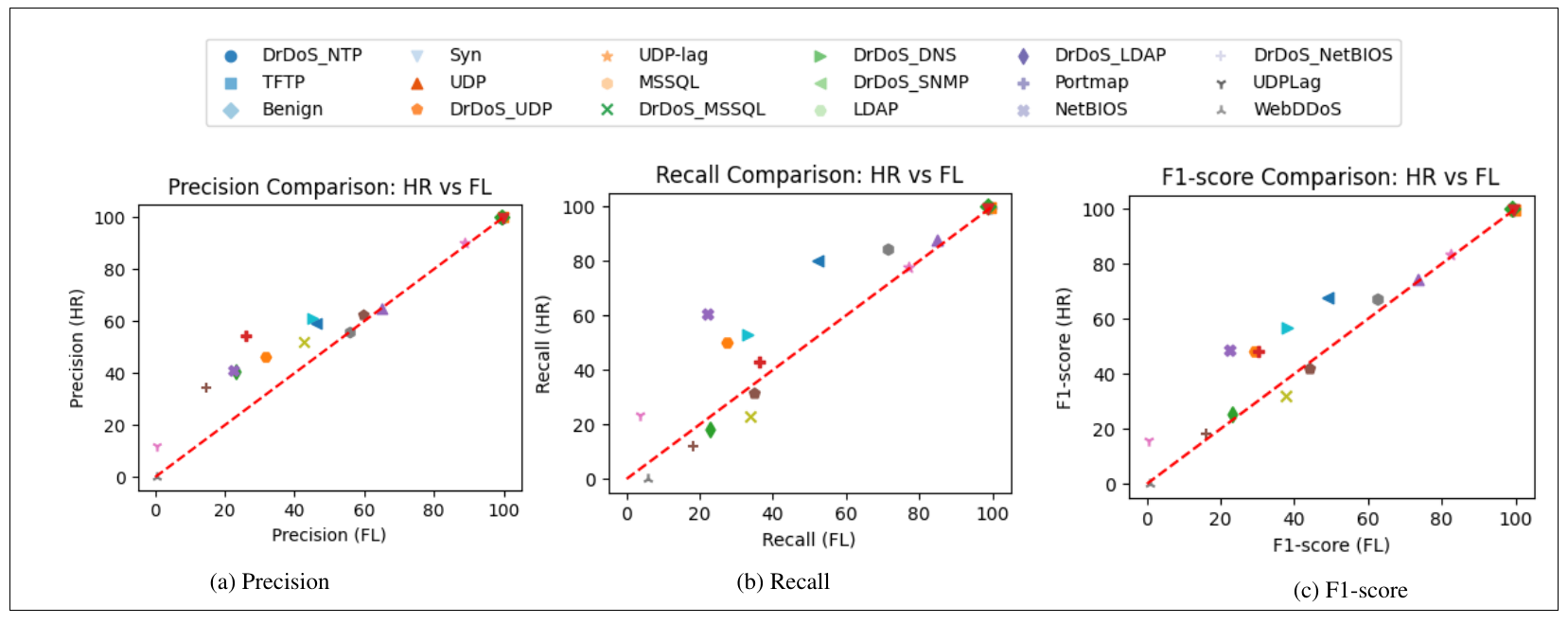}
    \caption{The difference of performance between HR and FL model on CIC-DDoS2019's attacks}
    \label{fig:cicddosattack}
\end{figure} 

\subsubsection{Model Comparison: False Negative Rate}

In this section, we examine the false negative rates of the top-performing classifiers in both the HR and FL models across the majority of the dataset. In the context of cybersecurity, misidentifying an attack as benign poses a greater risk than erroneously classifying a benign activity as an attack. Therefore, we assess the effectiveness of the HR model and FL model by analyzing the number of attacks incorrectly identified as benign. In this case, we also choose NSL-KDD, UNSW-NB15 and CIC-DDoS2019 datasets and results for other datasets are provided in the supplementary documents.  

 \begin{figure}[!htbp]
   \centering
    \includegraphics[scale = .72]{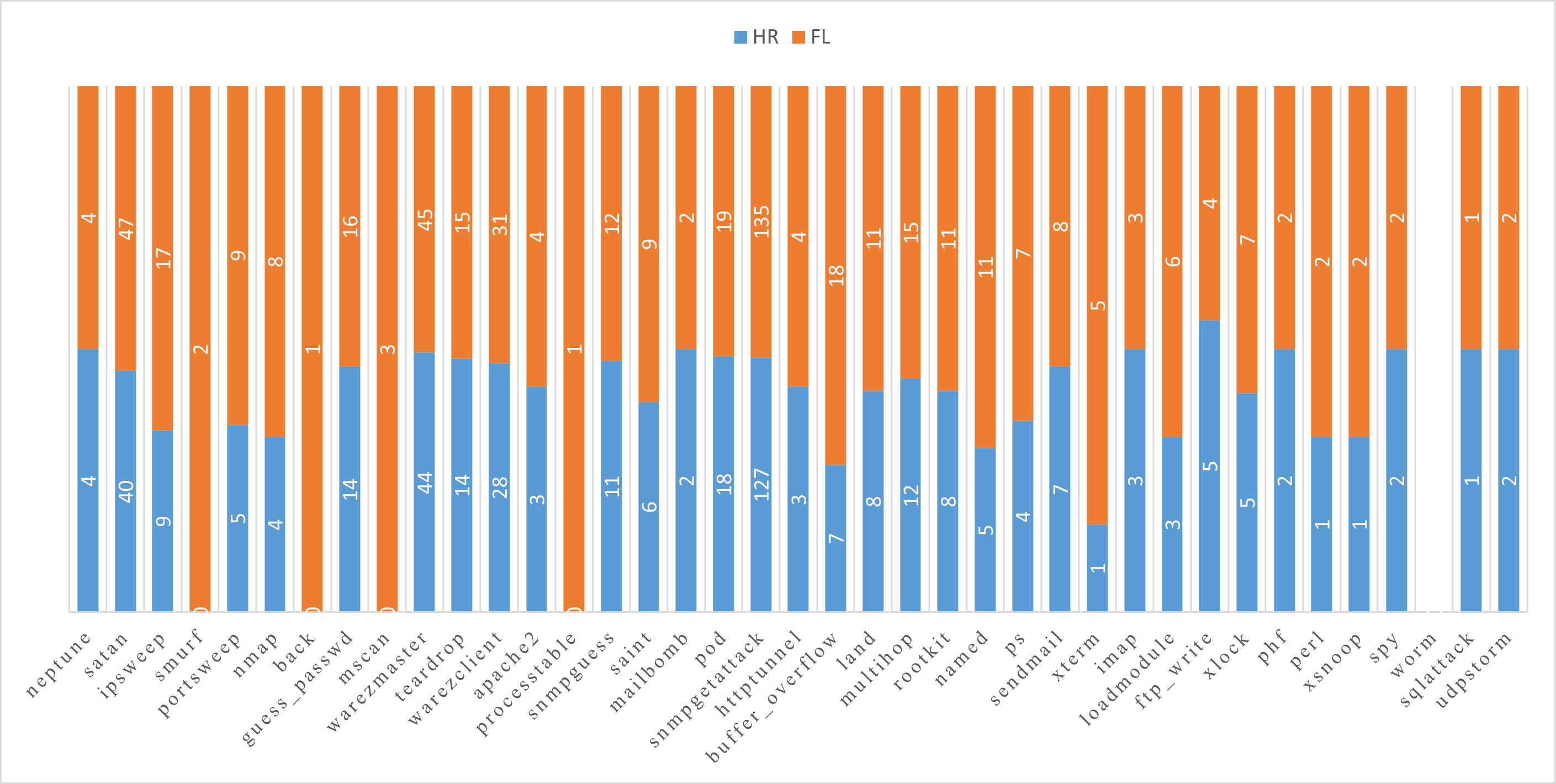}
    \caption{Number of attacks classified as normal (false negative classification of attacks) in the NSL-KDD dataset}
    \label{fig:nslkddnormalattack}
\end{figure}

\begin{figure}[!htbp]
    \centering
    \subfloat[UNSW-NB15]{\includegraphics[scale=.60]{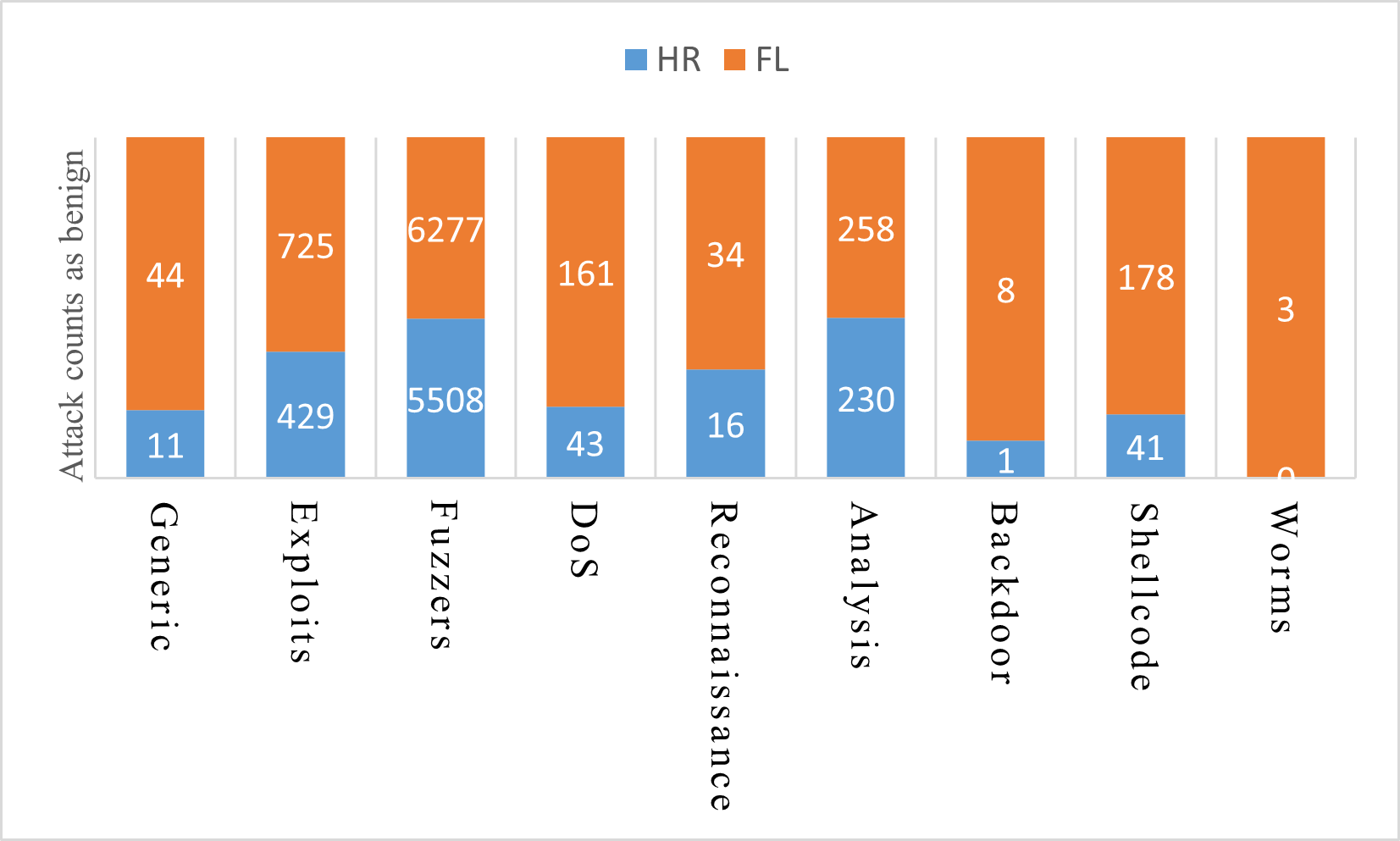}} 
     \subfloat[CIC-DDoS2019]{\includegraphics[scale=.60]{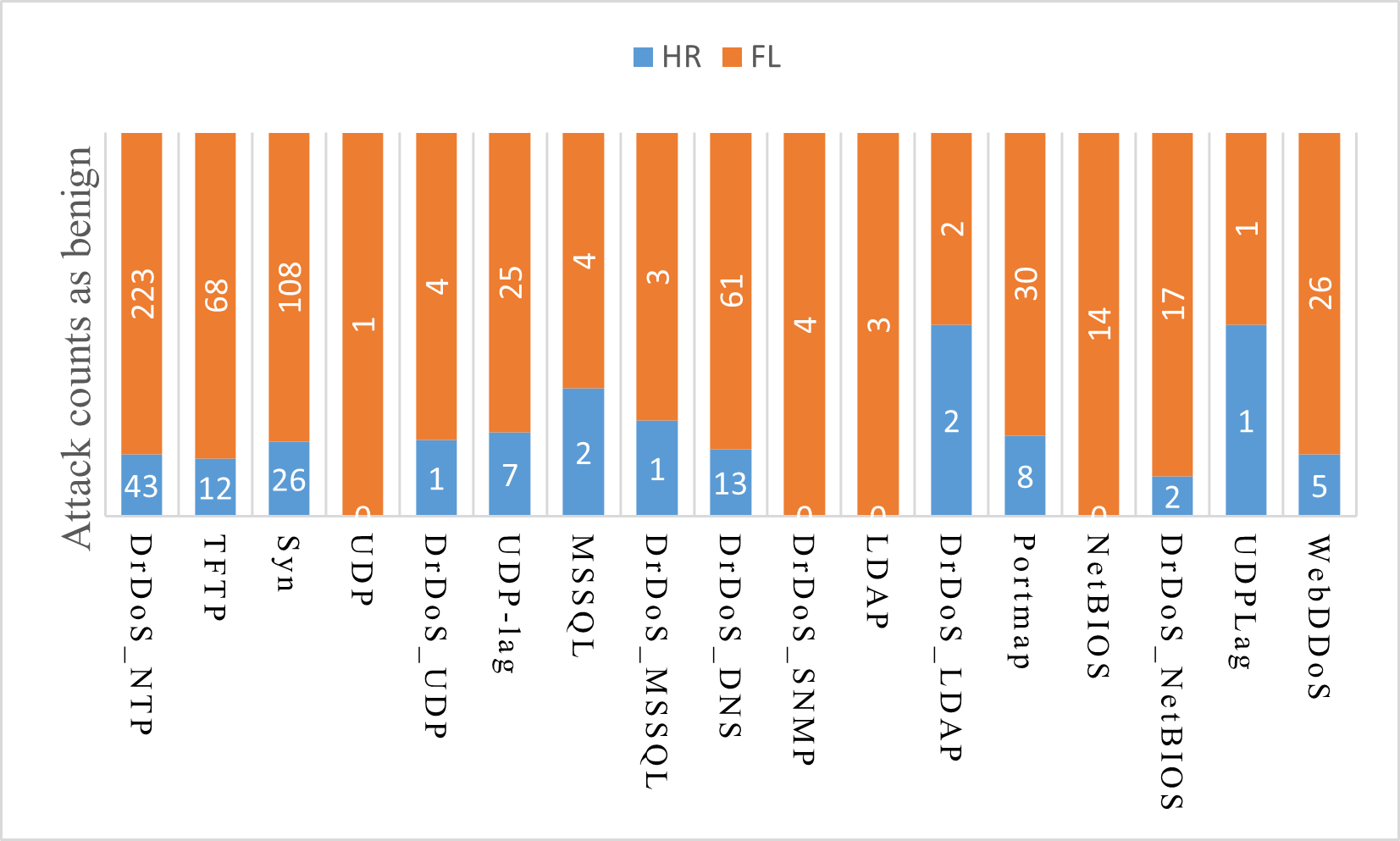}}   
    \caption{The false negative rate of HR and FL model}
    \label{fig:normalattack}
\end{figure}

NSL-KDD: Figure \ref{fig:nslkddnormalattack}, \ref{fig:normalattack} (a) and (b) represents the number of attacks that are being identified as benign categories (normal) for both the HR model (ET) and the FL model (ET). The results show that the HR model produces a lower number of attacks being identified as normal, which indicates a better performance of the HR model in distinguishing between normal and malicious activities. As seen in all plots, the number of attacks falsely classified as normal by the HR model is consistently less than those by the FL model. For example, for CIC-DDoS2019 dataset depicted in Figure \ref{fig:normalattack} (c), the HR model (HGBC) and FL model (HGBC) demonstrate notable differences in the number of attacks identified as benign for several attack types. Specifically, significant differences can be observed for the following attacks: DrDoS\_NTP (43 vs. 223), TFTP (12 vs. 68), Syn (26 vs. 108), UDP-lag (7 vs. 25), DrDoS\_DNS (13 vs. 61), Portmap (8 vs. 30), NetBIOS (0 vs. 14), DrDoS\_NetBIOS (2 vs. 17), and WebDDoS (5 vs. 26).

\begin{figure}[!htbp]
    \centering
    \includegraphics[scale = .45]{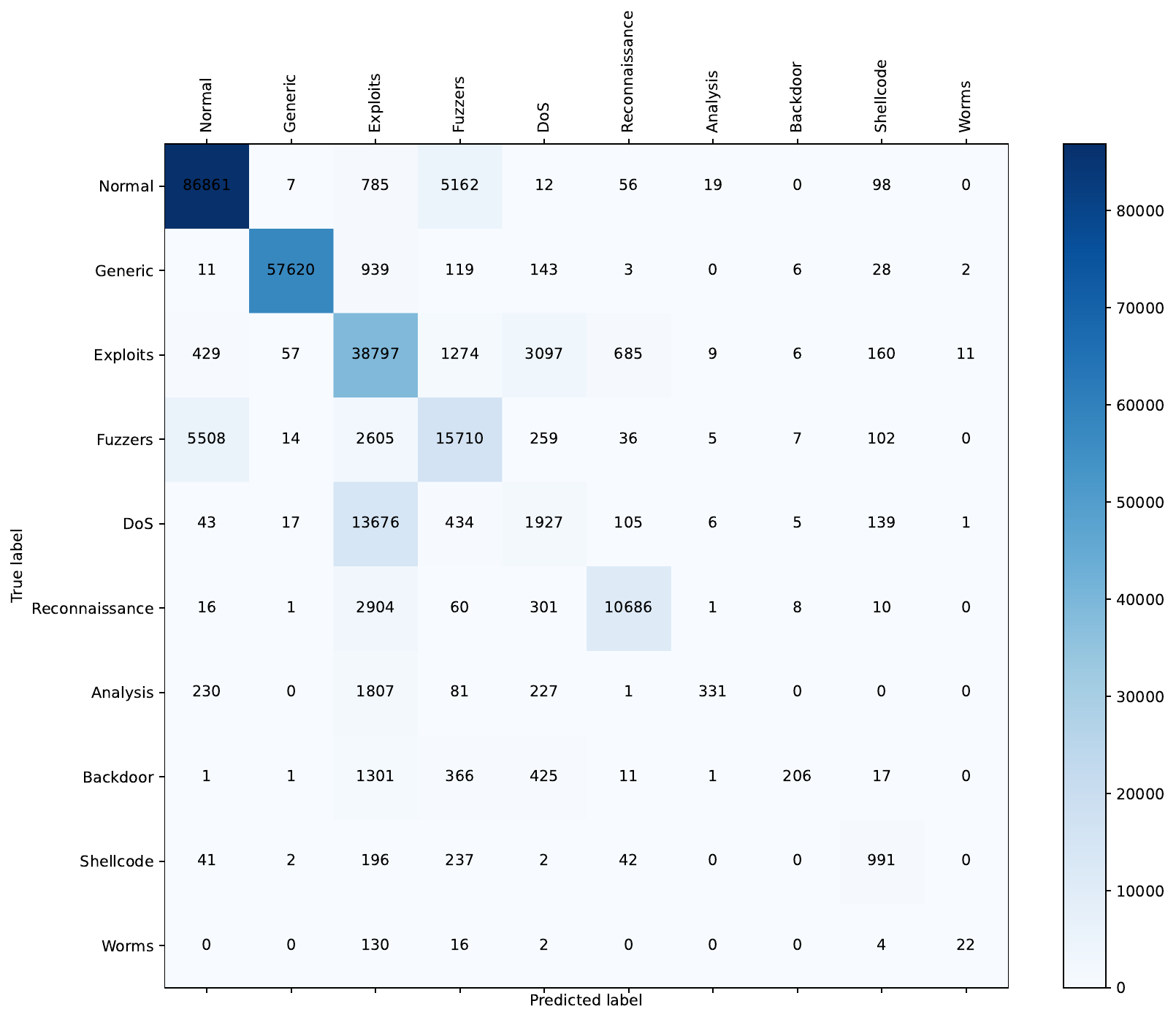}
    \caption{The confusion matrix of the best classifiers in HR (RF) model at level 3 on UNSW-NB15 dataset}
    \label{fig:hrrfunswnb15}
\end{figure}
\begin{figure}[!htbp]
    \centering
    \includegraphics[scale = .450]{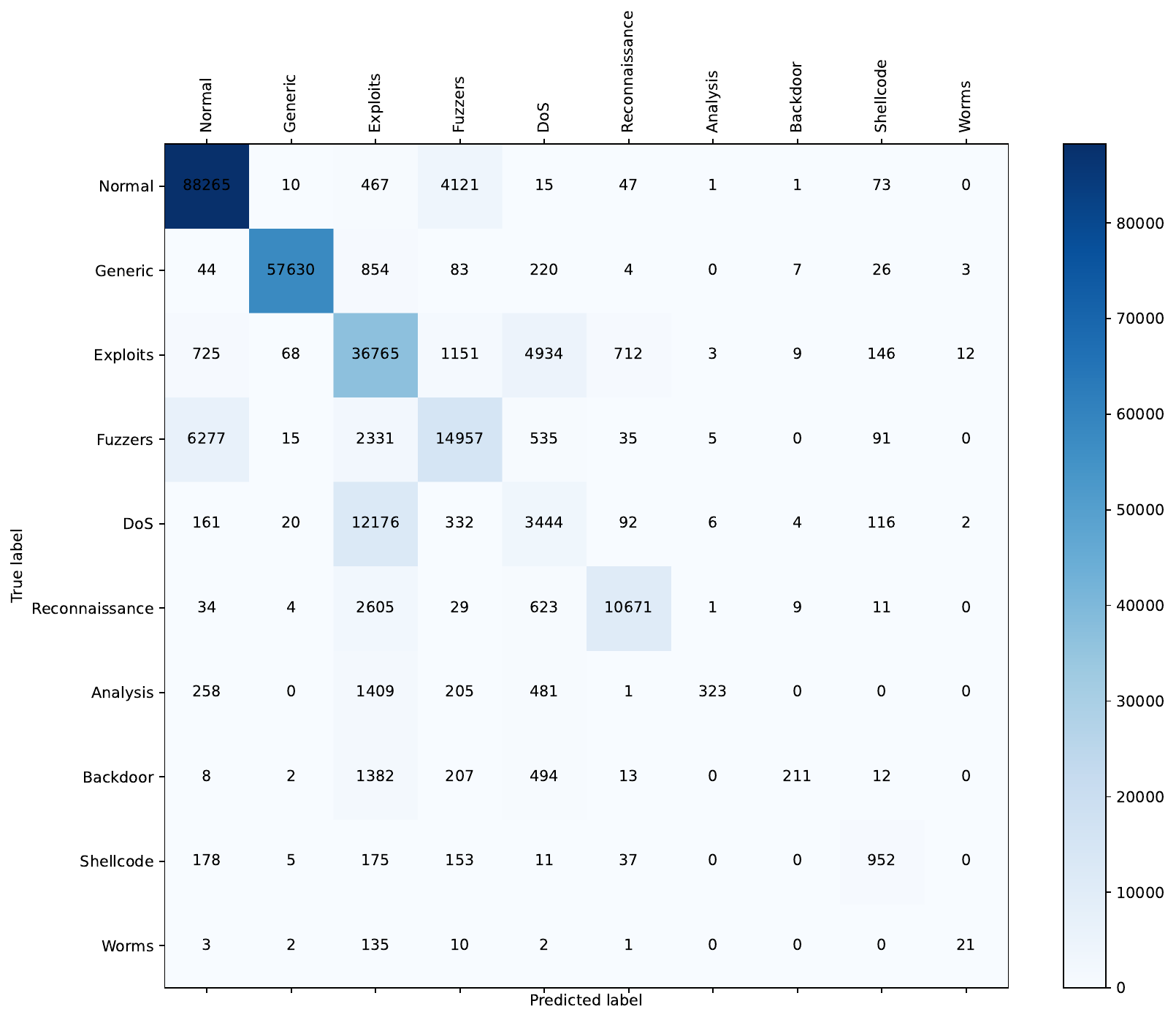}
    \caption{The confusion matrix of the best classifiers in FL (RF) model at level 3 on UNSW-NB15 dataset}
    \label{fig:flrfunswnb15}
\end{figure}
\begin{figure}[!htbp]
    \centering
    \includegraphics[scale = .350]{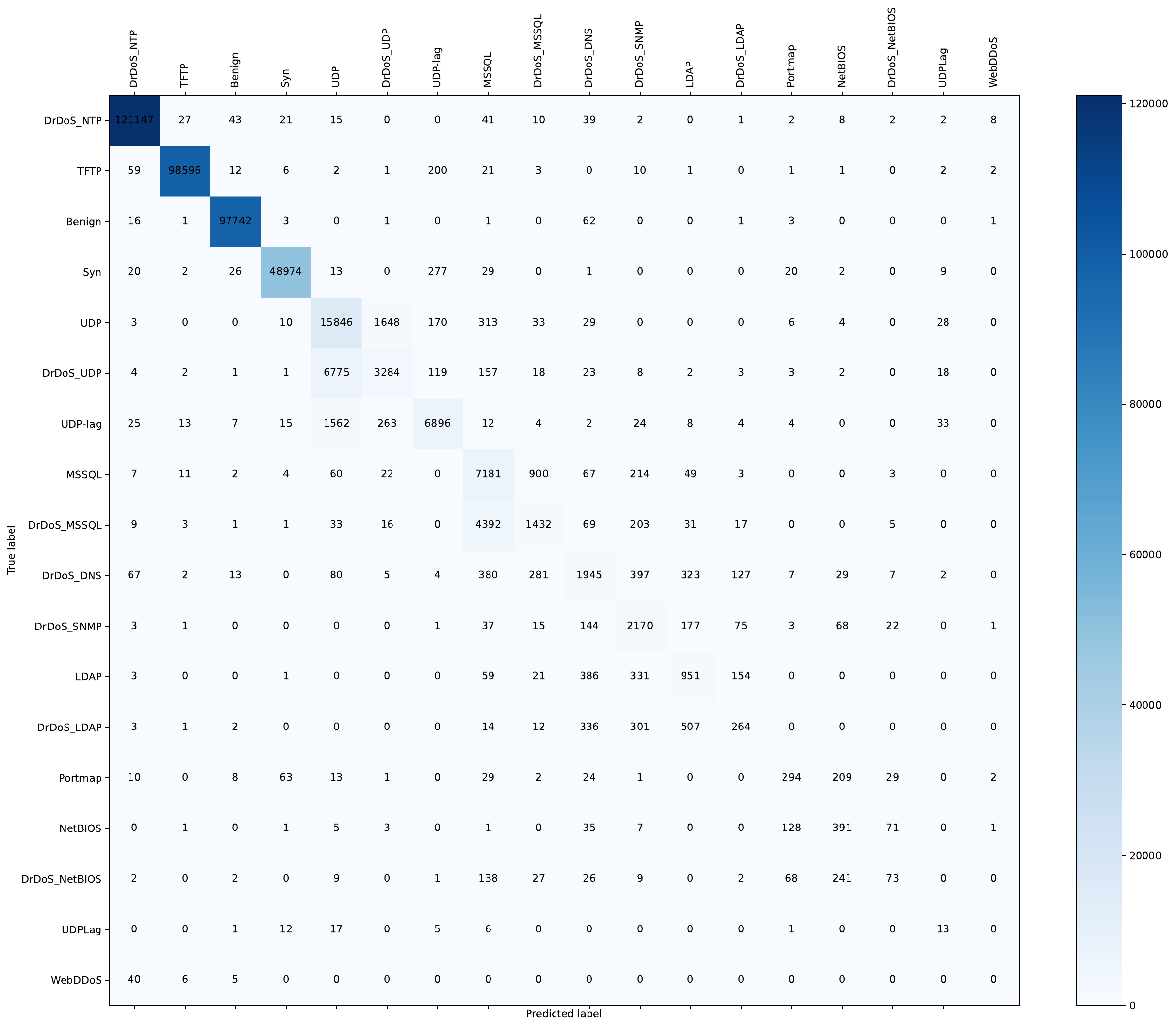}
    \caption{The confusion matrix of the best classifiers in HR (RF) model at level 3 on CIC-DDoS2019 dataset}
    \label{fig:hrhgbcCICDDOS}
\end{figure}
\begin{figure}[!htbp]
    \centering
    \includegraphics[scale = .350]{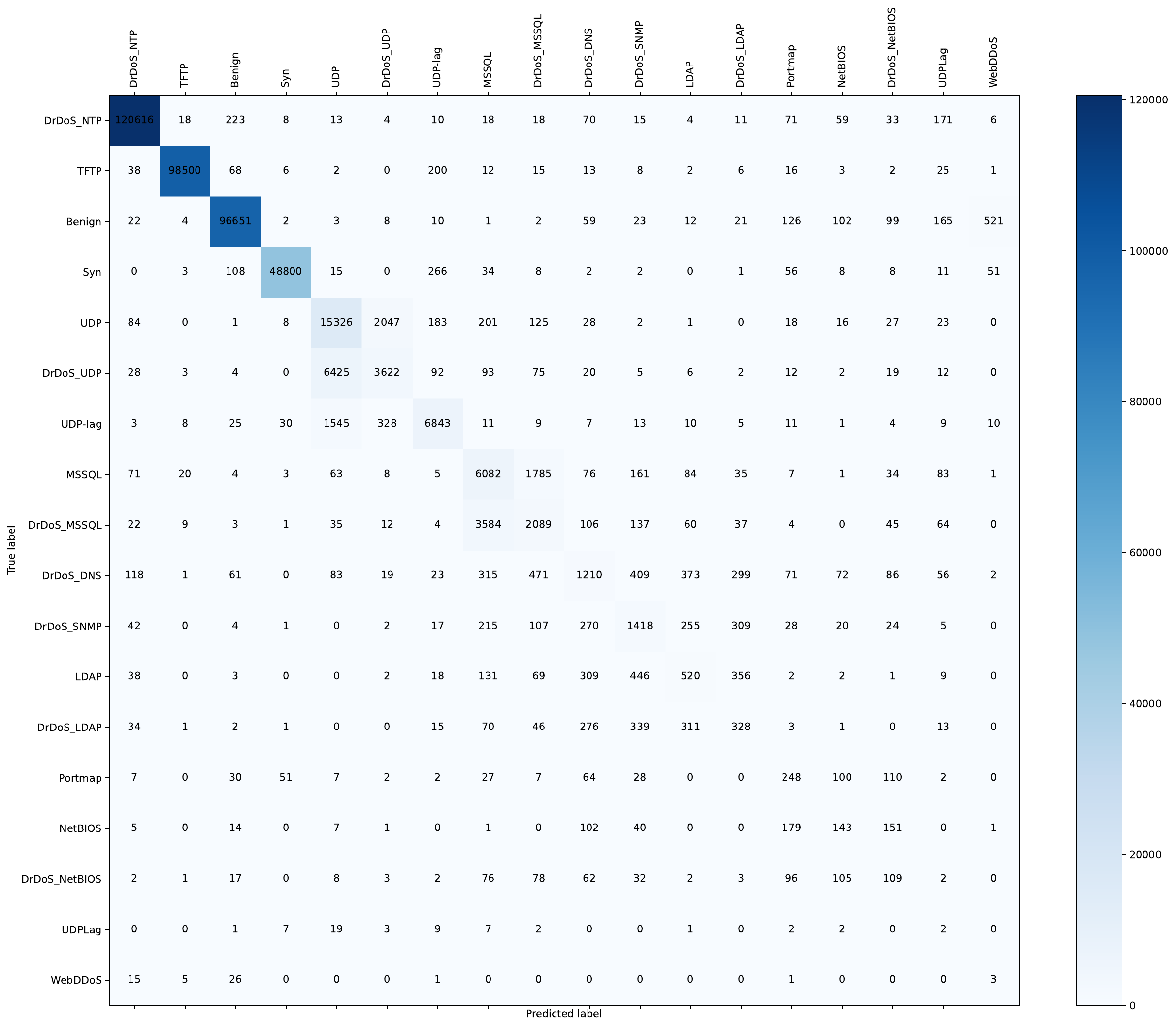}
    \caption{The confusion matrix of the best classifiers in FL (RF) model at level 3 on CIC-DDoS2019 dataset}
    \label{fig:flhgbcCICDDOS}
\end{figure}

Figure \ref{fig:hrrfunswnb15}, \ref{fig:flrfunswnb15}, \ref{fig:hrhgbcCICDDOS}, and \ref{fig:flhgbcCICDDOS} depict the confusion matrices of UNSW-NB15 and CIC-DDoS2019 datasets for both HR and FL models, serving as representatives.

\subsubsection{Model Comparison: False Positive}

\begin{figure}[!htbp]
   \centering
    \includegraphics[scale = .72]{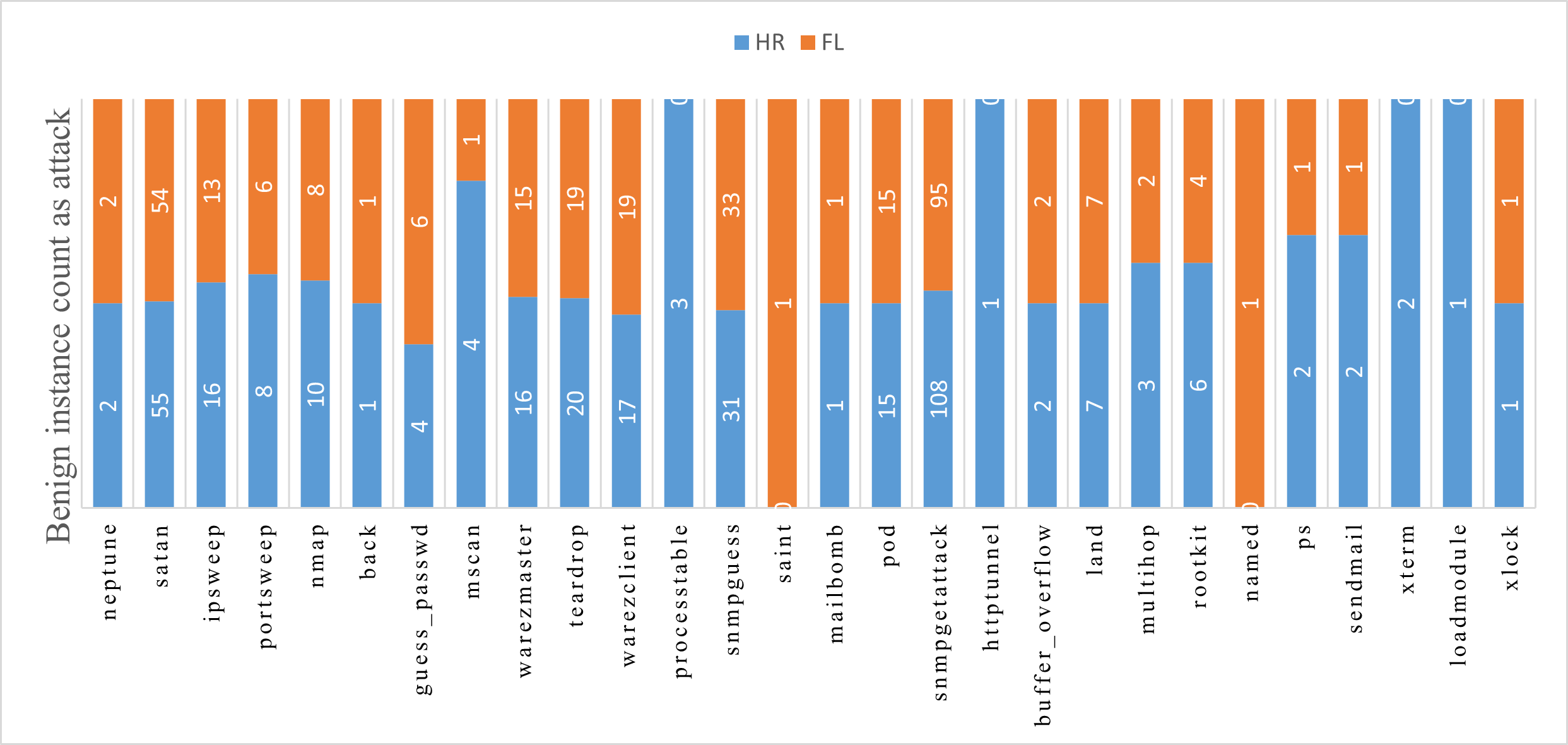}
    \caption{The false positive rate of HR and FL model on NSL-KDD's attacks}
    \label{fig:falsepositivenslkdd}
\end{figure}

\begin{figure}[!htbp]
    \centering
    \subfloat[UNSW-NB15]{\includegraphics[scale=.550]{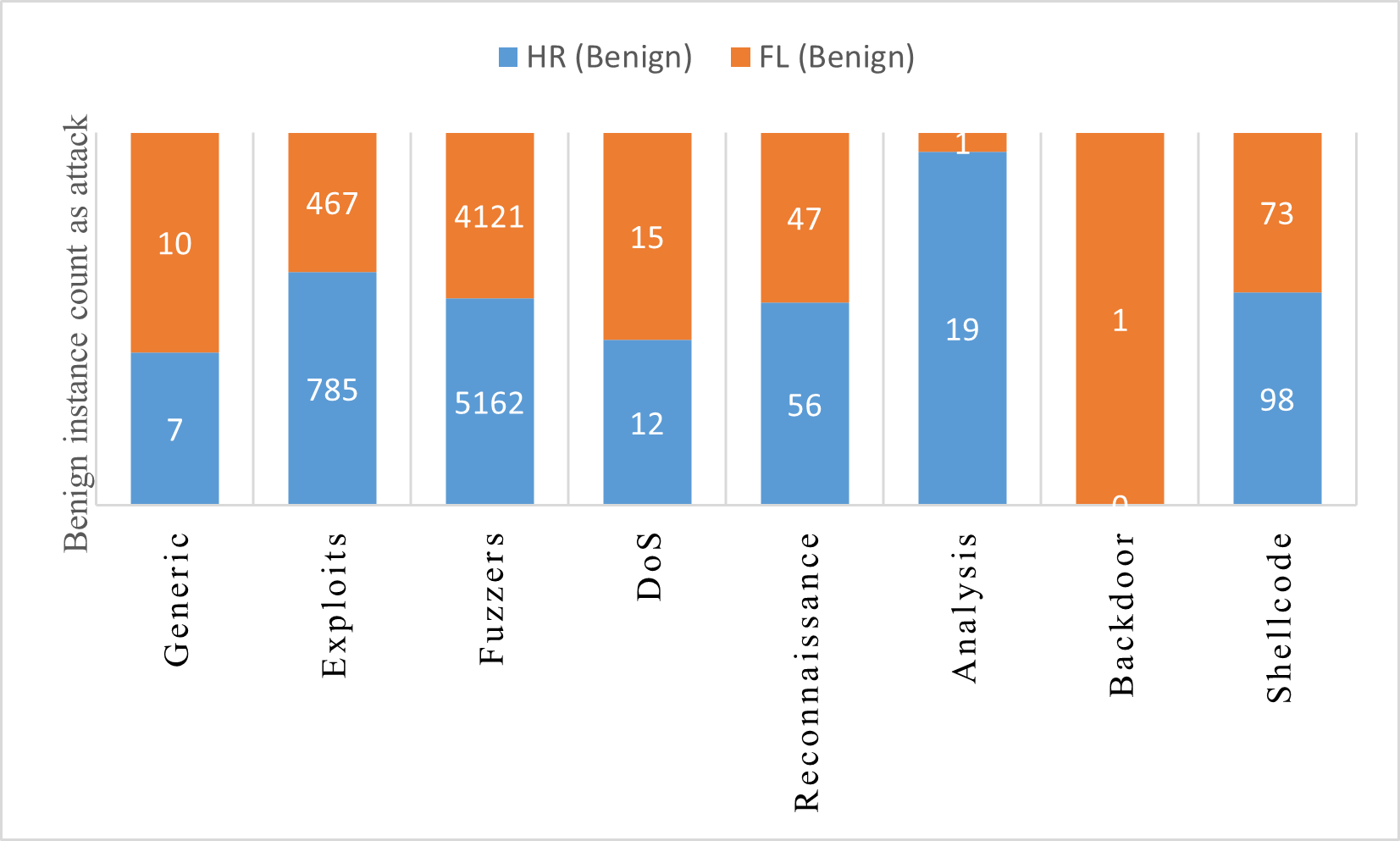}}
     \subfloat[CIC-DDoS2019]{\includegraphics[scale=.550]{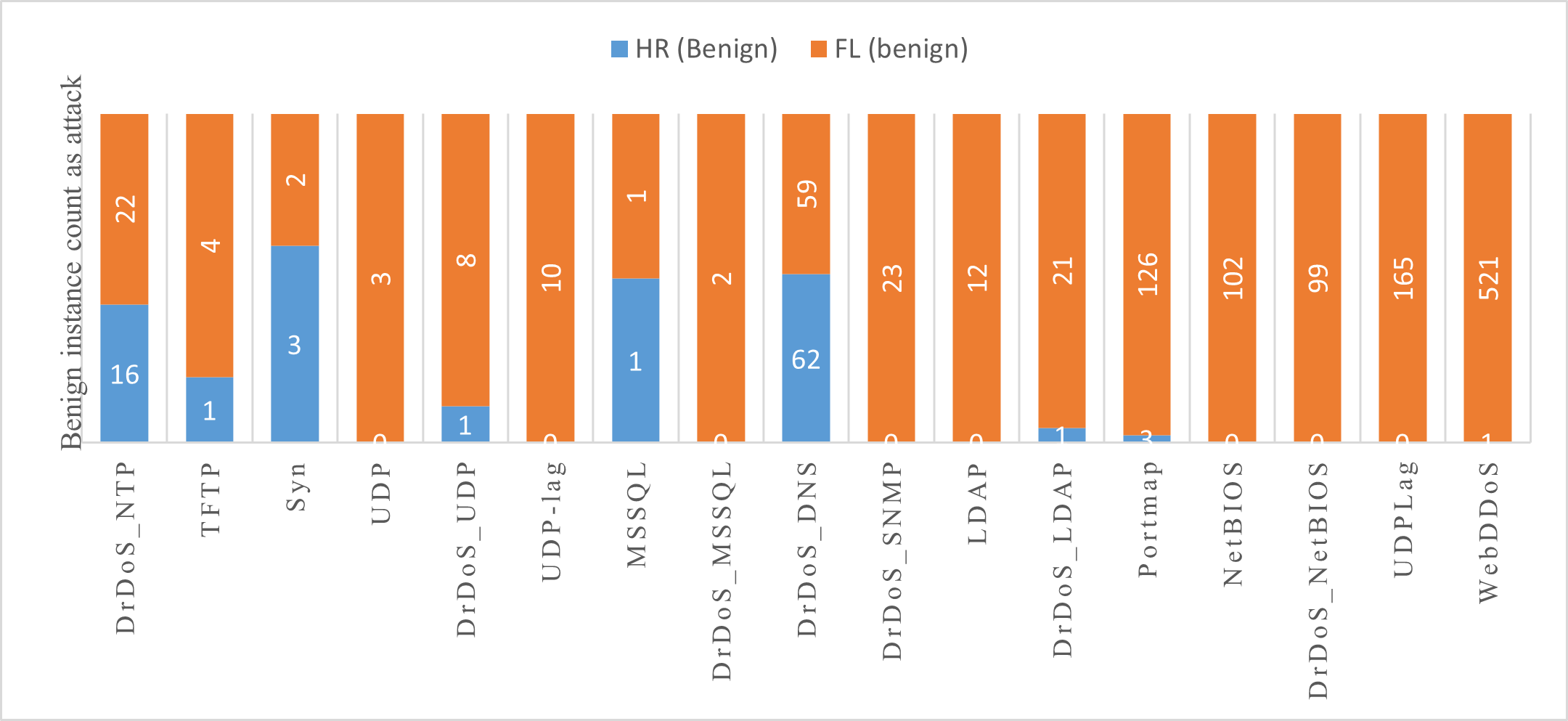}}       
    \caption{The false positive rate of HR and FL model}
    \label{fig:falsepositiverate}
\end{figure}

We present the result for the false positive rate in Figure \ref{fig:falsepositivenslkdd} and Figure \ref{fig:falsepositiverate} (a) and (b) for both models on NSL-KDD, UNSW-NB15, and CIC-DDoS2019 datasets respectively. All plots show that while the HR model demonstrates better performance in terms of false negative cases compared to the FL model, it also produces a higher number of false positives across most datasets except for the CIC-DDoS2019 dataset. However, in the context of intrusion detection systems, minimizing the false negative rate holds greater significance as a cyberattack can potentially lead to system disruption or shutdown. 

\subsection{Discussion} 

while choosing the HR model over the FL model, we need to consider some design aspects. The HR model is more complex and computationally expensive than the FL model. In the HR model, there involves more classifiers to be trained at different levels. The total number of algorithms required for the HR model can be calculated as: n of algorithms for n number of attack families (Level 2) + 2 (1 to classify attack families at Level 2 and 1 binary classifier to classify attack vs normal). On the other hand, in FL, only one classifier is trained. Therefore, running time of HR is generally a times more than that of FL. The prediction time will be generally increased by three times as prediction has be done in each of the three  Levels. Therefore, the HR model might not be deployed on devices with limited resources. The choice of model should be made based on the specific requirements of the task. Our studies show that the rate of attack instances being identified as normal instances is comparatively less in the HR model than the FL model. So, if detecting attack samples is the most important factor such as in case of cybersecurity, then the HR model is a good choice. However, if speed or resource constraints are important, then the FL model might be a better choice. Moreover, adapting to newly emerging attack types can be accomplished with greater ease in the hierarchical model compared to the flat (FL) approach. The hierarchical model necessitates retraining the classifier specifically for the new attack family, which involves working with a considerably smaller sample size compared to retraining the FL classifier on the entire dataset.


Boone et al. \cite{boone2022mask} made a point in their studies that that HR performs well when there are many classes at the bottom level, which can be the case in IDS as new attacks are emerging over time. Our studies showed that the HR approach did not significantly enhance accuracy compared to conventional flat multi-classification methods. This is because the current IDS datasets do not have a large number of target classes. Boone et al. \cite{boone2022mask} demonstrated that hierarchical classification does not yield better results than flat multi-classification methods, unless the target column comprises numerous classes and a well-defined taxonomy of the target classes is not adhered to. 


\section{Conclusions and Future Work}
\label{Conclusion}

In this studies, we investigate the performance of a three-level hierarchical and the conventional flat classification on ten contemporary IDS datasets using ten machine learning classifiers. The results show the HR model does not significantly improve over flat model in terms of weighted average precision, recall, and F1-score. However, the HR model identifies fewer number of attacks as benign compared to the FL model across most datasets. In terms of producing fewer false negative or detecting different attacks as benign, the HR model appears to perform better than the FL model which is important in the domain of cybersecurity. Moreover, our findings suggest that the performance of the model is significantly impacted when the classifier at the first level misclassifies the samples. We observe that the proportion of instances correctly identified at the first level tends to be greater, as machine learning algorithms typically exhibit greater accuracy when classifying binary targets. Our research demonstrates that multi-classification tasks frequently present difficulties for machine learning algorithms. 

In future, we can explore effective techniques to enhance the performance of the classifier at the first level since its misclassifications propagates down to the classification and significantly affect overall performance. We can include experimenting with novel classifiers or ensemble methods specifically tailored for IDS. There needs to analyze the importance of different features and assess their contribution to the hierarchical classification approach. We can focus on making hierarchical IDS models more interpretable and explainable. We can develop techniques to provide insights into why a particular decision or classification was made in hierarchical setting, which is crucial for understanding and trust in IDS systems.

\section*{Declarations}

\subsection*{Conflict of interest}
The authors have no conflicts of interest to declare that they are relevant to the content of this article.

\section*{Acknowledgments}
This material is based upon work supported by the Air Force Office of Scientific Research under award number FA2386-23-1-4003.

\section*{Author statements}
Md Ashraf Uddin: Conceptualization; Data curation; Implementation, Roles/Writing-original draft; and Writing, Visualization; Formal analysis.
Sunil Aryal: Funding acquisition; Investigation; Methodology; Project administration; Resources; Software; Supervision; Validation;  Roles/Writing-original draft; and Writing - review \& editing.
Mohamed Reda Bouadjenek: Conceptualization; Project administration;
Muna Al-Hawawreh: Review \& editing.
Md. Alamin Talukder: Data curation; Implementation; Visualization.

\bibliographystyle{els-article}
\bibliography{references}

\section{Supplementary Materials}

In this document, we provided the detailed description and results for 10 different classifiers on different benchmark IDS datasets. 

\subsection*{Methodology}

In this section, we define some mathematical symbols to describe our hierarchical model in detail below. We assume $X$ = Original dataset excluding the target class, $Y$ = Target column, $X_{train}$ = training set from $X$, $X_{test}$ = testing set from $X$, $Y_{train}$ = true value for $X_{train}$, $Y_{test}$ = true value for $X_{test}$, $X_1^{train}$ = training dataset for model $C_1$, $Y_1^{train}$ = true value for $X_1^{train}$ to train model $C_1$, $X_2^{train}$ = training set for model $C_2$, $Y_2^{train}$ = true value for $X_2^{train}$ to train the model $C_2$, $Y_1^{pred}$ = class values predicted by model $C_1$, $X_1^{test}$ = testing samples for model $C_1$, $Y_1^{test}$ = true value for testing model $C_1$, $X_2^{test}$ = testing set for model $C_2$, $Y_2^{test}$ = true value for model $C_2$, $Y_2^{pred}$ = class values predicted by model $C_2$. The training and testing phase of the hierarchical model presented in Figure \ref{fig:1} are described below.

\begin{itemize}
    \item Let $X$ be the original dataset that includes three target columns: level $1$ (binary target), level $2$ (category target), and level $3$ (subcategory target) and $Y$ be the target value of $X$, which includes various attack categories such as DoS, Trojan Horse, and Spyware depending on the dataset. The column $Y$ represents the true values for the level $3$ subcategory.
    \item We assign the training and testing sets for the root level classifier $(C_1)$, second level classifier $(C_2)$, and third level classifier $(C_3)$ as follows:  $[[X_1^{train}, Y_1^{train}]$ and $Y_1^{test}], [[X_2^{train}, Y_2^{train}]$ and $Y_2^{test}], [[X_3^{train}, Y_3^{train}]$ and $Y_3^{test}]$.
    \item We employ stratified cross-validation to train and test our hierarchical model at each level. The stratified function is applied to $X$ and $Y$. In each fold of the stratified cross-validation, we need to perform the following steps using $X_{train}$, $Y_{train}$, $X_{test}$, and $Y_{test}$. Here, $X_{train}$ and $Y_{train}$ represent the training set obtained from the stratified cross-validation function, and $X_{test}$ and $Y_{test}$ represent the testing set obtained from the stratified cross-validation function.
    
    \item We utilize stratified cross-validation on the datasets $X$ and $Y$, and in each fold, we conduct the subsequent steps.
    
     \begin{itemize}
         \item First, we train the classifier at the root or first level using $X_1^{train}$ and $Y_1^{train}$. The formation of $X_1^{train}$, $Y_1^{train}$, and $Y_1^{test}$ is as follows. To create $X_1^{train}$, we extract all the features from $X_{train}$ excluding the last three target columns. $Y_1^{train}$, which represents the true values for $X_1^{train}$, is created by extracting the binary target column also called level 1 from $X_{train}$.
         
         \item Next, we train the classifier at the second level using $X_2^{train}$ and $Y_2^{train}$. To create $X_2^{train}$, we remove all the instances labelled as normal or benign from $X_{train}$ and extract all the features excluding the level 1 (binary target) and level 2 category columns. $Y_2^{train}$ is formed by extracting the level $2$ column from $X_{train}$. $Y_2^{train}$ represents the true values for $X_2^{train}$. We create $Y_2^{train}$ by excluding the normal points from the level $2$ column of $X_{train}$.
         
         \item Next, we need to create training sets for the third level classifiers. The number of classifiers required at the third level depends on the dataset's level $2$.  In this case, we assume that two classifiers are needed to further identify the network attacks at the third level. We name these classifiers as $C_{31}$ and $C_{32}$, and the training sets for both classifiers are denoted as $X_{31}^{train}$, $Y_{31}^{train}$ and $X_{32}^{train}$, $Y_{32}^{train}$, respectively.
         
         \item To construct $X_{31}^{train}$, we extract the data instances labeled as $1$ in the level 2 category column of $X_{train}$ and remove the level $1$ and level $2$ columns. In this context, the label $1$ corresponds to a subcategory at level $2$. For constructing $Y_{31}^{train}$, we extract the data points from $Y_{train}$ that have a label of $1$ in the level $2$ column of $X_{train}$. To do this, we consider the indexes in the level $2$ column that are labeled as $1$ and retrieve the corresponding points from $Y_{train}$.
         
         \item  Similarly, for creating $X_{32}^{train}$, we extract the data points labeled as $2$ in the level 2 column of $X_{train}$ and exclude the level $1$ and level $2$ columns. Here, the label $2$ represents another subcategory at level $2$. For creating $Y_{32}^{train}$, we extract the data points from $Y_{train}$ that are labeled as $2$ in the level $2$ column of $X_{train}$. Again, we consider the indexes in the level $2$ column that are labeled as $2$ and select the corresponding points from $Y_{train}$.
         
     \end{itemize}
     
 \item During this phase, we evaluate the trained hierarchical model in each fold of the stratified cross-validation. To begin, we create $X_1^{test}$ by copying the $X_{test}$ data while excluding its level $1$, level $2$ and level $3$ columns. $Y_1^{test}$ remains the same as $Y_{test}$. The construction of $X_2^{test}$, $Y_2^{test}$, $X_3^{test}$, and $Y_3^{test}$ depends on the predictions made by the level 1, and level 2 classifiers.

      \begin{itemize}

        \item  $X_{31}^{test}$ represents the testing set for the third level classifier $(C_{31})$. This set is created based on the predictions made by the second level classifier. If the second level classifier $C_2$ identifies any points as $1$, the corresponding points in $X_{31}^{test}$ are predicted using the $C_{31}$ classifier.
        
        \item  Similarly, $X_{32}^{test}$ represents the testing set for the third level classifier $(C_{32})$, which is formed based on the predictions made by the second level classifier. If the second level classifier $C_2$ identifies any points as $2$, the corresponding points in $X_{32}^{test}$ are predicted using the $C_{32}$ classifier. 
          
        \item  Assuming $Y_1^{pred}$ represents the predicted outcomes from the root level classifier $(C_1)$, we iterate over the list of these predicted outcomes. If an outcome is classified as "normal," it is added to the $Y_2^{pred}$ and $Y_3^{pred}$ list. Otherwise, the corresponding instance is extracted from $X_1^{test}$, and put it in $X_2^{test}$, and this point is predicted by the second level classifier $(C_2)$. If the outcome from the second level classifier is a value other than $1$ or $2$, it is inserted into $Y_2^{pred}$ and $Y_3^{pred}$. However, if the outcome is 1, the corresponding point $X_3^{test}$ is extracted from $X_1^{test}$ and predicted by the third level classifier $C_{31}$. The outcome is then added to $Y_3^{pred}$. Similarly, if the outcome is 2, the corresponding point $X_3^{test}$ is extracted from $X_1^{test}$ and predicted by the third level classifier $C_{32}$. The outcome at this level is added to $Y_3^{pred}$.
          
      \end{itemize}    
     
     \item The performance of each fold is recorded, and the average performance scores are calculated for the stratified 10-fold cross-validation.
    
\end{itemize}

The algorithm of the hierarchical model, we implement in this article is presented in Algorithm \ref{alg:hierarchical-model}.

\begin{algorithm}
  \caption{Hierarchical Model Training and Testing}
  \label{alg:hierarchical-model}
  \SetAlgoLined
  \SetKwInOut{Input}{Input}
  \SetKwInOut{Output}{Output}

  \Input{Dataset $X$, Target column $Y$}
  \Output{Predicted outcomes $Y_2^{\text{pred}}$ and $Y_3^{\text{pred}}$}
  
  \textbf{Training Phase:}\\
  Split dataset into training and testing sets: $[[X_1^{train}, Y_1^{train}]$ and $Y_1^{test}]$, $[[X_2^{train}, Y_2^{train}]$ and $Y_2^{test}], [[X_{31}^{train}, Y_{31}^{train}]$ and $Y_{31}^{test}]$, and  $[[X_{32}^{train}, Y_{32}^{train}]$ and $Y_{32}^{test}]$ \;
  
  \For{each level in hierarchical model}{
    Apply stratified cross-validation on $X$ and $Y$\;
    Obtain training set $X_{\text{train}}$ and $Y_{\text{train}}$ from cross-validation\;
    \If{current level is root or first level}{
      Train classifier $C_1$ using $X_1^{train}$ and $Y_1^{train}$\;
    }\ElseIf{current level is second level}{
      Create $X_2^{train}$ by removing normal points from $X_{\text{train}}$\;
      Create $Y_2^{train}$ by extracting level 2 column from $X_{\text{train}}$\;
      Train classifier $C_2$ using $X_2^{train}$ and $Y_2^{train}$\;
    }\Else{
      Create multiple training sets $X_{31}^{train}$, $Y_{31}^{train}$, $X_{32}^{train}$, $Y_{32}^{train}$\;
      Build $X_{31}^{train}$ by extracting points labeled as 1 from $X_{\text{train}}$\;
      Build $Y_{31}^{train}$ by extracting corresponding points from $Y_{\text{train}}$\;
      Build $X_{32}^{train}$ by extracting points labeled as 2 from $X_{\text{train}}$\;
      Build $Y_{32}^{train}$ by extracting corresponding points from $Y_{\text{train}}$\;
      Train classifiers $C_{31}$ and $C_{32}$ using $X_{31}^{train}$, $Y_{31}^{train}$, $X_{32}^{train}$, $Y_{32}^{train}$\;
    }
  }
  
  \textbf{Testing Phase:}\\
  Obtain testing set $X_{\text{test}}$ and $Y_{\text{test}}$ from cross-validation\;
  Initialize empty prediction lists $Y_2^{\text{pred}}$ and $Y_3^{\text{pred}}$\;
  Obtain predicted outcome $Y_1^{\text{pred}}$ from classifier $C_1$\;
  
  \For{each outcome in $Y_1^{\text{pred}}$}{
    \If{outcome is normal}{
      Insert outcome into $Y_2^{\text{pred}}$\;
    }\Else{
      Extract corresponding point from $X_{\text{test}}$ as $X_2^{\text{test}}$\;
      Predict outcome using classifier $C_2$\;
      \If{outcome is not 1 or 2}{
        Insert outcome into $Y_2^{\text{pred}}$ and $Y_3^{\text{pred}}$\;
      }\ElseIf{outcome is 1}{
        Extract corresponding point from $X_{\text{test}}$ as $X_{31}^{\text{test}}$\;
        Predict outcome using classifier $C_{31}$\;
        Insert outcome into $Y_3^{\text{pred}}$\;
      }\ElseIf{outcome is 2}{
        Extract corresponding point from $X_{\text{test}}$ as $X_{32}^{\text{test}}$\;
        Predict outcome using classifier $C_{32}$\;
        Insert outcome into $Y_3^{\text{pred}}$\;
      }
    }
  }
\end{algorithm}

\subsection*{ Encoding of Dataset}

The following section provides an overview of the categorical features included in the original dataset, as well as the dataset's shape before and after one hot encoding. Table \ref{tab:datasetsubcategory} presents the distribution of every IDS dataset.


\begin{itemize}
  
    \item NSL-KDD: The NSL-KDD dataset has 148,517 rows and 43 columns. Upon performing one-hot encoding on the three categorical features (protocol\_type, service, and flag), the total number of features increases to 124, including two binary and categorical targets.
    
    \item UNSW-NB15: The original dataset contains 257,673 records and 45 fields. After performing one-hot encoding on the categorical features (proto, service, and state), the total number of fields increases to 199.
    
    \item CIC-IDS2017: The Canadian Institute for Cybersecurity developed CIC-IDS2017. The dataset includes user behaviour models that are protocol-agnostic through HTTP, HTTPS, FTP, SSH, and email. The dataset  consists of 692,703 samples and 78 features having four classes: 440,031 benign samples, 5,792 DoS SlowLoris samples, 5,499 DoS Slow Httptest samples, 231,073 DoS Hulk samples, 10,293 DoS GoldenEye samples, and 11 Heartbleed samples in the output class label. The dataset used in this experiment has no categorical features.  
    
    \item CIC-DDoS2019: The shape of the CIC-DDoS2019 dataset is 431,371 rows and 79 columns. The dataset used in this experiment has no categorical features. 
    
    \item Malmem2022: The dataset comprises 58,596 records and 57 numeric features. The distribution of these features is presented in Table \ref{tab:datasetsubcategory}. 
    
    \item CIC-Darknet2020: The dataset in the CIC-Darknet2020 dataset has 83 features and was labelled in two ways. Prior to pre-processing, we eliminated four characteristics from the dataset—Flow ID, Src IP, Dst IP, and Timestamp as those features do not aid in classification.
    
    \item XIIoTID: The XIIoTID dataset has an initial shape of (596017, 64), with two categorical features: protocol and service. Upon performing one-hot encoding on these features, the total number of features increases to 81.
    
    \item TON-IOT-IOT: After performing one hot encoding on the categorical features, the combined TON-IOT-IOT dataset has 29 features in total. We merged different IoT dataset following \cite{booij2021ton_iot}. The dataset contains four categorical features, namely temp\_condition, door\_state, sphone\_signal, and light\_status.
    
    \item TON-IOT-Network: We dropped src\_ip, dst\_ip, ts, dns\_query, http\_uri, and http\_user\_agent columns from the dataset. There are 21 categorical columns in the dataset. After applying one hot encoding, the total numbers of features stand at 132 including the target columns. TON-IOT-Network dataset used here has 461043 numbers of instances.  
    
    \item ISCX-URL2016: ISCX-URL2016 dataset consists of 57,000 URL instances having four kinds of classes such as Benign URLs (over 35,300), Spam URLs (over 12,000), Phishing URLs (Around 10,000), Malware URLs (over 11,500), and Defacement URLs (over 45,450). The dataset has 79 features, and the target class can be binary or multiple. The major features of this dataset are query length, domain tokens number, path token’s count. 
    
    \item BoT-IoT: The UNSW Canberra Centre for Cyber created the BoT-IoT dataset to identify malicious actions in IoT network traffic. More than 3.6 million records are there, each with 43 characteristics and a designation for the output class. There are 1,926,624 instances of DDoS attacks, 1,650,260 samples of DoS attacks, 91,082 samples of reconnaissance, 70 samples of theft, and 478 samples of regular operations. There are three categorical columns('flgs', 'proto', 'state') in the dataset. The total number of columns after performing one-hot encoding stands at 65 including the target columns. 

\end{itemize}

\subsection*{Hierarchical Structure of Datasets}

We define the hierarchical structure of several datasets according to the Cyber Kill Chain which consists of the following seven steps as depicted in Figure \ref{fig:killchain}:

\begin{figure}[!htbp]
    \centering
    \includegraphics[scale = .65]{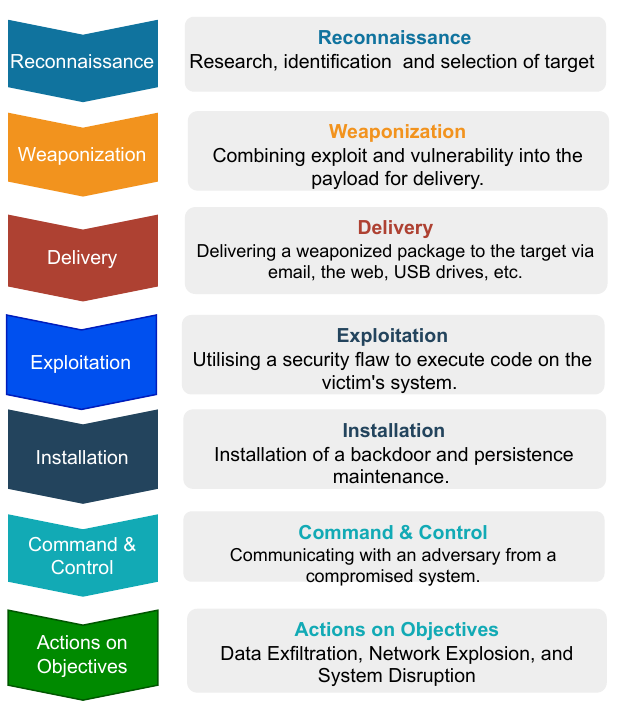}
    \caption{The Cyber kill chain framework}
    \label{fig:killchain}
\end{figure}

\begin{itemize}
    \item Reconnaissance: Attacker usually collects network topology about the target including IP address, employee identities, and potential security vulnerabilities in this kind of attack.
    \item Weaponization: The perpetrator utilizes the necessary tools or malware to exploit malicious payload such as a computer virus or Trojan horse. 
    \item Delivery:  The perpetrator delivers the malicious payload to the target, typically via phishing emails or drive-by downloads, malicious websites, physical media. 
    \item Exploitation: The attackers execute malicious payload in a target system to take advantages of a system’s vulnerability to gain unauthorized access to the target's system.
    \item Installation: To gain further access, the perpetrator installs malware on the target system.
    \item  Command and Control (C2):  The perpetrator establishes a communication channel with the compromised system, enabling them to issue commands, extract data, or carry out additional malicious actions.
    \item Actions on Objective:  The malicious actors execute their intended malicious actions that can vary based on their objectives. This might involve data exfiltration, data erasure, sabotage, or unauthorised access.
    
\end{itemize}

\begin{itemize}
    \item NSL-KDD dataset\cite{su2020bat} was designed to overcome some of the issues with KDD’99 dataset. This updated version of the KDD data set is still regarded as an effective benchmark dataset for researchers to compare different intrusion detection approaches. The NSL-KDD training and testing sets have a balanced quantity of records for benign and attack samples. The level 2 categories of this dataset are normal, DoS, probe, R2L and U2R. The original subcategories provided in the dataset are used as level 3 categories in our hierarchical approach.    



    \item CIC-DoS2017: Jazi et al. \cite{jazi2017detecting} classified the DoS attack as high volume or low volume attack in the CIC-DoS2017 attack. Figure \ref{fig:cicddos2017} illustrates that  We consider high volume and low volume attack categories as level 2 and the level 3 categories provided in the dataset is regarded as level 3 in our hierarchical model. 
    
    High volume attack: DoS attacks comprise high-volume attacks, also known as flooding attacks. Attackers send many requests to a victim to overwhelm the victim's resources and make the victim's website unavailable to legitimate users.
    
    Low volume attack:  Low-volume DoS attacks refer to a form of denial-of-service attack that disrupts a victim's service using small amounts of attack traffic. There are three primary low-volume DoS attack types: 1) Low-rate attacks transmit traffic in periodic short-duration. 2) Slow-rate attacks: These attacks exploit timing parameters on the server's side by transmitting and receiving traffic at a slower rate than anticipated. 3) One-shot attacks: These attacks cause harm to a target through a single connection/request. In the dataset, slow-rate attacks were further classified as slow send or slow receive. 

         \begin{figure}[!htbp]
                \centering
                \includegraphics[scale = .65]{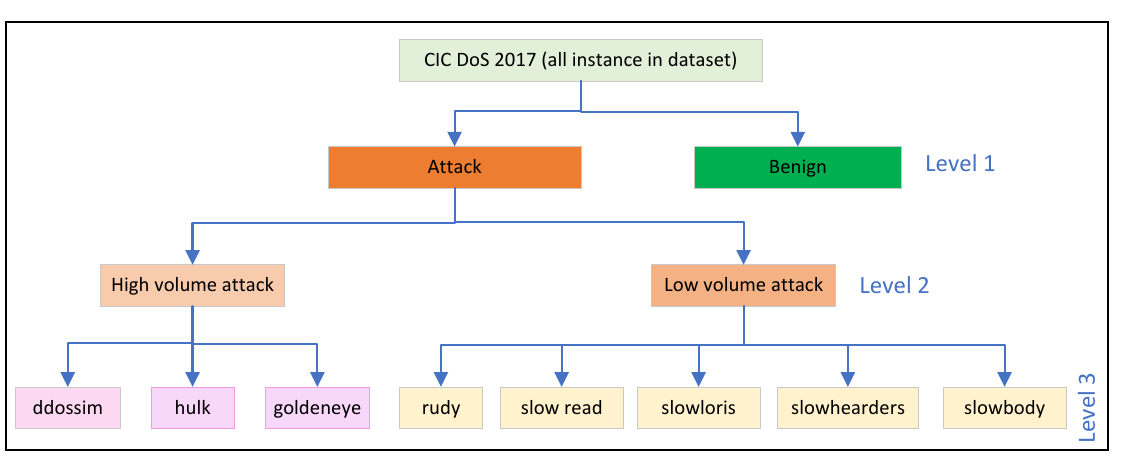}
                \caption{Hierarchical level of CIC-DDoS2017 dataset}
                \label{fig:cicddos2017}
        \end{figure}  

    \item \textbf{CIC-DDoS2019:} CIC-DDoS2019 includes new DDoS attacks that are carried out using TCP/UDP based protocols at the application layer. The creator of the dataset classified the taxonomy of DDoS attack as Reflection-based DDoS and Exploitation-based DDoS attacks. We consider reflection-based DDoS and exploitation-based DDoS attacks as level 2 categories of the dataset. Table \ref{tab:datasetsubcategory}  illustrates Level 2 and level 3 categories of this dataset-Level 2: E\_TCP, E\_UDP, R\_TCP, R\_TCP/UDP, and R\_UDP and Level 3: ( E\_TCP: Sync Flood), (E\_UDP: UDP-lag, UDP Flood), (R\_TCP: MySQL, SSDP), (R\_TCP/UDP: DNS, LDAP, NETBIOS, SNMP, PORTMAP), (R\_UDP: NTP, CharGen, TFTP).

    \item \textbf{ XIIoTID:} In our hierarchical model, we consider the level 2 and level 3 categories of XIIOTID dataset provided by Muna et al.\cite{9504604}. Table \ref{tab:datasetsubcategory} illustrates level 2 and level 3 categories of the XIIOTID dataset-Level 2: Expfiltration, Exploitation, RDOS, Lateral, Tampering, Reconnaissance, Weaponization, C\&C and Level 3: Exfiltration, Rshell, MITM, RDOS, MQTT, TCP, Modbus, Tampering, Scanning, Generic, Discovering, Fuzzing, BruteForce, Insider\_malicious, Dictionary and C\&C. To represent the attacks in the confusion matrix, we shorten the name of some attack categories of level 3 also called subcategory.   
    
    \item The dark web is an encrypted part of the internet that is inaccessible through regular search engines and requires the use of the Tor anonymous browsing platform. This represents only a small portion of the deep web, which encompasses the sections of the internet that are not indexed by search engines. The CICDarknet2020 dataset depicted in Figure \ref{fig:cicdarknet2020} is a two-layered approach, with the first layer generating both benign and malicious traffic from the internet dark web. The second layer includes the following network types of dark web: Audio-Stream, Browsing, Chat, Email, P2P, Transfer, Video-Stream, and VOIP. The creator integrated their previously developed datasets, ISCXTor2016 and ISCXVPN2016, merging VPN and Tor traffic into relevant Darknet categories to produce the representative dataset.
    In the dataset, Label 2 provides the four categories called VPN, TOR, non-VPN, and non-TOR whereas Label 3 specifies a subcategory within Label 2. Following \cite{gupta2022encrypted}, we lumped together the samples that did not have VPNs or TORs and labelled "normal" for the purposes of the experiment. The final labelling in the dataset is either "normal," "TOR," or "VPN".

        \begin{figure}[!htbp]
                \centering
                \includegraphics[scale = .65]{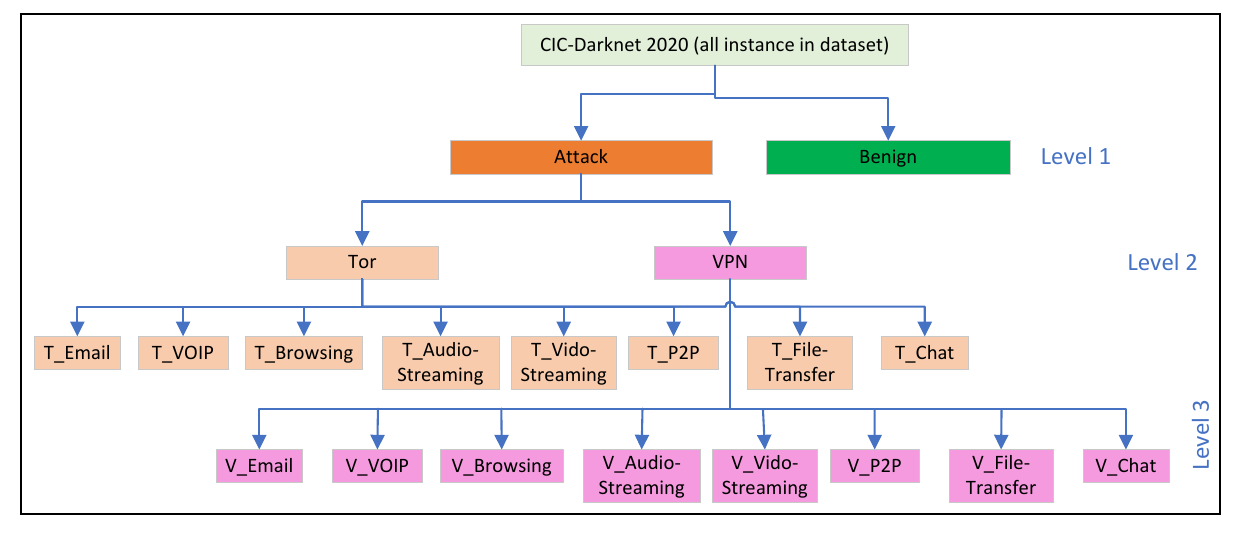}
                \caption{Hierarchical level of CIC-Darknet 2020 dataset}
                \label{fig:cicdarknet2020}
         \end{figure}

    \item Malmem2022: Obfuscated malware hides them to avoid detection and elimination using conventional anti-malware software. Malmem2022\cite{carrier2022detecting} is a simulated obfuscated dataset designed to be realistic as possible to train and test machine learning algorithms to detect obfuscated malware. The dataset is balanced one having level 2 categories: Spyware, Ransomware, and Trojan Horse. Table \ref{tab:datasetsubcategory} shows level 3 or subcategories of this dataset. 
    
    \item \textbf{TON-IOT-IOT\cite{alsaedi2020toniot}:} The creator of this dataset did not provide the subcategory. We considered the original level 2 as level 3 in our experiment and we created a new level 2 categories by grouping the original attack categories of the dataset according to "Kill Chain" framework. Figure \ref{fig:toniotiot} illustrates the level 2 and level 3 of the ToN-IoT-IoT dataset. We grouped all the samples of backdoor, XSS, Injection and MITM as Exploitation. A backdoor is a malicious software that enables unauthorised access to a computer system. Injection attacks involve inserting malicious code or commands into a target system to perform unauthorized actions, whereas XSS attacks involve injecting malicious scripts into web pages viewed by other users. While both Injection and XSS attacks involve the injection of malicious code, they target different vulnerabilities and have distinct goals.

         \begin{figure}[!htbp]
                \centering
                \includegraphics[scale = .65]{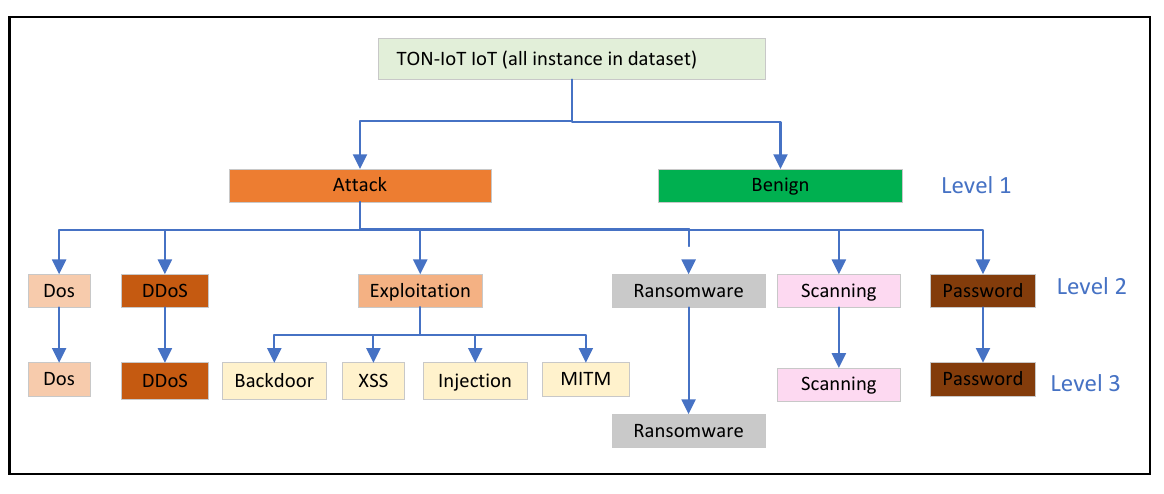}
                \caption{Hierarchical level of ToN IoT IoT dataset}
                \label{fig:toniotiot}
        \end{figure}  

    \item \textbf{TON-IOT-Network:} Like ToN-IoT-IoT dataset, the creator of ToN-IoT-Network dataset\cite{booij2021toniot} did not provide the subcategory. We considered the original level 2 as level 3 in our experiment and we created a new level 2 categories by grouping the original attack categories of the dataset according to "Kill Chain" framework. Like Figure \ref{fig:toniotiot}, the ToN-IoT-IoT dataset has the same level 2 categories and level 3 subcategories.

   \item ISCXURL2016: In WWW web, URLs serve as the primary mode of transport and attackers insert malware into users’ computer system through URL. The researchers focus on developing methods for blacklisting malicious URLs. Mamun et al. \cite{mamun2016detecting} formed a modern URL dataset that contains level 2 categories of URLs: benign URLs, spam URLs, phishing URLs, malware URLs and defacement URLs. We consider the level 2 category as level 3 as shown in Figure \ref{fig:iscxurl2016}. We formed level 2 category by grouping spam URLs and phishing URLs as delivery attack according to "Kill Chain" framework.

         \begin{figure}[!htbp]
                \centering
                \includegraphics[scale = .65]{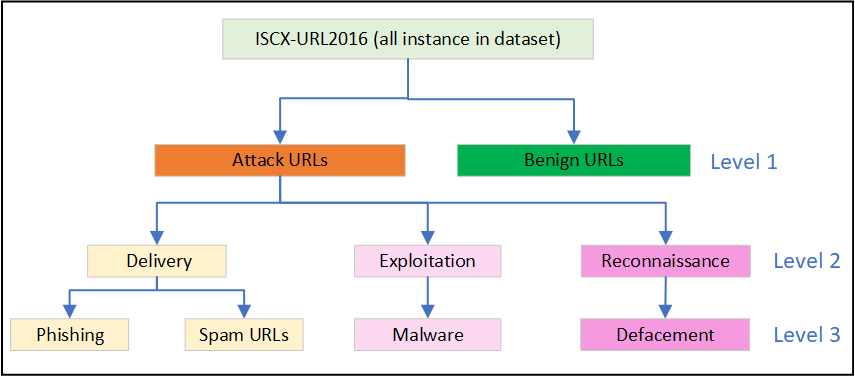}
                \caption{Hierarchical level of ISCXURL2016 dataset}
                \label{fig:iscxurl2016}
        \end{figure}

    \item BoT-IoT: Although many IDS datasets are available, very few datasets were designed on the Botnet scenarios. Koroniotis et al. \cite{koroniotis2019towards} introduced a new dataset called Bot-IoT to develop IDS to safeguard Internet of Things. They employed a practical testbed setting to address the limitations of existing datasets in terms of capturing full network details, precise labelling, and the variety of modern, complicated attacks. We used the level 2: Information gathering, Denial of Service, Information theft and level 3 categories provided in the datasets.
    
\end{itemize}

\subsection*{Machine Learning Algorithms}

This section, briefly discuss the machine learning algorithms used in this experiment to evaluate the effectiveness of the hierarchical classification approach.

\begin{itemize}

\item Gaussian Naive Bayes (NB)\cite{islam2022ggnb} that utilizes Bayes' theorem to calculate the probability of each class given a set of features assumes the features follow a Gaussian distribution. The algorithm employs the following equations: The prior probability $P(y)$ represents the probability of each class in the training data, which is calculated by dividing the count of instances with class $y$ by the total number of instances. The class conditional probability $P(x|y)$ assumes independence between features and calculates the probability of observing a feature value $x$ given a class $y$ based on the mean $\mu$ and variance $\sigma^2$ of the feature values for that class.

\begin{equation}
\label{NB}
    P(x|y) = \frac{1}{\sqrt{2\pi\sigma^2}} \exp\left(-\frac{(x-\mu)^2}{2\sigma^2}\right)
\end{equation}

Finally, the posterior probability $P(y|x)$ combines the prior probability and class conditional probability to determine the probability of each class given the observed feature values. To classify a new instance, the algorithm calculates the posterior probability for each class and selects the class with the highest probability. 
  
\item Decision Tree (DT)\cite{ahmim2019novel}: Decision tree builds a tree-like model to make decisions based on the provided data. This partitions the data into subsets based on different features and their thresholds, aiming to maximize the information gain or decrease in impurity at each node. DT can handle both categorical and numerical data, making them versatile for a wide range of applications. DT is capable of capturing non-linear relationships and interactions between features. Additionally, decision trees are robust to outliers and missing values, as it does not rely on the entire dataset for decision-making. 

\item Random Forest (RF): Random Forest is a popular Supervised Machine Learning method utilized for Regression and Classification problems~\cite{negandhi2019intrusion}. It generates decision trees from multiple samples and utilizes the majority vote to classify them and the average to compute regression. This classifier is preferred over a single decision tree because of its higher accuracy. It is primarily an ensemble approach based on bagging. The classifier performs as follows: Upon receiving D, the classifier creates a total of k D bootstrap samples, each labeled D$_i$. D$_i$ is sampled using D's replacement and contains the same number of tuples as D. Due to the use of sampling with replacement, some actual D tuples may not exist in D$_i$ at all, while others may appear many times. The classifier then constructs a decision tree for each D$_i$. As a result, a "forest" of k trees is created. Each tree provides one vote to classify an unknown tuple, X, by predicting its class. The final decision on X's class is given to the one with the most votes. The decision tree technique used in scikit-learn is CART (Classification and Regression Trees), and its tree induction is based on the Gini index. The Gini index for D is computed as follows:

\begin{equation}
  \label{equationbatch1}
       Gini(D) = 1- \sum_{i=1}^{k}p_i^{2}
\end{equation}

The probability that a tuple in D belongs to class C$_i$ is given by p$_i$. The better D was partitioned, the lower the index value. 

\item Extreme Tree (ET)\cite{kasongo2021advanced}: Extreme Trees, also known as Extremely Randomised Trees or Extra Trees, are a decision tree-based ensemble learning method. Extreme Trees, like conventional decision trees, make decisions based on features and their thresholds. However, Extreme Trees, unlike conventional decision trees, select the split points arbitrarily without contemplating the optimal thresholds. This randomisation reduces the model's variance and makes it less susceptible to overfitting. Extreme Trees bases its predictions on a majority vote or an average of the predictions from individual trees.

\item Histogram Gradient Boosting Classification\cite{louk2023dual}: The Histogram Gradient Boosting Classification (HGBC) algorithm is utilised for supervised classification tasks. Gradient Boosting Decision Tree (GBDT) variant that uses histograms to discretize features and approximate gradients during training. The foundation of HGBC is the construction of prediction-making decision trees based on the histogram categories. Each tree is trained using the gradient of the loss function with respect to the prior tree's predicted values. The final prediction is then derived by integrating the predictions of each tree. The HGBC algorithm can be mathematically described as follows:

Given a dataset D = {(x$_1$, y$_1$), ..., (x$_n$, y$_n$)}, where x$_i$ is a vector of features, and y$_i$ is the corresponding label, the HGBC algorithm seeks to learn a function f(x) that maps input features to the output label. The function f(x) is defined as a weighted sum of decision trees, where each tree h(x; $\theta_i$) is parameterized by a set of parameters $\theta_i$:

f(x) = $\sum_{i=1}^T w_i h(x; \theta_i)$

where T is the number of trees, w$_i$ is the weight assigned to tree i, and h(x; $\theta_i$) is the output of the i-th tree given the input x and its parameters $\theta_i$. The parameters $\theta_i$ are learned during the training phase by minimizing a loss function L(y, f(x)) that measures the difference between the predicted label f(x) and the true label y. The loss function is typically defined as the negative log-likelihood of the data or the mean squared error.

\item {Logistic Regression (LR)}:

Logistic regression \cite{shanthi2022genetic} is a classification method, rather than a regression technique, as stated by Kleinbaum (2002). This is commonly used for binary classification problems, where the output is either "Yes/No," "1\/0," or "True\/False." Logistic regression is particularly effective for linearly separable classes and is a straightforward classification technique that yields excellent results. However, linear regression is not effective for estimating the performance of a binary variable due to two factors:

    \begin{enumerate}
        \item Linear regression can predict values beyond the acceptable range, such as probabilities that fall outside the limit of 0 to 1.
        \item Dualistic experimental studies may not have residuals that are normally distributed around the predicted line because they can only have one of two possible outcomes for each trial.
    \end{enumerate}

The model is based on the logistic function, which generates an S-shaped curve that can take any real number between 0 and 1. The hypothesis for logistic regression is Z = AX + B, where h(x) is the output of the sigmoid function.

\begin{equation}
\begin{gathered}
\phi(z) =\frac{1}{1+e^{-z}} = \frac{e^z}{1+e^z}
\end{gathered}
\end{equation}

The model is called linear logistic regression because the logit transformation is a linear function of the parameters.

\begin{equation}
\begin{gathered}
logit \ p(x)=\log\left[\frac{p(x)}{1-p(x)}\right] = AX + B
\end{gathered}
\end{equation}

The logit p(x) has linear parameters, is continuous, and can vary from $-\infty$ to $\infty$ relying on the x range.

\item K-Nearest Neighbour (KNN)\cite{saleh2019hybrid}: Nearest Neighbours (KNN) assigns a class label to a new instance based on the majority vote of its K closest neighbours in the training data. K, the number of neighbours, is a crucial parameter that influences the performance of the KNN algorithm. A lesser value of K can lead to a more complex decision boundary, which may result in overfitting. Alternatively, a larger value of K can result in a smoother decision boundary, which might contribute to underfitting. 

\item Multi-Layer Perception (MLP)\cite{mahmod2015hybrid}: Multi-Layer Perceptron (MLP) networks, also known as feed-forward neural networks composed of multiple layers of interconnected structures, known as neurones, arranged sequentially. The input layer receives input features, while the output layer generates final predictions. Between the input and output layers, the hidden layers perform computations and learn intricate representations of the input data. Each neuron in the MLP applies an activation function to the weighted sum of its inputs, enabling the network to capture nonlinear relationships among the input features. The weights of the network are updated iteratively using backpropagation, which calculates gradients and modifies weights to minimise the difference between predicted and actual outputs. MLP networks are capable of modeling intricate relationships and learning intricate decision boundaries. 

\end{itemize}

\subsection*{Performance Analysis: HR vs FL}

In this section, we present the performance of HR and FL model in terms of accuracy, precision, recall and F1-score for 10 different datasets and 10 machine learning algorithms. 

\subsubsection*{Performance of HR model (Binary)}

\begin{figure}[!htbp]
   \centering
    \includegraphics[scale = .55]{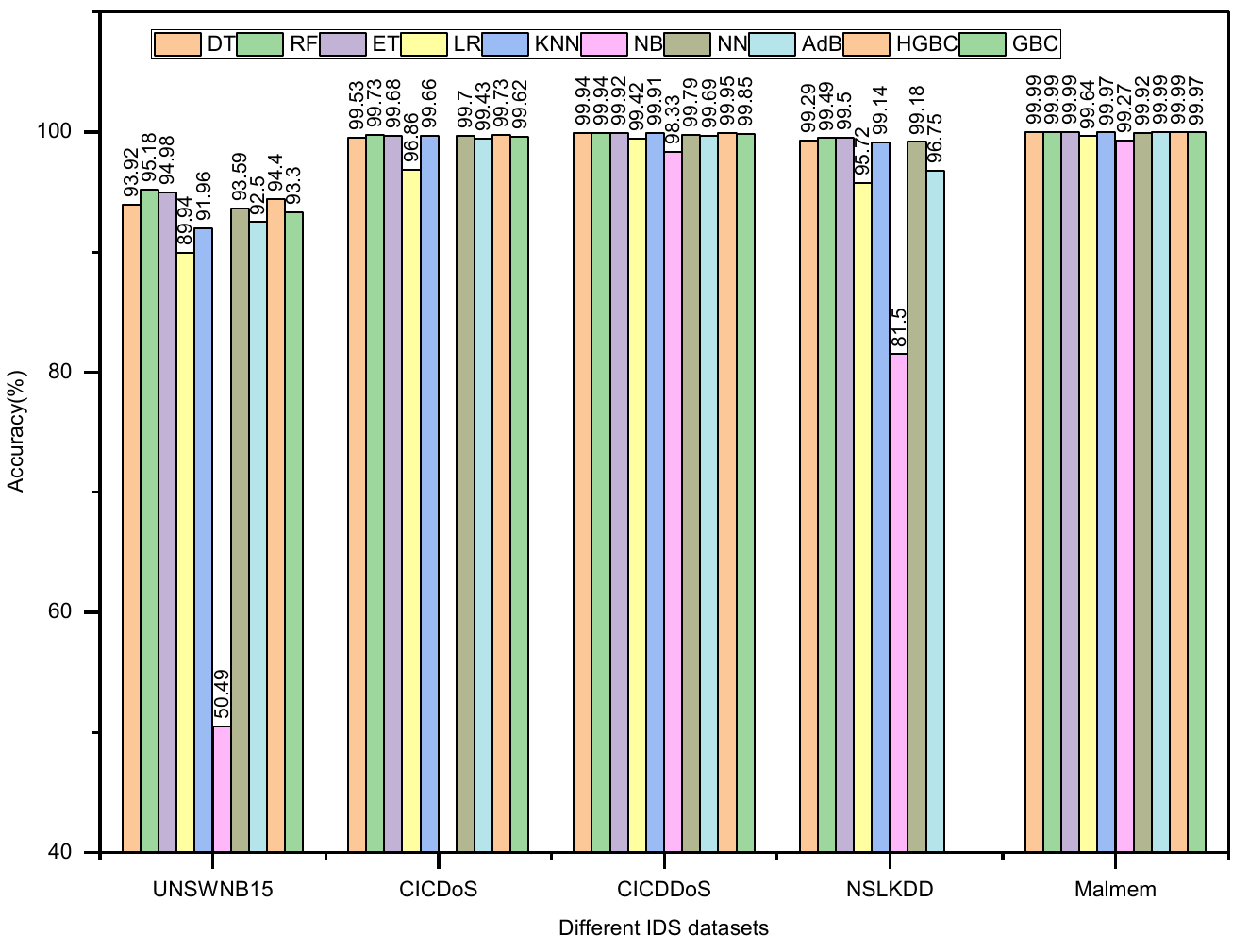}
    \caption{The accuracy of the 10 classifiers level 1 on 5 datasets }
    \label{fig:bhr10accuracy}
\end{figure}

\begin{figure}[!htbp]
   \centering
    \includegraphics[scale = .55]{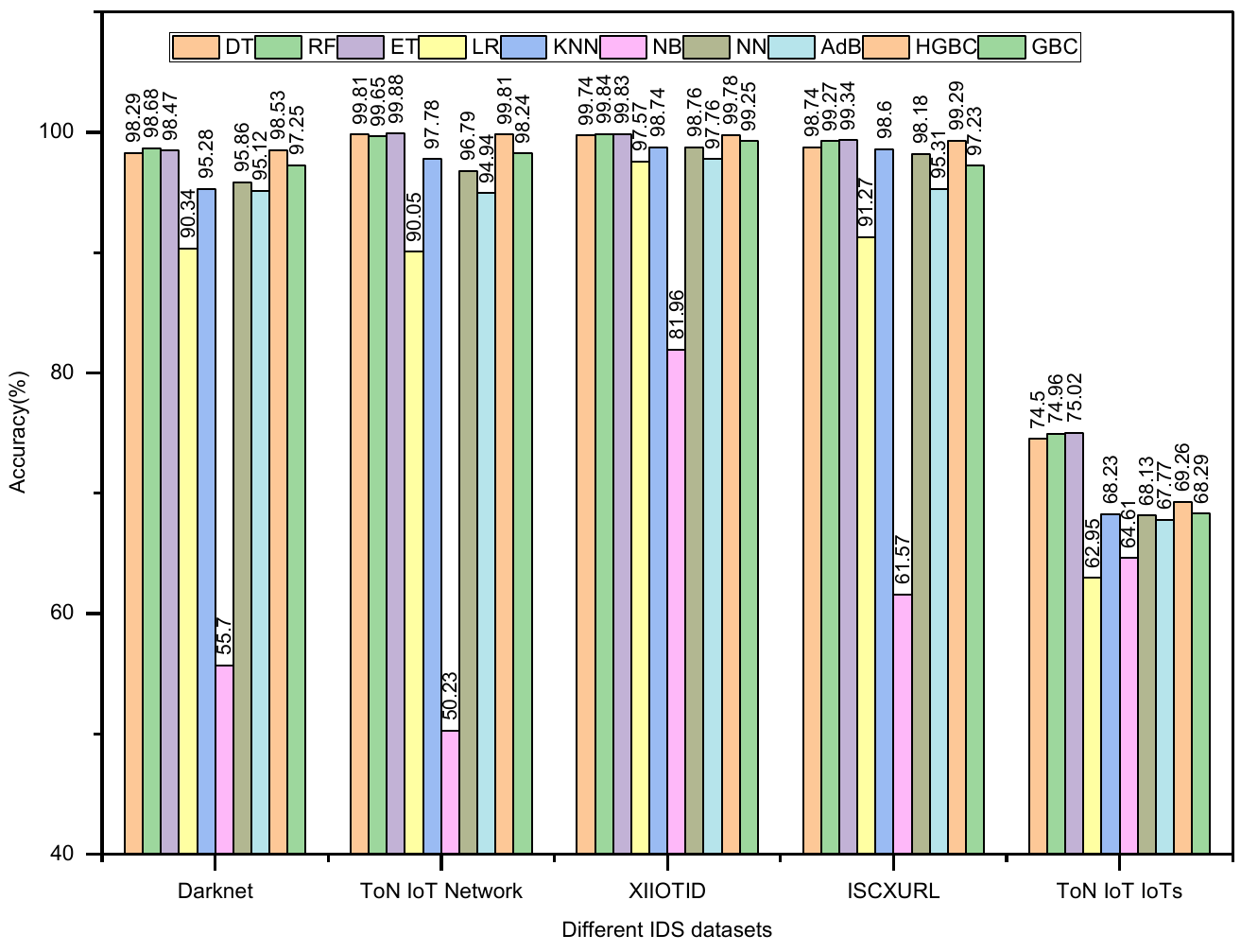}
    \caption{The accuracy of the 10 classifiers level 1 on 5 datasets }
    \label{fig:bhr20accuracy}
\end{figure}

Figure \ref{fig:bhr10accuracy} and \ref{fig:bhr20accuracy} present the accuracy of all the 10 classifiers for hierarchical model for level 1. For each dataset, we observe the following points: 

NSL-KDD: RF and ET classifiers perform well for this dataset. The accuracy of DT, RF, ET, KNN, and NN classifiers are all above 99.99\%. The accuracy of LR and AdB classifiers is above 95\% despite their marginally lower performance. With values near 81\% for accuracy, NB classifier has the lowest performance in terms of these metrics. 

UNSW-NB15: In level 1, notable improvements can be seen in RF and ET classifiers at the first level, where the hierarchical model achieves accuracy of 95.18\% and 94.98\%.

CIC-DoS2017: The highest accuracy achieved by the model is 99.73\%, attained by the RF and HGBC classifiers in the HR model. The lowest accuracy in the HR model is 38.91\%, achieved by the NB classifier.

CIC-DDoS2019: HR models achieve high accuracy across all classifiers, with values ranging from 98.33\% to 99.95\%. The classifiers with the highest accuracy are DT, RF, and HGBC. 

CIC-Darknet2020: The classification results show that most of the classifiers perform well with high accuracy. DT, RF, ET, and HGBC achieve accuracy scores above 98\%. LR and GBC achieve accuracy scores above 97\%. KNN and NN obtain accuracy scores above 95\%. However, NB stands out with a lower accuracy score of 55.7\%. 

ToN-IoT-Network: DT, RF, ET, and HGBC models achieve very high accuracy above 99\%. These classifiers perform better among other classifiers in accurately classifying the data.  Overall, the tree-based models (DT, RF, ET, HGBC) perform exceptionally well, achieving high accuracy, while the other models still show decent performance. In case of XIIOTID, the ensemble approaches such as RF, ET, HGBC, GBC demonstrate better performance. 

ISCXURL2016: RF, ET, and HGBC classifiers demonstrate particularly high performance with accuracy and precision above 99\%. However, LR and NB classifiers show lower performance compared to other classifiers in terms of accuracy.

ToN-IoT-IoTs: DT classifier performs the best, with an accuracy of 74.5\%. The RF classifier is a close second, with an accuracy of 74.96\%. 

In the following section, we present the results of the HR and FL models for different machine learning algorithms at levels 2 and 3. The green colour indicates that HR model performed better for a particular classifier. The yellow colour represents that the FL model performed better for a particular classifier. Gray colour indicates that both HR and FL demonstrated similar performance.  
\subsubsection*{Performance on NSL-KDD dataset}

Table \ref{tab:nslkdd} represents the classification performance of the HR model and the FL model on the NSL-KDD dataset at the last two levels: 2, and 3. The results are shown for eight different machine learning algorithms for NSL-KDD dataset. The metrics evaluated include Accuracy, Precision, Recall, and F1 score. Comparing the HR and FL models using the results in the table for level 2 and level 3, we can observe the following:

\begin{itemize}
    \item At Level 2: In the HR model, RF classifier achieves slightly higher accuracy (99.46\%) compared to RF (99.45\%) of the FL model. Similarly, the ET classifier also performs slightly better in the HR (99.47\%) than the ET (99.45\%) in the FL. Moreover, the HR model significantly outperforms the FL model in terms of accuracy for the NB classifier (80.67\%) and AdB classifier (92.83\%). However, for other classifiers, the FL model shows slightly higher accuracy compared to the HR model.
    
    \item At Level 3, both the HR and FL models show a significant drop in performance across all classifiers. This suggests that the classification task becomes more challenging and less accurate at deeper levels in the hierarchy. The HR model demonstrates superior performance across multiple classifiers, including RF with an accuracy of 99.27\%, ET with an accuracy of 99.29\%, KNN with an accuracy of 98.78\%, NB with an accuracy of 80.55\%, and AdB with an accuracy of 88.39\%. In comparison, the FL falls short in these classifiers. Notably, the HR model exhibits a significant performance gap over the FL model for NB and AdB classifiers.

\end{itemize}

The precision and recall values are generally high for both models and classifiers. The hierarchical model tends to have slightly higher precision and recall values compared to the flat model, indicating better performance in correctly identifying the positive class. The F1 score, which combines precision and recall, is also high for both models and classifiers. Similar to precision and recall, the HR model tends to have slightly higher F1 scores. This suggests that the hierarchical approach, where the data is split into sublevels, allows for better classification performance compared to the flat approach, where all the data is considered in a single level.

\begin{table}[!htbp]
\centering
\caption{Performance of HR and FL model on NSL-KDD dataset}
\label{tab:nslkdd}
\resizebox{\columnwidth}{!}{
\begin{tabular}{|c|l|l|l|l|l|l|l|l|l|}
\hline
Level & \multicolumn{1}{c|}{Classifier} & \multicolumn{1}{c|}{\begin{tabular}[c]{@{}c@{}}HR. \\ Accuracy\end{tabular}} & \multicolumn{1}{c|}{\begin{tabular}[c]{@{}c@{}}FL. \\      Accuracy\end{tabular}} & \multicolumn{1}{c|}{\begin{tabular}[c]{@{}c@{}}HR.\\      Precision\end{tabular}} & \multicolumn{1}{c|}{\begin{tabular}[c]{@{}c@{}}FL. \\ Precision\end{tabular}} & \multicolumn{1}{c|}{\begin{tabular}[c]{@{}c@{}}HR.\\       Recall\end{tabular}} & \multicolumn{1}{c|}{\begin{tabular}[c]{@{}c@{}}FL. \\      Recall\end{tabular}} & \multicolumn{1}{c|}{\begin{tabular}[c]{@{}c@{}}HR. \\      F1-score\end{tabular}} & \multicolumn{1}{c|}{\begin{tabular}[c]{@{}c@{}}FL. \\      F1-score\end{tabular}} \\ \hline
 & DT & 99.23 & \cellcolor[HTML]{FFEB9C}{\color[HTML]{9C5700} 99.33} & 99.24 & 99.33 & 99.23 & 99.33 & 99.24 & \cellcolor[HTML]{FFEB9C}{\color[HTML]{9C5700} 99.33} \\ \cline{2-10} 
 & RF & \cellcolor[HTML]{C6EFCE}{\color[HTML]{006100} 99.46} & 99.45 & 99.46 & 99.44 & 99.46 & 99.45 & \cellcolor[HTML]{C6EFCE}{\color[HTML]{006100} 99.46} & 99.44 \\ \cline{2-10} 
 & ET & \cellcolor[HTML]{C6EFCE}{\color[HTML]{006100} 99.47} & 99.45 & 99.46 & 99.45 & 99.47 & 99.45 & \cellcolor[HTML]{C6EFCE}{\color[HTML]{006100} 99.46} & 99.45 \\ \cline{2-10} 
 & LR & 95.43 & \cellcolor[HTML]{FFEB9C}{\color[HTML]{9C5700} 95.99} & 95.37 & 95.98 & 95.43 & 95.99 & 95.38 & \cellcolor[HTML]{FFEB9C}{\color[HTML]{9C5700} 95.97} \\ \cline{2-10} 
 & KNN & 99 & 99 & 99 & 98.98 & 99 & 99 & \cellcolor[HTML]{C6EFCE}{\color[HTML]{006100} 99} & 98.99 \\ \cline{2-10} 
 & NB & \cellcolor[HTML]{C6EFCE}{\color[HTML]{006100} 80.67} & 78.16 & 82.77 & 85.34 & 80.67 & 78.16 & 76.79 & \cellcolor[HTML]{FFEB9C}{\color[HTML]{9C5700} 78.13} \\ \cline{2-10} 
 & NN & 99.14 & \cellcolor[HTML]{FFEB9C}{\color[HTML]{9C5700} 99.18} & 99.13 & 99.17 & 99.14 & 99.18 & 99.13 & \cellcolor[HTML]{FFEB9C}{\color[HTML]{9C5700} 99.17} \\ \cline{2-10} 
\multirow{-8}{*}{2} & AdB & \cellcolor[HTML]{C6EFCE}{\color[HTML]{006100} 92.83} & 83.22 & 92.75 & 83.25 & 92.83 & 83.22 & \cellcolor[HTML]{C6EFCE}{\color[HTML]{006100} 92.76} & 82.43 \\ \hline
Level & \multicolumn{1}{c|}{Classifier} & \multicolumn{1}{c|}{\begin{tabular}[c]{@{}c@{}}HR.   \\ Accuracy\end{tabular}} & \multicolumn{1}{c|}{\begin{tabular}[c]{@{}c@{}}FL.   \\      Accuracy\end{tabular}} & \multicolumn{1}{c|}{\begin{tabular}[c]{@{}c@{}}HR.\\      Precision\end{tabular}} & \multicolumn{1}{c|}{\begin{tabular}[c]{@{}c@{}}FL.  \\  Precision\end{tabular}} & \multicolumn{1}{c|}{\begin{tabular}[c]{@{}c@{}}HR.\\       Recall\end{tabular}} & \multicolumn{1}{c|}{\begin{tabular}[c]{@{}c@{}}FL.   \\      Recall\end{tabular}} & \multicolumn{1}{c|}{\begin{tabular}[c]{@{}c@{}}HR.   \\      F1-score\end{tabular}} & \multicolumn{1}{c|}{\begin{tabular}[c]{@{}c@{}}FL.   \\      F1-score\end{tabular}} \\ \hline
 & DT & 99.01 & \cellcolor[HTML]{FFEB9C}{\color[HTML]{9C5700} 99.07} & 99.01 & 99.04 & 99.01 & 99.07 & 99.01 & \cellcolor[HTML]{FFEB9C}{\color[HTML]{9C5700} 99.05} \\ \cline{2-10} 
 & RF & \cellcolor[HTML]{C6EFCE}{\color[HTML]{006100} 99.27} & 99.25 & 99.22 & 99.17 & 99.27 & 99.25 & \cellcolor[HTML]{C6EFCE}{\color[HTML]{006100} 99.22} & 99.17 \\ \cline{2-10} 
 & ET & \cellcolor[HTML]{C6EFCE}{\color[HTML]{006100} 99.29} & 99.27 & 99.25 & 99.21 & 99.29 & 99.27 & \cellcolor[HTML]{C6EFCE}{\color[HTML]{006100} 99.26} & 99.23 \\ \cline{2-10} 
 & LR & 95.12 & \cellcolor[HTML]{FFEB9C}{\color[HTML]{9C5700} 97.16} & 93.84 & 96.66 & 95.12 & 97.16 & 94.29 & \cellcolor[HTML]{FFEB9C}{\color[HTML]{9C5700} 96.7} \\ \cline{2-10} 
 & KNN & \cellcolor[HTML]{C6EFCE}{\color[HTML]{006100} 98.78} & 98.77 & 98.68 & 98.66 & 98.78 & 98.77 & \cellcolor[HTML]{C6EFCE}{\color[HTML]{006100} 98.72} & 98.71 \\ \cline{2-10} 
 & NB & \cellcolor[HTML]{C6EFCE}{\color[HTML]{006100} 80.55} & 59.05 & 76.94 & 93.64 & 80.55 & 59.05 & \cellcolor[HTML]{C6EFCE}{\color[HTML]{006100} 75} & 67.62 \\ \cline{2-10} 
 & NN & 98.85 & \cellcolor[HTML]{FFEB9C}{\color[HTML]{9C5700} 98.93} & 98.79 & 98.86 & 98.85 & 98.93 & 98.77 & \cellcolor[HTML]{FFEB9C}{\color[HTML]{9C5700} 98.84} \\ \cline{2-10} 
\multirow{-8}{*}{3} & AdB & \cellcolor[HTML]{C6EFCE}{\color[HTML]{006100} 88.39} & 81.67 & 86.86 & 70.26 & 88.39 & 81.67 & \cellcolor[HTML]{C6EFCE}{\color[HTML]{006100} 87.24} & 75 \\ \hline
\end{tabular}
}
\end{table}

\subsubsection*{Performance on UNSW-NB15}

Table \ref{tab:unswnb15} present the performance of the HR and FL model for level 2 and level 3 on UNSW-NB 15 dataset in terms of accuracy, precision, recall and F1-score. After analyzing the results of the HR model and FL model for the last two levels ( Level 2, and Level 3) on the UNSW-NB15 dataset, we can find the following significant points:

\begin{itemize}
    \item At level 2, the HR model achieves accuracy of 83.44\% and 83.15\% for RF and ET, respectively, while the FL model achieves a bit higher accuracy of 83.59\% and 83.25\% for RF and ET. In the FL at level 2, most classifiers exhibit slightly higher accuracy compared to the HR model.
    
    \item However, at level 3, the HR model outperforms the FL model in five classifiers. Conversely, the FL model outperforms the HR model in another five classifiers. Both models consistently exhibit the same trend across multiple classifiers when considering the precision, recall and F1\-score.
    
\end{itemize}

The FL model exhibits a substantial drop in accuracy, precision, recall, and F1 score for NB and AdB, while the HR model maintains a relatively high performance for these two classifiers. The FL achieves relatively higher accuracy for the LR classifier, particularly at the second and third levels. However, the HR model shows relatively higher F1-score than FL model for six classifiers. In both cases (Level 2 and Level 3), the differences in accuracy and F1 score between the HR and FL models are minimal. Therefore, the performance advantage between the two models is negligible.

\begin{table}[!htbp]
\centering
\caption{Performance of HR and FL model on UNSW-NB15}
\label{tab:unswnb15}
\begin{tabular}{|c|l|l|l|l|l|l|l|l|l|}
\hline
Level & \multicolumn{1}{c|}{Classifier} & \multicolumn{1}{c|}{\begin{tabular}[c]{@{}c@{}}HR. \\ Accuracy\end{tabular}} & \multicolumn{1}{c|}{\begin{tabular}[c]{@{}c@{}}FL. \\      Accuracy\end{tabular}} & \multicolumn{1}{c|}{\begin{tabular}[c]{@{}c@{}}HR.\\      Precision\end{tabular}} & \multicolumn{1}{c|}{\begin{tabular}[c]{@{}c@{}}FL. \\ Precision\end{tabular}} & \multicolumn{1}{c|}{\begin{tabular}[c]{@{}c@{}}HR.\\       Recall\end{tabular}} & \multicolumn{1}{c|}{\begin{tabular}[c]{@{}c@{}}FL. \\      Recall\end{tabular}} & \multicolumn{1}{c|}{\begin{tabular}[c]{@{}c@{}}HR. \\      F1-score\end{tabular}} & \multicolumn{1}{c|}{\begin{tabular}[c]{@{}c@{}}FL. \\      F1-score\end{tabular}} \\ \hline
 & DT & 81.59 & \cellcolor[HTML]{FFEB9C}{\color[HTML]{9C5700} 81.72} & 80.67 & 80.83 & 81.59 & 81.72 & 80.71 & \cellcolor[HTML]{FFEB9C}{\color[HTML]{9C5700} 80.84} \\ \cline{2-10} 
 & RF & 83.44 & \cellcolor[HTML]{FFEB9C}{\color[HTML]{9C5700} 83.59} & 82.52 & 82.52 & 83.44 & 83.59 & 82.36 & \cellcolor[HTML]{FFEB9C}{\color[HTML]{9C5700} 82.39} \\ \cline{2-10} 
 & ET & 83.15 & \cellcolor[HTML]{FFEB9C}{\color[HTML]{9C5700} 83.25} & 82.1 & 82.1 & 83.15 & 83.25 & 81.91 & \cellcolor[HTML]{FFEB9C}{\color[HTML]{9C5700} 81.92} \\ \cline{2-10} 
 & LR & 76.68 & \cellcolor[HTML]{FFEB9C}{\color[HTML]{9C5700} 77.46} & 77.7 & 77.31 & 76.68 & 77.46 & 75.41 & \cellcolor[HTML]{FFEB9C}{\color[HTML]{9C5700} 76} \\ \cline{2-10} 
 & KNN & \cellcolor[HTML]{C6EFCE}{\color[HTML]{006100} 78.61} & 78.41 & 77.65 & 76.97 & 78.61 & 78.41 & \cellcolor[HTML]{C6EFCE}{\color[HTML]{006100} 77.85} & 77.36 \\ \cline{2-10} 
 & NB & \cellcolor[HTML]{C6EFCE}{\color[HTML]{006100} 39.51} & 21.91 & 36.34 & 35.64 & 39.51 & 21.91 & \cellcolor[HTML]{C6EFCE}{\color[HTML]{006100} 25.64} & 20.3 \\ \cline{2-10} 
 & NN & 81.86 & \cellcolor[HTML]{FFEB9C}{\color[HTML]{9C5700} 82.19} & 80.64 & 81.28 & 81.86 & 82.19 & 80.04 & \cellcolor[HTML]{FFEB9C}{\color[HTML]{9C5700} 80.21} \\ \cline{2-10} 
 & AdB & \cellcolor[HTML]{C6EFCE}{\color[HTML]{006100} 72.15} & 69.49 & 75.23 & 69.24 & 72.15 & 69.49 & \cellcolor[HTML]{C6EFCE}{\color[HTML]{006100} 71.86} & 68.72 \\ \cline{2-10} 
 & HGBC & \cellcolor[HTML]{C6EFCE}{\color[HTML]{006100} 83.51} & 82.77 & 84.21 & 83.14 & 83.51 & 82.77 & \cellcolor[HTML]{C6EFCE}{\color[HTML]{006100} 81.25} & 80.37 \\ \cline{2-10} 
\multirow{-10}{*}{2} & GBC & 82.21 & \cellcolor[HTML]{FFEB9C}{\color[HTML]{9C5700} 82.34} & 83.14 & 81.94 & 82.21 & 82.34 & \cellcolor[HTML]{C6EFCE}{\color[HTML]{006100} 80.11} & 80.03 \\ \hline
Level & \multicolumn{1}{c|}{Classifier} & \multicolumn{1}{c|}{\begin{tabular}[c]{@{}c@{}}HR.   \\ Accuracy\end{tabular}} & \multicolumn{1}{c|}{\begin{tabular}[c]{@{}c@{}}FL.   \\      Accuracy\end{tabular}} & \multicolumn{1}{c|}{\begin{tabular}[c]{@{}c@{}}HR.\\      Precision\end{tabular}} & \multicolumn{1}{c|}{\begin{tabular}[c]{@{}c@{}}FL. \\ Precision\end{tabular}} & \multicolumn{1}{c|}{\begin{tabular}[c]{@{}c@{}}HR.\\ Recall\end{tabular}} & \multicolumn{1}{c|}{\begin{tabular}[c]{@{}c@{}}FL.   \\ Recall\end{tabular}} & \multicolumn{1}{c|}{\begin{tabular}[c]{@{}c@{}}HR.   \\      F1-score\end{tabular}} & \multicolumn{1}{c|}{\begin{tabular}[c]{@{}c@{}}FL.   \\      F1-score\end{tabular}} \\ \hline
 & DT & 80.88 & \cellcolor[HTML]{FFEB9C}{\color[HTML]{9C5700} 81.05} & 80.1 & 81.2 & 80.88 & 81.05 & 79.7 & \cellcolor[HTML]{FFEB9C}{\color[HTML]{9C5700} 80.7} \\ \cline{2-10} 
 & RF & 82.72 & \cellcolor[HTML]{FFEB9C}{\color[HTML]{9C5700} 82.76} & 82.41 & 82.75 & 82.72 & 82.76 & 81.29 & \cellcolor[HTML]{FFEB9C}{\color[HTML]{9C5700} 81.74} \\ \cline{2-10} 
 & ET & \cellcolor[HTML]{C6EFCE}{\color[HTML]{006100} 82.36} & 82.35 & 81.92 & 82.58 & 82.36 & 82.35 & 80.75 & \cellcolor[HTML]{FFEB9C}{\color[HTML]{9C5700} 81.56} \\ \cline{2-10} 
 & LR & 74.59 & \cellcolor[HTML]{FFEB9C}{\color[HTML]{9C5700} 76.6} & 75.44 & 73.84 & 74.59 & 76.6 & 72.82 & \cellcolor[HTML]{FFEB9C}{\color[HTML]{9C5700} 73.46} \\ \cline{2-10} 
 & KNN & \cellcolor[HTML]{C6EFCE}{\color[HTML]{006100} 77.1} & 76.8 & 75.54 & 75.91 & 77.1 & 76.8 & 75.88 & \cellcolor[HTML]{FFEB9C}{\color[HTML]{9C5700} 76.1} \\ \cline{2-10} 
 & NB & \cellcolor[HTML]{C6EFCE}{\color[HTML]{006100} 39.34} & 20.4 & 33.56 & 35 & 39.34 & 20.4 & \cellcolor[HTML]{C6EFCE}{\color[HTML]{006100} 25.3} & 18.28 \\ \cline{2-10} 
 & NN & 80.71 & \cellcolor[HTML]{FFEB9C}{\color[HTML]{9C5700} 81.03} & 79.47 & 79.74 & 80.71 & 81.03 & 78.38 & \cellcolor[HTML]{FFEB9C}{\color[HTML]{9C5700} 78.9} \\ \cline{2-10} 
 & AdB & \cellcolor[HTML]{C6EFCE}{\color[HTML]{006100} 69.39} & 55.51 & 73.76 & 63.05 & 69.39 & 55.51 & \cellcolor[HTML]{C6EFCE}{\color[HTML]{006100} 69.76} & 58.37 \\ \cline{2-10} 
 & HGBC & \cellcolor[HTML]{C6EFCE}{\color[HTML]{006100} 82.56} & 81.94 & 83.67 & 81.89 & 82.56 & 81.94 & \cellcolor[HTML]{C6EFCE}{\color[HTML]{006100} 79.89} & 79.18 \\ \cline{2-10} 
\multirow{-10}{*}{3} & GBC & 81.23 & \cellcolor[HTML]{FFEB9C}{\color[HTML]{9C5700} 81.79} & 82.58 & 81.25 & 81.23 & 81.79 & 78.75 & \cellcolor[HTML]{FFEB9C}{\color[HTML]{9C5700} 79.5} \\ \hline
\end{tabular}
\end{table}

\subsubsection*{Performance on CIC-DoS2017}

Table \ref{tab:cicdos} represents the classification performance of the HR (hierarchical) model and the FL (flat) model on the CIC-DoS2017 dataset at three levels: 2, and 3. The results are shown for ten different machine learning algorithms. Upon analyzing the provided data in Table \ref{tab:cicdos}, here are some key observations and comparisons of the results:

Level 2:
\begin{itemize}

    \item At level 2, the HR model shows slightly higher accuracy compared to the FL model across various classifiers, including RF, NB, NN, AdB, and HGBC, with accuracy of 99.71\%, 30.37\%, 99.61\%, 98.48\%, and 99.65\% respectively. The lowest performing classifier for both models is NB.
    \item  In terms of F1-score, both the HR model and FL model achieve similar scores for several classifiers such as DT, RF, ET, and KNN. However, when considering NB, NN, AdB, and HGBC classifiers, the HR model demonstrates better performance compared to the FL model. There is a significant performance gap, particularly for NB and AdB, indicating that the HR model outperforms the FL model at this level in the CIC-DoS2017 dataset. 
    
\end{itemize}
Level 3:
\begin{itemize}
    \item At level 3, the HR model exhibits consistently high accuracy and well-balanced precision, recall, and F1-score across various classifiers, with RF (99.66\%), NB (35.84\%), AdB (98.85\%), HGBC (99.66\%), and GBC (99.56\%) achieving considerable performance. Conversely, the FL model outperforms the HR model for the remaining five classifiers.
    
    \item When considering the F1-score, both models demonstrate similar performance for two classifiers, DT (99.70\%) and ET (99.65\%). However, the HR outperforms the FL model for the other five classifiers. Thus, in terms of F1-score, the HR model excels over the FL model at this level. 
\end{itemize}

Overall, the HR model outperforms the FL model in terms of accuracy, precision, recall, and F1-score across different levels and classifiers on CIC-DoS2017 dataset.

\begin{table}[!htbp]
\centering
\caption{Performance of Hierarchical and Flat model on CIC-DoS2017}
\label{tab:cicdos}
\begin{tabular}{|c|l|l|l|l|l|l|l|l|l|}
\hline
Level & \multicolumn{1}{c|}{Classifier} & \multicolumn{1}{c|}{\begin{tabular}[c]{@{}c@{}}HR. \\ Accuracy\end{tabular}} & \multicolumn{1}{c|}{\begin{tabular}[c]{@{}c@{}}FL. \\      Accuracy\end{tabular}} & \multicolumn{1}{c|}{\begin{tabular}[c]{@{}c@{}}HR.\\      Precision\end{tabular}} & \multicolumn{1}{c|}{\begin{tabular}[c]{@{}c@{}}FL.\\  Precision\end{tabular}} & \multicolumn{1}{c|}{\begin{tabular}[c]{@{}c@{}}HR.\\       Recall\end{tabular}} & \multicolumn{1}{c|}{\begin{tabular}[c]{@{}c@{}}FL. \\      Recall\end{tabular}} & \multicolumn{1}{c|}{\begin{tabular}[c]{@{}c@{}}HR. \\      F1-score\end{tabular}} & \multicolumn{1}{c|}{\begin{tabular}[c]{@{}c@{}}FL. \\      F1-score\end{tabular}} \\ \hline
 & DT & 99.49 & 99.49 & 99.49 & 99.49 & 99.49 & 99.49 & \cellcolor[HTML]{F2F2F2}{\color[HTML]{FA7D00} \textbf{99.49}} & \cellcolor[HTML]{F2F2F2}{\color[HTML]{FA7D00} \textbf{99.49}} \\ \cline{2-10} 
 & RF & \cellcolor[HTML]{C6EFCE}{\color[HTML]{006100} 99.71} & 99.7 & 99.71 & 99.7 & 99.71 & 99.7 & \cellcolor[HTML]{F2F2F2}{\color[HTML]{FA7D00} \textbf{99.7}} & \cellcolor[HTML]{F2F2F2}{\color[HTML]{FA7D00} \textbf{99.7}} \\ \cline{2-10} 
 & ET & 99.65 & 99.65 & 99.65 & 99.65 & 99.65 & 99.65 & \cellcolor[HTML]{F2F2F2}{\color[HTML]{FA7D00} \textbf{99.65}} & \cellcolor[HTML]{FFEB9C}{\color[HTML]{9C5700} 99.65} \\ \cline{2-10} 
 & LR & 96.65 & \cellcolor[HTML]{FFEB9C}{\color[HTML]{9C5700} 96.74} & 96.48 & 96.55 & 96.65 & 96.74 & 96.2 & \cellcolor[HTML]{F2F2F2}{\color[HTML]{FA7D00} \textbf{96.33}} \\ \cline{2-10} 
 & KNN & 99.62 & \cellcolor[HTML]{FFEB9C}{\color[HTML]{9C5700} 99.63} & 99.62 & 99.62 & 99.62 & 99.63 & \cellcolor[HTML]{F2F2F2}{\color[HTML]{FA7D00} \textbf{99.62}} & 99.62 \\ \cline{2-10} 
 & NB & \cellcolor[HTML]{C6EFCE}{\color[HTML]{006100} 36.34} & 30.37 & 93.8 & 93.86 & 36.34 & 30.37 & \cellcolor[HTML]{C6EFCE}{\color[HTML]{006100} 49.01} & 42.08 \\ \cline{2-10} 
 & NN & \cellcolor[HTML]{C6EFCE}{\color[HTML]{006100} 99.62} & 99.61 & 99.62 & 99.61 & 99.62 & 99.61 & \cellcolor[HTML]{C6EFCE}{\color[HTML]{006100} 99.62} & 99.61 \\ \cline{2-10} 
 & AdB & \cellcolor[HTML]{C6EFCE}{\color[HTML]{006100} 99.34} & 98.48 & 99.33 & 98.49 & 99.34 & 98.48 & \cellcolor[HTML]{C6EFCE}{\color[HTML]{006100} 99.33} & 98.47 \\ \cline{2-10} 
 & HGBC & \cellcolor[HTML]{C6EFCE}{\color[HTML]{006100} 99.71} & 99.65 & 99.7 & 99.65 & 99.71 & 99.65 & \cellcolor[HTML]{C6EFCE}{\color[HTML]{006100} 99.7} & \cellcolor[HTML]{FFEB9C}{\color[HTML]{9C5700} 99.65} \\ \cline{2-10} 
\multirow{-10}{*}{2} & GBC & 99.6 & \cellcolor[HTML]{FFEB9C}{\color[HTML]{9C5700} 99.62} & 99.6 & 99.62 & 99.6 & 99.62 & \cellcolor[HTML]{F2F2F2}{\color[HTML]{FA7D00} \textbf{99.6}} & \cellcolor[HTML]{F2F2F2}{\color[HTML]{FA7D00} \textbf{99.62}} \\ \hline
Level & \multicolumn{1}{c|}{Classifier} & \multicolumn{1}{c|}{\begin{tabular}[c]{@{}c@{}}HR.   \\ Accuracy\end{tabular}} & \multicolumn{1}{c|}{\begin{tabular}[c]{@{}c@{}}FL.   \\      Accuracy\end{tabular}} & \multicolumn{1}{c|}{\begin{tabular}[c]{@{}c@{}}HR.\\      Precision\end{tabular}} & \multicolumn{1}{c|}{\begin{tabular}[c]{@{}c@{}}FL. \\ Precision\end{tabular}} & \multicolumn{1}{c|}{\begin{tabular}[c]{@{}c@{}}HR.\\       Recall\end{tabular}} & \multicolumn{1}{c|}{\begin{tabular}[c]{@{}c@{}}FL.   \\      Recall\end{tabular}} & \multicolumn{1}{c|}{\begin{tabular}[c]{@{}c@{}}HR. \\      F1-score\end{tabular}} & \multicolumn{1}{c|}{\begin{tabular}[c]{@{}c@{}}FL. \\      F1-score\end{tabular}} \\ \hline
 & DT & 99.43 & 99.43 & 99.43 & 99.43 & 99.43 & 99.43 & \cellcolor[HTML]{F2F2F2}{\color[HTML]{FA7D00} \textbf{99.43}} & \cellcolor[HTML]{F2F2F2}{\color[HTML]{FA7D00} \textbf{99.43}} \\ \cline{2-10} 
 & RF & \cellcolor[HTML]{C6EFCE}{\color[HTML]{006100} 99.66} & 99.65 & 99.66 & 99.65 & 99.66 & 99.65 & \cellcolor[HTML]{C6EFCE}{\color[HTML]{006100} 99.66} & 99.64 \\ \cline{2-10} 
 & ET & 99.6 & \cellcolor[HTML]{FFEB9C}{\color[HTML]{9C5700} 99.61} & 99.59 & 99.6 & 99.6 & 99.61 & \cellcolor[HTML]{F2F2F2}{\color[HTML]{FA7D00} \textbf{99.6}} & \cellcolor[HTML]{F2F2F2}{\color[HTML]{FA7D00} \textbf{99.6}} \\ \cline{2-10} 
 & LR & 96.37 & \cellcolor[HTML]{FFEB9C}{\color[HTML]{9C5700} 96.63} & 95.81 & 96 & 96.37 & 96.63 & 95.71 & \cellcolor[HTML]{FFEB9C}{\color[HTML]{9C5700} 96.04} \\ \cline{2-10} 
 & KNN & 99.54 & \cellcolor[HTML]{FFEB9C}{\color[HTML]{9C5700} 99.56} & 99.53 & 99.55 & 99.54 & 99.56 & 99.53 & \cellcolor[HTML]{FFEB9C}{\color[HTML]{9C5700} 99.55} \\ \cline{2-10} 
 & NB & \cellcolor[HTML]{C6EFCE}{\color[HTML]{006100} 35.84} & 27.68 & 94.18 & 94.81 & 35.84 & 27.68 & \cellcolor[HTML]{C6EFCE}{\color[HTML]{006100} 49.5} & 38.99 \\ \cline{2-10} 
 & NN & 99.51 & 99.51 & 99.49 & 99.5 & 99.51 & 99.51 & 99.49 & \cellcolor[HTML]{FFEB9C}{\color[HTML]{9C5700} 99.5} \\ \cline{2-10} 
 & AdB & \cellcolor[HTML]{C6EFCE}{\color[HTML]{006100} 98.85} & 77.89 & 98.82 & 93.66 & 98.85 & 77.89 & \cellcolor[HTML]{C6EFCE}{\color[HTML]{006100} 98.82} & 84.89 \\ \cline{2-10} 
 & HGBC & \cellcolor[HTML]{C6EFCE}{\color[HTML]{006100} 99.66} & 98.63 & 99.66 & 98.71 & 99.66 & 98.63 & \cellcolor[HTML]{C6EFCE}{\color[HTML]{006100} 99.66} & 98.66 \\ \cline{2-10} 
\multirow{-10}{*}{3} & GBC & \cellcolor[HTML]{C6EFCE}{\color[HTML]{006100} 99.56} & 99.34 & 99.55 & 99.31 & 99.56 & 99.34 & \cellcolor[HTML]{C6EFCE}{\color[HTML]{006100} 99.55} & 99.31 \\ \hline
\end{tabular}
\end{table}

\subsubsection*{Performance on CIC-DDoS2019 dataset}

Table \ref{tab:cicddos} represents the performance of the HR model and the FL model for Level 2 and Level 3 on CIC-DDoS2019 dataset. Comparing the HR and FL models on the CIC-DDoS 2019dataset, we can analyze the values from the provided Table \ref{tab:cicddos}. 

Level 2:

\begin{itemize}
    \item At level 2, both the HR and FL models exhibit a slight decrease in performance across most classifiers at this level. However, the HR model demonstrates superior accuracy compared to the FL model for the following classifiers: DT (96.56\%), ET (96.4\%), LR (91.91\%), NB (88.22\%), AdB (92.96\%), and HGBC (97.05\%). Conversely, the FL model outperforms the HR model for the KNN (95.86\%), NN (96.37\%), and GBC (96.68\%) classifiers. The highest accuracy achieved by both models is 96.81, attained by the RF classifier.
    
    \item The F1-score for DT (96.7\%) and RF (96.98\%) are comparable for both the HR and FL models. However, in terms of F1-score, the HR model outperforms the FL model for most of the classifiers. These findings suggest that the HR model is slightly superior to the FL model on the CIC-DDoS dataset.

\end{itemize}

Level 3:

\begin{itemize}

    \item At level 3, both the HR and FL models show a further decrease in performance compared to levels 2 across most classifiers. However, the HR model achieves higher accuracy for eight out of the ten classifiers. The FL model slightly outperforms the HR model for only two classifiers: DT (93.24\%) and NN (93.34\%). 
    
    \item The HR model demonstrates better performance than the FL model in terms of F1-score for most of the classifiers. The HR model achieves the same F1-score as the FL model for RF (92.86\%) and NN (92.02\%). 
    
\end{itemize}

Overall, based on the provided data in Table \ref{tab:cicddos}, the HR model consistently outperforms the FL model across all levels and classifiers. The HR model achieves higher accuracy, precision, recall, and F1-score compared to the FL model. The HR and FL models perform similarly, with the HR models generally having a slight edge in terms of accuracy. Therefore, the HR model is the better performer for the CIC-DDoS2019 dataset.

\begin{table}[!htbp]
\centering
\caption{Performance of Hierarchical and Flat model on CIC-DDoS2019}
\label{tab:cicddos}
\begin{tabular}{|c|l|l|l|l|l|l|l|l|l|}
\hline
Level & \multicolumn{1}{c|}{Classifier} & \multicolumn{1}{c|}{\begin{tabular}[c]{@{}c@{}}HR.\\  Accuracy\end{tabular}} & \multicolumn{1}{c|}{\begin{tabular}[c]{@{}c@{}}FL. \\      Accuracy\end{tabular}} & \multicolumn{1}{c|}{\begin{tabular}[c]{@{}c@{}}HR.\\      Precision\end{tabular}} & \multicolumn{1}{c|}{\begin{tabular}[c]{@{}c@{}}FL. \\ Precision\end{tabular}} & \multicolumn{1}{c|}{\begin{tabular}[c]{@{}c@{}}HR.\\       Recall\end{tabular}} & \multicolumn{1}{c|}{\begin{tabular}[c]{@{}c@{}}FL. \\      Recall\end{tabular}} & \multicolumn{1}{c|}{\begin{tabular}[c]{@{}c@{}}HR. \\      F1-score\end{tabular}} & \multicolumn{1}{c|}{\begin{tabular}[c]{@{}c@{}}FL. \\      F1-score\end{tabular}} \\ \hline
 & DT & \cellcolor[HTML]{C6EFCE}{\color[HTML]{006100} 96.56} & 96.55 & 97.03 & 97.03 & 96.56 & 96.55 & \cellcolor[HTML]{F2F2F2}{\color[HTML]{FA7D00} \textbf{96.7}} & \cellcolor[HTML]{F2F2F2}{\color[HTML]{FA7D00} \textbf{96.7}} \\ \cline{2-10} 
 & RF & 96.81 & 96.81 & 97.39 & 97.39 & 96.81 & 96.81 & \cellcolor[HTML]{F2F2F2}{\color[HTML]{FA7D00} \textbf{96.98}} & \cellcolor[HTML]{F2F2F2}{\color[HTML]{FA7D00} \textbf{96.98}} \\ \cline{2-10} 
 & ET & \cellcolor[HTML]{C6EFCE}{\color[HTML]{006100} 96.4} & 96.39 & 96.63 & 96.62 & 96.4 & 96.39 & \cellcolor[HTML]{C6EFCE}{\color[HTML]{006100} 96.49} & 96.47 \\ \cline{2-10} 
 & LR & \cellcolor[HTML]{C6EFCE}{\color[HTML]{006100} 91.91} & 91.17 & 90.76 & 88.43 & 91.91 & 91.17 & \cellcolor[HTML]{C6EFCE}{\color[HTML]{006100} 91.02} & 89.03 \\ \cline{2-10} 
 & KNN & 95.85 & \cellcolor[HTML]{FFEB9C}{\color[HTML]{9C5700} 95.86} & 96.08 & 96.1 & 95.85 & 95.86 & 95.94 & \cellcolor[HTML]{FFEB9C}{\color[HTML]{9C5700} 95.95} \\ \cline{2-10} 
 & NB & \cellcolor[HTML]{C6EFCE}{\color[HTML]{006100} 88.22} & 82.84 & 83.05 & 83.6 & 88.22 & 82.84 & \cellcolor[HTML]{C6EFCE}{\color[HTML]{006100} 84.95} & 82.11 \\ \cline{2-10} 
 & NN & 96.34 & \cellcolor[HTML]{FFEB9C}{\color[HTML]{9C5700} 96.37} & 96.95 & 97.02 & 96.34 & 96.37 & 96.51 & \cellcolor[HTML]{FFEB9C}{\color[HTML]{9C5700} 96.55} \\ \cline{2-10} 
 & AdB & \cellcolor[HTML]{C6EFCE}{\color[HTML]{006100} 92.96} & 89.08 & 92.24 & 88.33 & 92.96 & 89.08 & \cellcolor[HTML]{C6EFCE}{\color[HTML]{006100} 92.47} & 88.55 \\ \cline{2-10} 
 & HGBC & \cellcolor[HTML]{C6EFCE}{\color[HTML]{006100} 97.05} & 97.03 & 97.34 & 97.33 & 97.05 & 97.03 & \cellcolor[HTML]{C6EFCE}{\color[HTML]{006100} 97.14} & 97.13 \\ \cline{2-10} 
\multirow{-10}{*}{2} & GBC & 96.67 & \cellcolor[HTML]{FFEB9C}{\color[HTML]{9C5700} 96.68} & 97.34 & 97.34 & 96.67 & 96.68 & 96.85 & \cellcolor[HTML]{FFEB9C}{\color[HTML]{9C5700} 96.86} \\ \hline
Level & \multicolumn{1}{c|}{Classifier} & \multicolumn{1}{c|}{\begin{tabular}[c]{@{}c@{}}HR. \\   Accuracy\end{tabular}} & \multicolumn{1}{c|}{\begin{tabular}[c]{@{}c@{}}FL.   \\      Accuracy\end{tabular}} & \multicolumn{1}{c|}{\begin{tabular}[c]{@{}c@{}}HR.\\      Precision\end{tabular}} & \multicolumn{1}{c|}{\begin{tabular}[c]{@{}c@{}}FL.   \\ Precision\end{tabular}} & \multicolumn{1}{c|}{\begin{tabular}[c]{@{}c@{}}HR.\\       Recall\end{tabular}} & \multicolumn{1}{c|}{\begin{tabular}[c]{@{}c@{}}FL.   \\      Recall\end{tabular}} & \multicolumn{1}{c|}{\begin{tabular}[c]{@{}c@{}}HR.   \\      F1-score\end{tabular}} & \multicolumn{1}{c|}{\begin{tabular}[c]{@{}c@{}}FL.   \\      F1-score\end{tabular}} \\ \hline
 & DT & 93.21 & \cellcolor[HTML]{FFEB9C}{\color[HTML]{9C5700} 93.24} & 92.8 & 92.84 & 93.21 & 93.24 & 92.61 & \cellcolor[HTML]{FFEB9C}{\color[HTML]{9C5700} 92.65} \\ \cline{2-10} 
 & RF & \cellcolor[HTML]{C6EFCE}{\color[HTML]{006100} 93.58} & 93.54 & 93.3 & 93.24 & 93.58 & 93.54 & \cellcolor[HTML]{F2F2F2}{\color[HTML]{FA7D00} \textbf{92.86}} & \cellcolor[HTML]{F2F2F2}{\color[HTML]{FA7D00} \textbf{92.85}} \\ \cline{2-10} 
 & ET & \cellcolor[HTML]{C6EFCE}{\color[HTML]{006100} 93.13} & 93.05 & 92.93 & 92.89 & 93.13 & 93.05 & \cellcolor[HTML]{C6EFCE}{\color[HTML]{006100} 92.88} & 92.86 \\ \cline{2-10} 
 & LR & \cellcolor[HTML]{C6EFCE}{\color[HTML]{006100} 89.27} & 86.6 & 87.89 & 83.57 & 89.27 & 86.6 & \cellcolor[HTML]{C6EFCE}{\color[HTML]{006100} 87.89} & 83.32 \\ \cline{2-10} 
 & KNN & \cellcolor[HTML]{C6EFCE}{\color[HTML]{006100} 92.74} & 92.66 & 92.55 & 92.62 & 92.74 & 92.66 & 92.35 & \cellcolor[HTML]{FFEB9C}{\color[HTML]{9C5700} 92.39} \\ \cline{2-10} 
 & NB & \cellcolor[HTML]{C6EFCE}{\color[HTML]{006100} 79.33} & 56.49 & 83.42 & 86.41 & 79.33 & 56.49 & \cellcolor[HTML]{C6EFCE}{\color[HTML]{006100} 79.6} & 63.71 \\ \cline{2-10} 
 & NN & 93.25 & \cellcolor[HTML]{FFEB9C}{\color[HTML]{9C5700} 93.34} & 92.68 & 93.38 & 93.25 & 93.34 & \cellcolor[HTML]{F2F2F2}{\color[HTML]{FA7D00} \textbf{92.02}} & \cellcolor[HTML]{F2F2F2}{\color[HTML]{FA7D00} \textbf{92.2}} \\ \cline{2-10} 
 & AdB & \cellcolor[HTML]{C6EFCE}{\color[HTML]{006100} 89.53} & 71.77 & 88.23 & 77.03 & 89.53 & 71.77 & \cellcolor[HTML]{C6EFCE}{\color[HTML]{006100} 88.25} & 73.5 \\ \cline{2-10} 
 & HGBC & \cellcolor[HTML]{C6EFCE}{\color[HTML]{006100} 94.4} & 93.31 & 94.42 & 93.59 & 94.4 & 93.31 & \cellcolor[HTML]{C6EFCE}{\color[HTML]{006100} 94.03} & 93.27 \\ \cline{2-10} 
\multirow{-10}{*}{3} & GBC & \cellcolor[HTML]{C6EFCE}{\color[HTML]{006100} 93.75} & 92.67 & 93.56 & 93.05 & 93.75 & 92.67 & \cellcolor[HTML]{C6EFCE}{\color[HTML]{006100} 92.53} & 91.8 \\ \hline
\end{tabular}
\end{table}

\subsubsection*{Performance on Malmem2022}

Table \ref{tab:malmem} shows the classification performance for HR and FL models of 10 classifiers for level 2, and 3. The following are the key observations:

Level 2:

\begin{itemize}
    \item In general, both models perform similarly across most classifiers. The accuracy comparison between the HR and FL models at level 2 shows that there is no clear trend of one model consistently outperforming the other. In some cases, the FL model achieves slightly higher accuracy, while in other cases, the HR model performs slightly better. However, there are some variations in performance. Notable differences include: HR model achieves higher accuracy for LR (70.99\%), AdB (78.66\%), HGBC (87.22\%), and GBC (82.61\%) classifiers. KNN (80.52\%) and NB (80.52\%) classifiers have identical accuracy for both models. The HR and FL models demonstrate comparable performance, with slight variations observed across different classifiers.
    
    \item The F1 scores of the HR and FL models for 10 classifiers indicate that FL exhibits better performance across most classifiers. The HR model performs better than the FL model for AdB (78.6\%), HGBC (87.2\%) and GBC (82.63\%) classifiers in terms of F1 score. AdB(HR: 78.6\%, FL: 66.92\%) classifier shows a slightly higher F1- score for the HR model. KNN (80.43\%) classifier has identical F1-score for both models.
    
\end{itemize}

Level 3:

\begin{itemize}
    \item The accuracy values for both the HR and FL models vary across the classifiers. However, in general, the accuracy of the FL model is slightly higher than the HR model for most classifiers.
    \item There is the same trend of F1-score like accuracy between HR and FL. The F1-score for FL model are generally better across the most classifiers. 
    
    \end{itemize}

Overall, the FL model tends to exhibit slightly better accuracy and precision compared to the HR model for most classifiers. However, there is no clear distinction between the models in terms of recall and F1 score. 

\begin{table}[!htbp]
\centering
\caption{Performance of Hierarchical and Flat model on Malmem2022 dataset}
\label{tab:malmem}
\begin{tabular}{|c|c|c|c|c|c|c|c|c|c|}
\hline
Level & Classifier & \begin{tabular}[c]{@{}c@{}}HR. \\ Accuracy\end{tabular} & \begin{tabular}[c]{@{}c@{}}FL. \\      Accuracy\end{tabular} & \begin{tabular}[c]{@{}c@{}}HR.\\      Precision\end{tabular} & \begin{tabular}[c]{@{}c@{}}FL. \\ Precision\end{tabular} & \begin{tabular}[c]{@{}c@{}}HR.\\       Recall\end{tabular} & \begin{tabular}[c]{@{}c@{}}FL. \\      Recall\end{tabular} & \begin{tabular}[c]{@{}c@{}}HR. \\      F1-score\end{tabular} & \begin{tabular}[c]{@{}c@{}}FL. \\      F1-score\end{tabular} \\ \hline
 & DT & 85.14 & \cellcolor[HTML]{FFEB9C}{\color[HTML]{9C5700} 85.21} & 85.15 & 85.22 & 85.14 & 85.21 & 85.15 & \cellcolor[HTML]{FFEB9C}{\color[HTML]{9C5700} 85.22} \\ \cline{2-10} 
 & RF & 87.9 & \cellcolor[HTML]{FFEB9C}{\color[HTML]{9C5700} 87.95} & 87.89 & 87.95 & 87.9 & 87.95 & 87.89 & \cellcolor[HTML]{FFEB9C}{\color[HTML]{9C5700} 87.95} \\ \cline{2-10} 
 & ET & 86.99 & \cellcolor[HTML]{FFEB9C}{\color[HTML]{9C5700} 87.03} & 86.99 & 87.03 & 86.99 & 87.03 & 86.98 & \cellcolor[HTML]{FFEB9C}{\color[HTML]{9C5700} 87.03} \\ \cline{2-10} 
 & LR & \cellcolor[HTML]{C6EFCE}{\color[HTML]{006100} 70.99} & 70.9 & 71.28 & 71.15 & 70.99 & 70.9 & 70.98 & \cellcolor[HTML]{FFEB9C}{\color[HTML]{9C5700} 70.85} \\ \cline{2-10} 
 & KNN & 80.52 & 80.52 & 80.52 & 80.52 & 80.52 & 80.52 & \cellcolor[HTML]{F2F2F2}{\color[HTML]{FA7D00} \textbf{80.43}} & \cellcolor[HTML]{F2F2F2}{\color[HTML]{FA7D00} \textbf{80.43}} \\ \cline{2-10} 
 & NB & 68.31 & \cellcolor[HTML]{FFEB9C}{\color[HTML]{9C5700} 68.33} & 73.03 & 73.1 & 68.31 & 68.33 & 63.85 & \cellcolor[HTML]{FFEB9C}{\color[HTML]{9C5700} 63.95} \\ \cline{2-10} 
 & NN & 76.58 & \cellcolor[HTML]{FFEB9C}{\color[HTML]{9C5700} 76.81} & 76.69 & 77.32 & 76.58 & 76.81 & 76.6 & \cellcolor[HTML]{FFEB9C}{\color[HTML]{9C5700} 76.74} \\ \cline{2-10} 
 & AdB & \cellcolor[HTML]{C6EFCE}{\color[HTML]{006100} 78.66} & 68.31 & 78.86 & 70.41 & 78.66 & 68.31 & \cellcolor[HTML]{C6EFCE}{\color[HTML]{006100} 78.6} & 66.92 \\ \cline{2-10} 
 & HGBC & \cellcolor[HTML]{C6EFCE}{\color[HTML]{006100} 87.22} & 87.17 & 87.2 & 87.15 & 87.22 & 87.17 & \cellcolor[HTML]{C6EFCE}{\color[HTML]{006100} 87.2} & 87.15 \\ \cline{2-10} 
\multirow{-10}{*}{2} & GBC & \cellcolor[HTML]{C6EFCE}{\color[HTML]{006100} 82.61} & 82.45 & 82.77 & 82.56 & 82.61 & 82.45 & \cellcolor[HTML]{C6EFCE}{\color[HTML]{006100} 82.63} & 82.46 \\ \hline
Level & Classifier & \begin{tabular}[c]{@{}c@{}}HR.  \\  Accuracy\end{tabular} & \begin{tabular}[c]{@{}c@{}}FL.   \\      Accuracy\end{tabular} & \begin{tabular}[c]{@{}c@{}}HR.\\      Precision\end{tabular} & \begin{tabular}[c]{@{}c@{}}FL.  \\  Precision\end{tabular} & \begin{tabular}[c]{@{}c@{}}HR.\\       Recall\end{tabular} & \begin{tabular}[c]{@{}c@{}}FL.   \\      Recall\end{tabular} & \begin{tabular}[c]{@{}c@{}}HR.   \\      F1-score\end{tabular} & \begin{tabular}[c]{@{}c@{}}FL.   \\      F1-score\end{tabular} \\ \hline
 & DT & 72.8 & \cellcolor[HTML]{FFEB9C}{\color[HTML]{9C5700} 73.93} & 72.79 & 73.89 & 72.8 & 73.93 & 72.78 & \cellcolor[HTML]{FFEB9C}{\color[HTML]{9C5700} 73.9} \\ \cline{2-10} 
 & RF & 75.77 & \cellcolor[HTML]{FFEB9C}{\color[HTML]{9C5700} 76.32} & 75.62 & 76.2 & 75.77 & 76.32 & 75.57 & \cellcolor[HTML]{FFEB9C}{\color[HTML]{9C5700} 76.17} \\ \cline{2-10} 
 & ET & 74.79 & \cellcolor[HTML]{FFEB9C}{\color[HTML]{9C5700} 75.11} & 74.63 & 74.96 & 74.79 & 75.11 & 74.65 & \cellcolor[HTML]{FFEB9C}{\color[HTML]{9C5700} 74.99} \\ \cline{2-10} 
 & LR & 57.57 & \cellcolor[HTML]{FFEB9C}{\color[HTML]{9C5700} 58.08} & 58.47 & 59.19 & 57.57 & 58.08 & \cellcolor[HTML]{C6EFCE}{\color[HTML]{006100} 56.77} & 56.61 \\ \cline{2-10} 
 & KNN & 66.38 & \cellcolor[HTML]{FFEB9C}{\color[HTML]{9C5700} 66.6} & 66.56 & 66.84 & 66.38 & 66.6 & 66.12 & \cellcolor[HTML]{FFEB9C}{\color[HTML]{9C5700} 66.39} \\ \cline{2-10} 
 & NB & 55.16 & \cellcolor[HTML]{FFEB9C}{\color[HTML]{9C5700} 56.53} & 59.29 & 62.13 & 55.16 & 56.53 & 53.05 & \cellcolor[HTML]{FFEB9C}{\color[HTML]{9C5700} 54.23} \\ \cline{2-10} 
 & NN & 61.22 & \cellcolor[HTML]{FFEB9C}{\color[HTML]{9C5700} 62.25} & 63.1 & 63.68 & 61.22 & 62.25 & 61.27 & \cellcolor[HTML]{FFEB9C}{\color[HTML]{9C5700} 61.8} \\ \cline{2-10} 
 & AdB & \cellcolor[HTML]{C6EFCE}{\color[HTML]{006100} 63.74} & 56.12 & 63.46 & 58.76 & 63.74 & 56.12 & \cellcolor[HTML]{C6EFCE}{\color[HTML]{006100} 63.17} & 55.2 \\ \cline{2-10} 
 & HGBC & 75.56 & \cellcolor[HTML]{FFEB9C}{\color[HTML]{9C5700} 76.33} & 75.46 & 76.29 & 75.56 & 76.33 & 75.22 & \cellcolor[HTML]{FFEB9C}{\color[HTML]{9C5700} 76.07} \\ \cline{2-10} 
\multirow{-10}{*}{3} & GBC & 69.86 & \cellcolor[HTML]{FFEB9C}{\color[HTML]{9C5700} 71.35} & 69.77 & 71.4 & 69.86 & 71.35 & 69.45 & \cellcolor[HTML]{FFEB9C}{\color[HTML]{9C5700} 70.76} \\ \hline
\end{tabular}
\end{table}

\subsubsection*{Performance Analysis on CIC-Darknet2020}

Table \ref{tab:cicdarknet} illustrates the performance results of HR and FL model on CIC-Darknet2020 dataset for level 2 and level 3. At level 2, both the HR and FL models show similar trends in the classification results.

Level 2: 

\begin{itemize}
    \item For the HR model, DT, RF, and ET achieve accuracy scores above 98\%. LR and GBC achieve accuracy scores above 97\%. KNN and NN achieve accuracy scores above 95\%.
    \item For the FL model, the accuracy scores are comparable to the HR model, with most classifiers achieving scores above 98\%. The performance of LR and GBC is similar to the HR model, with accuracy scores above 97\%. KNN and NN achieve accuracy scores above 95\%.
    \item The NB classifier performs poorly in both models, with accuracy scores of 55.04\% in the HR model and 36.07\% in the FL model. This indicates that NB is not well-suited for the given dataset or classification task. Both the HR (hierarchical) and FL (flat) models exhibit similar trends for performance metrics such as F1-score. 
    
\end{itemize}

Overall, both models demonstrate good performance at level 2, with several classifiers achieving high accuracy, precision, recall, and F1 scores. The HR and FL models show similar results, indicating that the HR approach does not significantly outperform the flat model at this level.

Level 3:

\begin{itemize}
    \item For level 3, both the HR and FL models demonstrate varying performance across different classifiers. The accuracy, precision, recall, and F1-score vary for each classifier. However, the HR model generally shows higher accuracy and precision compared to the FL model for RF, ET, NB, AdB, HGBC, and GBC. The FL model performs better for DT (HR:97.77\%, FL: 97.81\%), LR(HR: 86.52\%, FL: 86.91\%), KNN (HR: 93.51\%, FL: 93.63\%), and NN (HR: 94.32\%, FL: 94.59\%) in terms of accuracy. 
    \item For the F1-score metric in level 3, the FL model exhibits slightly better performance than the HR model in three classifiers: DT (HR:97.75\%, FL: 97.79\%), KNN (HR: 93.42\%, FL: 93.43\%), and NN (HR: 94.07\%, FL: 94.22\%). However, the HR model demonstrates higher F1-score for the remaining classifiers.

\end{itemize}
  
 Overall, the HR model tends to perform better than the FL model in terms of accuracy and error metrics across various classifiers at level 3.

\begin{table}[!htbp]
\centering
\caption{Performance of Hierarchical and Flat model on CIC-Darknet2020}
\label{tab:cicdarknet}
\begin{tabular}{|c|l|l|l|l|l|l|l|l|l|}
\hline
Level & \multicolumn{1}{c|}{Classifier} & \multicolumn{1}{c|}{\begin{tabular}[c]{@{}c@{}}HR. \\ Accuracy\end{tabular}} & \multicolumn{1}{c|}{\begin{tabular}[c]{@{}c@{}}FL. \\      Accuracy\end{tabular}} & \multicolumn{1}{c|}{\begin{tabular}[c]{@{}c@{}}HR.\\      Precision\end{tabular}} & \multicolumn{1}{c|}{\begin{tabular}[c]{@{}c@{}}FL. \\ Precision\end{tabular}} & \multicolumn{1}{c|}{\begin{tabular}[c]{@{}c@{}}HR.\\       Recall\end{tabular}} & \multicolumn{1}{c|}{\begin{tabular}[c]{@{}c@{}}FL. \\      Recall\end{tabular}} & \multicolumn{1}{c|}{\begin{tabular}[c]{@{}c@{}}HR. \\      F1-score\end{tabular}} & \multicolumn{1}{c|}{\begin{tabular}[c]{@{}c@{}}FL. \\      F1-score\end{tabular}} \\ \hline
 & DT & 98.25 & \cellcolor[HTML]{FFEB9C}{\color[HTML]{9C5700} 98.26} & 98.25 & 98.26 & 98.25 & 98.26 & 98.25 & \cellcolor[HTML]{FFEB9C}{\color[HTML]{9C5700} 98.26} \\ \cline{2-10} 
 & RF & \cellcolor[HTML]{C6EFCE}{\color[HTML]{006100} 98.66} & 98.64 & 98.65 & 98.63 & 98.66 & 98.64 & \cellcolor[HTML]{C6EFCE}{\color[HTML]{006100} 98.65} & 98.63 \\ \cline{2-10} 
 & ET & 98.45 & \cellcolor[HTML]{FFEB9C}{\color[HTML]{9C5700} 98.46} & 98.45 & 98.46 & 98.45 & 98.46 & 98.45 & \cellcolor[HTML]{FFEB9C}{\color[HTML]{9C5700} 98.46} \\ \cline{2-10} 
 & LR & 90.18 & \cellcolor[HTML]{FFEB9C}{\color[HTML]{9C5700} 90.31} & 89.95 & 90.11 & 90.18 & 90.31 & 90.05 & \cellcolor[HTML]{FFEB9C}{\color[HTML]{9C5700} 90.19} \\ \cline{2-10} 
 & KNN & 95.2 & 95.2 & 95.24 & 95.23 & 95.2 & 95.2 & \cellcolor[HTML]{C6EFCE}{\color[HTML]{006100} 95.22} & 95.21 \\ \cline{2-10} 
 & NB & \cellcolor[HTML]{C6EFCE}{\color[HTML]{006100} 55.04} & 36.07 & 83.04 & 80.96 & 55.04 & 36.07 & \cellcolor[HTML]{C6EFCE}{\color[HTML]{006100} 60.29} & 38.81 \\ \cline{2-10} 
 & NN & 95.81 & \cellcolor[HTML]{FFEB9C}{\color[HTML]{9C5700} 95.85} & 95.82 & 95.88 & 95.81 & 95.85 & 95.81 & \cellcolor[HTML]{FFEB9C}{\color[HTML]{9C5700} 95.86} \\ \cline{2-10} 
 & AdB & \cellcolor[HTML]{C6EFCE}{\color[HTML]{006100} 95.08} & 93.94 & 94.99 & 93.78 & 95.08 & 93.94 & \cellcolor[HTML]{C6EFCE}{\color[HTML]{006100} 95.02} & 93.56 \\ \cline{2-10} 
 & HGBC & 98.52 & \cellcolor[HTML]{FFEB9C}{\color[HTML]{9C5700} 98.71} & 98.51 & 98.71 & 98.52 & 98.71 & 98.51 & \cellcolor[HTML]{FFEB9C}{\color[HTML]{9C5700} 98.71} \\ \cline{2-10} 
\multirow{-10}{*}{2} & GBC & 97.23 & \cellcolor[HTML]{FFEB9C}{\color[HTML]{9C5700} 97.55} & 97.2 & 97.53 & 97.23 & 97.55 & 97.2 & \cellcolor[HTML]{FFEB9C}{\color[HTML]{9C5700} 97.53} \\ \hline
Level & \multicolumn{1}{c|}{Classifier} & \multicolumn{1}{c|}{\begin{tabular}[c]{@{}c@{}}HR. \\ Accuracy\end{tabular}} & \multicolumn{1}{c|}{\begin{tabular}[c]{@{}c@{}}FL.   \\      Accuracy\end{tabular}} & \multicolumn{1}{c|}{\begin{tabular}[c]{@{}c@{}}HR.\\      Precision\end{tabular}} & \multicolumn{1}{c|}{\begin{tabular}[c]{@{}c@{}}FL. \\ Precision\end{tabular}} & \multicolumn{1}{c|}{\begin{tabular}[c]{@{}c@{}}HR.\\       Recall\end{tabular}} & \multicolumn{1}{c|}{\begin{tabular}[c]{@{}c@{}}FL.   \\      Recall\end{tabular}} & \multicolumn{1}{c|}{\begin{tabular}[c]{@{}c@{}}HR.   \\      F1-score\end{tabular}} & \multicolumn{1}{c|}{\begin{tabular}[c]{@{}c@{}}FL.   \\      F1-score\end{tabular}} \\ \hline
 & DT & 97.77 & \cellcolor[HTML]{FFEB9C}{\color[HTML]{9C5700} 97.81} & 97.74 & 97.78 & 97.77 & 97.81 & 97.75 & \cellcolor[HTML]{FFEB9C}{\color[HTML]{9C5700} 97.79} \\ \cline{2-10} 
 & RF & \cellcolor[HTML]{C6EFCE}{\color[HTML]{006100} 98.27} & 98.24 & 98.23 & 98.2 & 98.27 & 98.24 & \cellcolor[HTML]{C6EFCE}{\color[HTML]{006100} 98.14} & 98.11 \\ \cline{2-10} 
 & ET & \cellcolor[HTML]{C6EFCE}{\color[HTML]{006100} 98} & 97.99 & 97.92 & 97.89 & 98 & 97.99 & \cellcolor[HTML]{C6EFCE}{\color[HTML]{006100} 97.95} & 97.93 \\ \cline{2-10} 
 & LR & 86.52 & \cellcolor[HTML]{FFEB9C}{\color[HTML]{9C5700} 86.91} & 84.89 & 83.83 & 86.52 & 86.91 & \cellcolor[HTML]{C6EFCE}{\color[HTML]{006100} 84.88} & 84.21 \\ \cline{2-10} 
 & KNN & 93.51 & \cellcolor[HTML]{FFEB9C}{\color[HTML]{9C5700} 93.63} & 93.53 & 93.47 & 93.51 & 93.63 & 93.42 & \cellcolor[HTML]{FFEB9C}{\color[HTML]{9C5700} 93.43} \\ \cline{2-10} 
 & NB & \cellcolor[HTML]{C6EFCE}{\color[HTML]{006100} 44.45} & 5.68 & 80.85 & 83.41 & 44.45 & 5.68 & \cellcolor[HTML]{C6EFCE}{\color[HTML]{006100} 54.95} & 3.93 \\ \cline{2-10} 
 & NN & 94.32 & \cellcolor[HTML]{FFEB9C}{\color[HTML]{9C5700} 94.59} & 94.24 & 94.39 & 94.32 & 94.59 & 94.07 & \cellcolor[HTML]{FFEB9C}{\color[HTML]{9C5700} 94.22} \\ \cline{2-10} 
 & AdB & \cellcolor[HTML]{C6EFCE}{\color[HTML]{006100} 91.34} & 86.05 & 91.42 & 82.51 & 91.34 & 86.05 & \cellcolor[HTML]{C6EFCE}{\color[HTML]{006100} 91.05} & 83.36 \\ \cline{2-10} 
 & HGBC & \cellcolor[HTML]{C6EFCE}{\color[HTML]{006100} 98.2} & 95.32 & 98.17 & 95.69 & 98.2 & 95.32 & \cellcolor[HTML]{C6EFCE}{\color[HTML]{006100} 98.08} & 95.41 \\ \cline{2-10} 
\multirow{-10}{*}{3} & GBC & \cellcolor[HTML]{C6EFCE}{\color[HTML]{006100} 96.89} & 87.19 & 96.79 & 94.31 & 96.89 & 87.19 & \cellcolor[HTML]{C6EFCE}{\color[HTML]{006100} 96.63} & 89.98 \\ \hline
\end{tabular}
\end{table}

\subsubsection*{Performance on ToN-IoT-Network}

Table \ref{tab:toniotnetwork} present the performance of the HR and FL model at level 2 and 3 on ToN-IoT-Network dataset in terms of accuracy, precision, recall and F1-score. To compare the HR and FL models at Level 2, we can analyze the performance of the classifiers. We present the summary of the results as follows:

Level 2:

\begin{itemize}

    \item DT (99.4\%), RF (99.32\%), and ET (99.53\%) classifiers perform similarly for both the HR and FL models, achieving high accuracy, precision, recall, and F1 scores above 99\%. These classifiers demonstrate robust performance in accurately classifying the data for both models. For NB, AdB (HR: 82.12\%, FL: 54.43\% ) and HGBC (HR: 99.53\%, FL: 99.48\%), the HR model performs slightly better  than FL model. NB (HR 25.85\%, FL:17.48\%) classifier, however, exhibits significantly lower performance for both models and AdB classifier performs relatively better for the HR model compared to the FL model. This achieves higher accuracy and F1 scores for the HR model, indicating that it can effectively classify the HR model data compared to the FL model data.

    \item We can observe the same trend between the two models for other performance metrics such as precision, recall and F1-score. 
    
\end{itemize}

Overall, the HR and FL models show similar performance across most classifiers. The decision tree-based classifiers (DT, RF, ET) perform consistently well, while the performance of other classifiers varies slightly between the HR and FL models. For Level 3, the HR and FL models were evaluated using ten classifiers. Here's a concise analysis of the results:

Level 3:

\begin{itemize}
    \item  DT (HR: 99.23\%, FL: 99.22\%), and RF (HR: 99.15\%, FL: 99.14\%) demonstrate slightly better accuracy for the HR model. ET ( HR: 99.36\%, FL: 99.36\%) classifier exhibits similar performance for both the HR and FL models, with high accuracy, precision, recall, and F1 scores above 99\%. These classifiers demonstrate strong classification capabilities for Level 3 subcategories. HGBC (HR: 99.39\%, FL: 99.03\%) classifier shows the highest accuracy among all the classifiers for the HR model. NB classifier performs poorly for both models, with low accuracy and F1 scores. This might not be well-suited for this specific dataset and Level 3 subcategories.

    \item The F1-score shows a consistent trend between the HR and FL models, indicating similar performance. This suggests that the HR model exhibits superior performance at the third level.
    
\end{itemize}

Overall, the decision tree-based classifiers (DT, RF, ET) consistently perform well across both models. The other classifiers show slightly varying performance between the HR and FL models but generally maintain satisfactory accuracy and F1 scores. Both models demonstrate comparable performance across various metrics such as accuracy, precision, recall, and F1-score, indicating no significant difference in their performance on this dataset.

\begin{table}[!htbp]
\centering
\caption{Performance of Hierarchical and Flat model on ToN-IoT-Network dataset}
\label{tab:toniotnetwork}
\begin{tabular}{|c|l|l|l|l|l|l|l|l|l|}
\hline
Level & \multicolumn{1}{c|}{Classifier} & \multicolumn{1}{c|}{\begin{tabular}[c]{@{}c@{}}HR. \\ Accuracy\end{tabular}} & \multicolumn{1}{c|}{\begin{tabular}[c]{@{}c@{}}FL. \\      Accuracy\end{tabular}} & \multicolumn{1}{c|}{\begin{tabular}[c]{@{}c@{}}HR.\\      Precision\end{tabular}} & \multicolumn{1}{c|}{\begin{tabular}[c]{@{}c@{}}FL. \\ Precision\end{tabular}} & \multicolumn{1}{c|}{\begin{tabular}[c]{@{}c@{}}HR.\\       Recall\end{tabular}} & \multicolumn{1}{c|}{\begin{tabular}[c]{@{}c@{}}FL. \\      Recall\end{tabular}} & \multicolumn{1}{c|}{\begin{tabular}[c]{@{}c@{}}HR. \\      F1-score\end{tabular}} & \multicolumn{1}{c|}{\begin{tabular}[c]{@{}c@{}}FL. \\      F1-score\end{tabular}} \\ \hline
 & DT & 99.4 & 99.4 & 99.4 & 99.4 & 99.4 & 99.4 & \cellcolor[HTML]{F2F2F2}{\color[HTML]{FA7D00} \textbf{99.4}} & 99.4 \\ \cline{2-10} 
 & RF & 99.32 & 99.32 & 99.32 & 99.32 & 99.32 & 99.32 & \cellcolor[HTML]{F2F2F2}{\color[HTML]{FA7D00} \textbf{99.31}} & 99.31 \\ \cline{2-10} 
 & ET & 99.53 & 99.53 & 99.53 & 99.53 & 99.53 & 99.53 & \cellcolor[HTML]{F2F2F2}{\color[HTML]{FA7D00} \textbf{99.53}} & 99.53 \\ \cline{2-10} 
 & LR & 83.21 & \cellcolor[HTML]{FFEB9C}{\color[HTML]{9C5700} 85.11} & 81.92 & 83.36 & 83.21 & 85.11 & 81.48 & \cellcolor[HTML]{FFEB9C}{\color[HTML]{9C5700} 83.52} \\ \cline{2-10} 
 & KNN & 95.45 & \cellcolor[HTML]{FFEB9C}{\color[HTML]{9C5700} 95.57} & 95.54 & 95.62 & 95.45 & 95.57 & 95.48 & \cellcolor[HTML]{FFEB9C}{\color[HTML]{9C5700} 95.59} \\ \cline{2-10} 
 & NB & \cellcolor[HTML]{C6EFCE}{\color[HTML]{006100} 25.85} & 17.48 & 71.59 & 69.32 & 25.85 & 17.48 & \cellcolor[HTML]{C6EFCE}{\color[HTML]{006100} 30.47} & 18.09 \\ \cline{2-10} 
 & NN & 93.15 & \cellcolor[HTML]{FFEB9C}{\color[HTML]{9C5700} 93.29} & 93.2 & 93.24 & 93.15 & 93.29 & 93.15 & \cellcolor[HTML]{FFEB9C}{\color[HTML]{9C5700} 93.25} \\ \cline{2-10} 
 & AdB & \cellcolor[HTML]{C6EFCE}{\color[HTML]{006100} 82.12} & 54.43 & 83.4 & 72.21 & 82.12 & 54.43 & \cellcolor[HTML]{C6EFCE}{\color[HTML]{006100} 82.03} & 59.2 \\ \cline{2-10} 
 & HGBC & \cellcolor[HTML]{C6EFCE}{\color[HTML]{006100} 99.53} & 99.48 & 99.53 & 99.48 & 99.53 & 99.48 & \cellcolor[HTML]{C6EFCE}{\color[HTML]{006100} 99.53} & 99.48 \\ \cline{2-10} 
\multirow{-10}{*}{2} & GBC & 97.69 & \cellcolor[HTML]{FFEB9C}{\color[HTML]{9C5700} 98.38} & 97.78 & 98.39 & 97.69 & 98.38 & 97.7 & \cellcolor[HTML]{FFEB9C}{\color[HTML]{9C5700} 98.37} \\ \hline
Level & \multicolumn{1}{c|}{Classifier} & \multicolumn{1}{c|}{\begin{tabular}[c]{@{}c@{}}HR.   \\ Accuracy\end{tabular}} & \multicolumn{1}{c|}{\begin{tabular}[c]{@{}c@{}}FL.   \\      Accuracy\end{tabular}} & \multicolumn{1}{c|}{\begin{tabular}[c]{@{}c@{}}HR.\\      Precision\end{tabular}} & \multicolumn{1}{c|}{\begin{tabular}[c]{@{}c@{}}FL. \\ Precision\end{tabular}} & \multicolumn{1}{c|}{\begin{tabular}[c]{@{}c@{}}HR.\\       Recall\end{tabular}} & \multicolumn{1}{c|}{\begin{tabular}[c]{@{}c@{}}FL.   \\      Recall\end{tabular}} & \multicolumn{1}{c|}{\begin{tabular}[c]{@{}c@{}}HR.   \\      F1-score\end{tabular}} & \multicolumn{1}{c|}{\begin{tabular}[c]{@{}c@{}}FL.   \\      F1-score\end{tabular}} \\ \hline
 & DT & \cellcolor[HTML]{C6EFCE}{\color[HTML]{006100} 99.23} & 99.22 & 99.23 & 99.22 & 99.23 & 99.22 & \cellcolor[HTML]{C6EFCE}{\color[HTML]{006100} 99.23} & 99.22 \\ \cline{2-10} 
 & RF & \cellcolor[HTML]{C6EFCE}{\color[HTML]{006100} 99.15} & 99.14 & 99.2 & 99.15 & 99.15 & 99.14 & \cellcolor[HTML]{C6EFCE}{\color[HTML]{006100} 99.16} & 99.14 \\ \cline{2-10} 
 & ET & 99.36 & 99.36 & 99.38 & 99.37 & 99.36 & 99.36 & \cellcolor[HTML]{F2F2F2}{\color[HTML]{FA7D00} \textbf{99.37}} & \cellcolor[HTML]{F2F2F2}{\color[HTML]{FA7D00} \textbf{99.37}} \\ \cline{2-10} 
 & LR & 82.9 & \cellcolor[HTML]{FFEB9C}{\color[HTML]{9C5700} 85.44} & 81.96 & 83.51 & 82.9 & 85.44 & 81.29 & \cellcolor[HTML]{FFEB9C}{\color[HTML]{9C5700} 83.47} \\ \cline{2-10} 
 & KNN & 95.12 & \cellcolor[HTML]{FFEB9C}{\color[HTML]{9C5700} 95.32} & 95.22 & 95.29 & 95.12 & 95.32 & 95.14 & \cellcolor[HTML]{FFEB9C}{\color[HTML]{9C5700} 95.27} \\ \cline{2-10} 
 & NB & \cellcolor[HTML]{C6EFCE}{\color[HTML]{006100} 25.08} & 18.1 & 73.86 & 72.29 & 25.08 & 18.1 & \cellcolor[HTML]{C6EFCE}{\color[HTML]{006100} 29.75} & 17.38 \\ \cline{2-10} 
 & NN & 92.85 & \cellcolor[HTML]{FFEB9C}{\color[HTML]{9C5700} 93.1} & 92.88 & 92.93 & 92.85 & 93.1 & 92.8 & \cellcolor[HTML]{FFEB9C}{\color[HTML]{9C5700} 92.97} \\ \cline{2-10} 
 & AdB & \cellcolor[HTML]{C6EFCE}{\color[HTML]{006100} 79.59} & 71.23 & 78.49 & 68.85 & 79.59 & 71.23 & \cellcolor[HTML]{C6EFCE}{\color[HTML]{006100} 78.03} & 67.37 \\ \cline{2-10} 
 & HGBC & \cellcolor[HTML]{C6EFCE}{\color[HTML]{006100} 99.39} & 99.03 & 99.41 & 99.04 & 99.39 & 99.03 & \cellcolor[HTML]{C6EFCE}{\color[HTML]{006100} 99.39} & 99.03 \\ \cline{2-10} 
\multirow{-10}{*}{3} & GBC & 97.45 & \cellcolor[HTML]{FFEB9C}{\color[HTML]{9C5700} 98.19} & 97.64 & 98.18 & 97.45 & 98.19 & 97.5 & \cellcolor[HTML]{FFEB9C}{\color[HTML]{9C5700} 98.17} \\ \hline
\end{tabular}
\end{table}

\subsubsection*{Performance on XIIOTID}

Table \ref{tab:xiiotid} shows the comparison of the FL and HR models for Level 2 in terms of accuracy, precision, recall and F1-score, we observe the following:

Level 2:

\begin{itemize}
    \item Both models achieve similar accuracy for three classifiers: ET, KNN and GBC. Both models achieve high accuracy, precision, recall, and F1-score across multiple classifiers on XIIOTID dataset. 
    \item RF, ET, and HGBC show consistent high performance for both models. The NB and AdB classifiers perform significantly better for the HR model, with higher accuracy, precision, recall, and F1-score.

\end{itemize}

Overall, the FL and HR models demonstrate similar performance for most classifiers at Level 2, but there are some variations in certain metrics and classifiers, particularly with the NB and AdB classifiers.

Level 3:

Table \ref{tab:xiiotid} compares the performance of HR and FL models at level 3 using various classifiers. Both models achieve high accuracy, with most classifiers scoring above 99\%.

\begin{itemize}

    \item In terms of precision, recall, and F1-score, the HR model generally performs well across five classifiers, consistently achieving high values. Both models achieve the same accuracy for KNN classifier(98.69\%). The FL model also demonstrates good performance, although some classifiers show slight variations compared to the HR model.
    \item Notably, the NB (49.42\%) and AdB (70.03\%) classifiers stand out as having lower accuracy, precision, recall, and F1-score in the FL model compared to the HR model. This indicates that the HR model outperforms the FL model for NB classification.  
\end{itemize}

Overall, the results highlight the effectiveness of both HR and FL models at level 3, with the HR model generally exhibiting superior performance across various classifiers.

\begin{table}[!htbp]
\centering
\caption{Performance of Hierarchical and Flat model on XIIOTID dataset}
\label{tab:xiiotid}
\begin{tabular}{|c|c|c|c|c|c|c|c|c|c|}
\hline
Level & Classifier & \begin{tabular}[c]{@{}c@{}}HR. \\ Accuracy\end{tabular} & \begin{tabular}[c]{@{}c@{}}FL. \\      Accuracy\end{tabular} & \begin{tabular}[c]{@{}c@{}}HR.\\      Precision\end{tabular} & \begin{tabular}[c]{@{}c@{}}FL. \\ Precision\end{tabular} & \begin{tabular}[c]{@{}c@{}}HR.\\       Recall\end{tabular} & \begin{tabular}[c]{@{}c@{}}FL. \\      Recall\end{tabular} & \begin{tabular}[c]{@{}c@{}}HR. \\      F1-score\end{tabular} & \begin{tabular}[c]{@{}c@{}}FL. \\      F1-score\end{tabular} \\ \hline
 & DT & 99.72 & \cellcolor[HTML]{FFEB9C}{\color[HTML]{9C5700} 99.74} & 99.72 & 99.74 & 99.72 & 99.74 & 99.72 & \cellcolor[HTML]{FFEB9C}{\color[HTML]{9C5700} 99.74} \\ \cline{2-10} 
 & RF & 99.83 & \cellcolor[HTML]{FFEB9C}{\color[HTML]{9C5700} 99.84} & 99.83 & 99.84 & 99.83 & 99.84 & 99.83 & \cellcolor[HTML]{FFEB9C}{\color[HTML]{9C5700} 99.84} \\ \cline{2-10} 
 & ET & \cellcolor[HTML]{F2F2F2}{\color[HTML]{FA7D00} \textbf{99.83}} & \cellcolor[HTML]{F2F2F2}{\color[HTML]{FA7D00} \textbf{99.83}} & 99.83 & 99.83 & 99.83 & 99.83 & \cellcolor[HTML]{F2F2F2}{\color[HTML]{FA7D00} \textbf{99.83}} & \cellcolor[HTML]{F2F2F2}{\color[HTML]{FA7D00} \textbf{99.83}} \\ \cline{2-10} 
 & LR & 97.25 & \cellcolor[HTML]{FFEB9C}{\color[HTML]{9C5700} 97.45} & 97.29 & 97.5 & 97.25 & 97.45 & 97.2 & \cellcolor[HTML]{FFEB9C}{\color[HTML]{9C5700} 97.4} \\ \cline{2-10} 
 & KNN & \cellcolor[HTML]{F2F2F2}{\color[HTML]{FA7D00} \textbf{98.71}} & \cellcolor[HTML]{F2F2F2}{\color[HTML]{FA7D00} \textbf{98.71}} & 98.71 & 98.71 & 98.71 & 98.71 & \cellcolor[HTML]{F2F2F2}{\color[HTML]{FA7D00} \textbf{98.71}} & \cellcolor[HTML]{F2F2F2}{\color[HTML]{FA7D00} \textbf{98.71}} \\ \cline{2-10} 
 & NB & \cellcolor[HTML]{C6EFCE}{\color[HTML]{006100} 72.71} & 54.24 & 87.44 & 83.76 & 72.71 & 54.24 & \cellcolor[HTML]{C6EFCE}{\color[HTML]{006100} 78.15} & 60.34 \\ \cline{2-10} 
 & NN & 98.73 & \cellcolor[HTML]{FFEB9C}{\color[HTML]{9C5700} 98.83} & 98.73 & 98.83 & 98.73 & 98.83 & 98.73 & \cellcolor[HTML]{FFEB9C}{\color[HTML]{9C5700} 98.82} \\ \cline{2-10} 
 & AdB & \cellcolor[HTML]{C6EFCE}{\color[HTML]{006100} 86.14} & 68.29 & 82.01 & 52.31 & 86.14 & 68.29 & \cellcolor[HTML]{C6EFCE}{\color[HTML]{006100} 82.03} & 57.79 \\ \cline{2-10} 
 & HGBC & \cellcolor[HTML]{C6EFCE}{\color[HTML]{006100} 99.61} & 99.46 & 99.62 & 99.51 & 99.61 & 99.46 & \cellcolor[HTML]{C6EFCE}{\color[HTML]{006100} 99.62} & 99.48 \\ \cline{2-10} 
\multirow{-10}{*}{2} & GBC & \cellcolor[HTML]{F2F2F2}{\color[HTML]{FA7D00} \textbf{99.23}} & \cellcolor[HTML]{F2F2F2}{\color[HTML]{FA7D00} \textbf{99.23}} & 99.23 & 99.29 & 99.23 & 99.23 & 99.23 & \cellcolor[HTML]{FFEB9C}{\color[HTML]{9C5700} 99.25} \\ \hline
Level & Classifier & \begin{tabular}[c]{@{}c@{}}HR. \\ Accuracy\end{tabular} & \begin{tabular}[c]{@{}c@{}}FL. \\      Accuracy\end{tabular} & \begin{tabular}[c]{@{}c@{}}HR.\\      Precision\end{tabular} & \begin{tabular}[c]{@{}c@{}}FL.  \\  Precision\end{tabular} & \begin{tabular}[c]{@{}c@{}}HR.\\       Recall\end{tabular} & \begin{tabular}[c]{@{}c@{}}FL.   \\      Recall\end{tabular} & \begin{tabular}[c]{@{}c@{}}HR.   \\      F1-score\end{tabular} & \begin{tabular}[c]{@{}c@{}}FL.   \\      F1-score\end{tabular} \\ \hline
 & DT & 99.72 & \cellcolor[HTML]{FFEB9C}{\color[HTML]{9C5700} 99.73} & 99.72 & 99.74 & 99.72 & 99.73 & 99.72 & \cellcolor[HTML]{FFEB9C}{\color[HTML]{9C5700} 99.73} \\ \cline{2-10} 
 & RF & 99.83 & \cellcolor[HTML]{FFEB9C}{\color[HTML]{9C5700} 99.84} & 99.83 & 99.83 & 99.83 & 99.84 & 99.82 & \cellcolor[HTML]{FFEB9C}{\color[HTML]{9C5700} 99.83} \\ \cline{2-10} 
 & ET & \cellcolor[HTML]{C6EFCE}{\color[HTML]{006100} 99.83} & 99.82 & 99.83 & 99.82 & 99.83 & 99.82 & \cellcolor[HTML]{F2F2F2}{\color[HTML]{FA7D00} \textbf{99.82}} & \cellcolor[HTML]{F2F2F2}{\color[HTML]{FA7D00} \textbf{99.82}} \\ \cline{2-10} 
 & LR & 96.73 & \cellcolor[HTML]{FFEB9C}{\color[HTML]{9C5700} 96.76} & 96.59 & 96.69 & 96.73 & 96.76 & 96.41 & \cellcolor[HTML]{FFEB9C}{\color[HTML]{9C5700} 96.45} \\ \cline{2-10} 
 & KNN & \cellcolor[HTML]{F2F2F2}{\color[HTML]{FA7D00} \textbf{98.69}} & \cellcolor[HTML]{F2F2F2}{\color[HTML]{FA7D00} \textbf{98.69}} & 98.67 & 98.67 & 98.69 & 98.69 & \cellcolor[HTML]{F2F2F2}{\color[HTML]{FA7D00} \textbf{98.67}} & \cellcolor[HTML]{F2F2F2}{\color[HTML]{FA7D00} \textbf{98.67}} \\ \cline{2-10} 
 & NB & \cellcolor[HTML]{C6EFCE}{\color[HTML]{006100} 70.79} & 49.42 & 85.3 & 90.01 & 70.79 & 49.42 & \cellcolor[HTML]{C6EFCE}{\color[HTML]{006100} 75.27} & 51.23 \\ \cline{2-10} 
 & NN & 98.73 & \cellcolor[HTML]{FFEB9C}{\color[HTML]{9C5700} 98.81} & 98.7 & 98.78 & 98.73 & 98.81 & 98.7 & \cellcolor[HTML]{FFEB9C}{\color[HTML]{9C5700} 98.79} \\ \cline{2-10} 
 & AdB & \cellcolor[HTML]{C6EFCE}{\color[HTML]{006100} 85.13} & 70.03 & 82.67 & 52.3 & 85.13 & 70.03 & \cellcolor[HTML]{C6EFCE}{\color[HTML]{006100} 82.33} & 58.56 \\ \cline{2-10} 
 & HGBC & \cellcolor[HTML]{C6EFCE}{\color[HTML]{006100} 99.61} & 99.47 & 99.62 & 99.53 & 99.61 & 99.47 & \cellcolor[HTML]{C6EFCE}{\color[HTML]{006100} 99.61} & 99.5 \\ \cline{2-10} 
\multirow{-10}{*}{3} & GBC & \cellcolor[HTML]{C6EFCE}{\color[HTML]{006100} 99.23} & 89.05 & 99.21 & 95.33 & 99.23 & 89.05 & \cellcolor[HTML]{C6EFCE}{\color[HTML]{006100} 99.18} & 91.26 \\ \hline
\end{tabular}
\end{table}

\subsubsection*{Performance on ISCXURL2016}

Table \ref{tab:iscxurl} compares the HR and FL models for Level 2 categories, we can observe the following:

Level 2:
\begin{itemize}
    \item Both models achieve relatively high accuracy, precision, recall, and F1-score across different classifiers. RF (HR: 98.27\%, FL: 98.19\%), ET (HR: 98.42\%, FL: 98.41\%), and HGBC (HR: 98.55\%, FL: 98.64\%) consistently perform well for both models. DT, RF, ET, KNN, NN and AdB classifiers show higher performance in terms of accuracy, precision, recall, and F1-score for HR models. On the other hand, LR, NB, HGBC and GBC obtain higher performances for the FL model. 
    \item LR, KNN, NN, and AdB classifiers exhibit comparable performance between the HR and FL models. The NB classifier performs better for the FL model, with higher accuracy, precision, recall, and F1-score.

\end{itemize}

In summary, the HR and FL models demonstrate comparable performance for most classifiers at Level 2. However, there are variations in certain metrics and classifiers, particularly with the NB classifier, and differences in prediction errors for LR, NB, AdB, and GBC classifiers between the two models.

Level 3:

\begin{itemize}
    \item The HR model exhibits a slightly higher accuracy and F1-score compared to the FL model for classifiers such as DT ( HR: 95.85\%, FL: 95.47\%), RF (HR: 97.83\%, FL:97.82\%), ET (HR: 98.04\%, FL: 98\%), KNN(HR: 95.03\%, FL: 94.89\%), NN(HR: 94.84\%, FL: 94.36\%), and AdB(HR: 83.38\%, FL:	65.94\%). However, the accuracy difference between these classifiers and the FL model is not significant, except for the AdB classifier. 
   \item Both the HR model and FL model achieve their highest performance with the HGBC classifier. The HR model achieves an accuracy of 98.23\% with HGBC, while the FL model achieves an even higher accuracy of 98.46\% with the same classifier.
\end{itemize}

\begin{table}[!htbp]
\centering
\caption{Performance of Hierarchical and Flat model on ISCXURL2016}
\label{tab:iscxurl}
\begin{tabular}{|c|c|c|c|c|c|c|c|c|c|}
\hline
Level & Classifier & \begin{tabular}[c]{@{}c@{}}HR. \\ Accuracy\end{tabular} & \begin{tabular}[c]{@{}c@{}}FL. \\      Accuracy\end{tabular} & \begin{tabular}[c]{@{}c@{}}HR.\\      Precision\end{tabular} & \begin{tabular}[c]{@{}c@{}}FL. \\ Precision\end{tabular} & \begin{tabular}[c]{@{}c@{}}HR.\\       Recall\end{tabular} & \begin{tabular}[c]{@{}c@{}}FL. \\      Recall\end{tabular} & \begin{tabular}[c]{@{}c@{}}HR. \\      F1-score\end{tabular} & \begin{tabular}[c]{@{}c@{}}FL. \\      F1-score\end{tabular} \\ \hline
 & DT & \cellcolor[HTML]{C6EFCE}{\color[HTML]{006100} 96.48} & 96.27 & 96.49 & 96.28 & 96.48 & 96.27 & \cellcolor[HTML]{C6EFCE}{\color[HTML]{006100} 96.48} & 96.26 \\ \cline{2-10} 
 & RF & \cellcolor[HTML]{C6EFCE}{\color[HTML]{006100} 98.27} & 98.19 & 98.28 & 98.2 & 98.27 & 98.19 & \cellcolor[HTML]{C6EFCE}{\color[HTML]{006100} 98.27} & 98.2 \\ \cline{2-10} 
 & ET & \cellcolor[HTML]{C6EFCE}{\color[HTML]{006100} 98.42} & 98.41 & 98.43 & 98.41 & 98.42 & 98.41 & \cellcolor[HTML]{C6EFCE}{\color[HTML]{006100} 98.42} & 98.41 \\ \cline{2-10} 
 & LR & 76.1 & \cellcolor[HTML]{FFEB9C}{\color[HTML]{9C5700} 76.51} & 75.82 & 76.13 & 76.1 & 76.51 & 75.81 & \cellcolor[HTML]{FFEB9C}{\color[HTML]{9C5700} 75.96} \\ \cline{2-10} 
 & KNN & \cellcolor[HTML]{C6EFCE}{\color[HTML]{006100} 95.77} & 95.74 & 95.78 & 95.76 & 95.77 & 95.74 & \cellcolor[HTML]{C6EFCE}{\color[HTML]{006100} 95.77} & 95.74 \\ \cline{2-10} 
 & NB & 44.54 & \cellcolor[HTML]{FFEB9C}{\color[HTML]{9C5700} 52.95} & 58.28 & 62.74 & 44.54 & 52.95 & 40.47 & \cellcolor[HTML]{FFEB9C}{\color[HTML]{9C5700} 49.01} \\ \cline{2-10} 
 & NN & \cellcolor[HTML]{C6EFCE}{\color[HTML]{006100} 95.4} & 95.11 & 95.4 & 95.12 & 95.4 & 95.11 & \cellcolor[HTML]{C6EFCE}{\color[HTML]{006100} 95.4} & 95.11 \\ \cline{2-10} 
 & AdB & \cellcolor[HTML]{C6EFCE}{\color[HTML]{006100} 84.22} & 75.9 & 84.07 & 75.48 & 84.22 & 75.9 & \cellcolor[HTML]{C6EFCE}{\color[HTML]{006100} 84.07} & 75.51 \\ \cline{2-10} 
 & HGBC & 98.55 & \cellcolor[HTML]{FFEB9C}{\color[HTML]{9C5700} 98.64} & 98.55 & 98.64 & 98.55 & 98.64 & 98.55 & \cellcolor[HTML]{FFEB9C}{\color[HTML]{9C5700} 98.64} \\ \cline{2-10} 
\multirow{-10}{*}{2} & GBC & 93.43 & \cellcolor[HTML]{FFEB9C}{\color[HTML]{9C5700} 94.25} & 93.47 & 94.29 & 93.43 & 94.25 & 93.43 & \cellcolor[HTML]{FFEB9C}{\color[HTML]{9C5700} 94.25} \\ \hline
Level & Classifier & \begin{tabular}[c]{@{}c@{}}HR.   \\ Accuracy\end{tabular} & \begin{tabular}[c]{@{}c@{}}FL.   \\      Accuracy\end{tabular} & \begin{tabular}[c]{@{}c@{}}HR.\\      Precision\end{tabular} & \begin{tabular}[c]{@{}c@{}}FL.   \\ Precision\end{tabular} & \begin{tabular}[c]{@{}c@{}}HR.\\       Recall\end{tabular} & \begin{tabular}[c]{@{}c@{}}FL.   \\      Recall\end{tabular} & \begin{tabular}[c]{@{}c@{}}HR.   \\      F1-score\end{tabular} & \begin{tabular}[c]{@{}c@{}}FL.   \\      F1-score\end{tabular} \\ \hline
 & DT & \cellcolor[HTML]{C6EFCE}{\color[HTML]{006100} 95.85} & 95.47 & 95.83 & 95.45 & 95.85 & 95.47 & \cellcolor[HTML]{C6EFCE}{\color[HTML]{006100} 95.83} & 95.45 \\ \cline{2-10} 
 & RF & \cellcolor[HTML]{C6EFCE}{\color[HTML]{006100} 97.83} & 97.82 & 97.87 & 97.85 & 97.83 & 97.82 & \cellcolor[HTML]{C6EFCE}{\color[HTML]{006100} 97.84} & 97.83 \\ \cline{2-10} 
 & ET & \cellcolor[HTML]{C6EFCE}{\color[HTML]{006100} 98.04} & 98 & 98.08 & 98.02 & 98.04 & 98 & \cellcolor[HTML]{C6EFCE}{\color[HTML]{006100} 98.05} & 98.01 \\ \cline{2-10} 
 & LR & 74.04 & \cellcolor[HTML]{FFEB9C}{\color[HTML]{9C5700} 76.77} & 74.15 & 76.64 & 74.04 & 76.77 & 73.89 & \cellcolor[HTML]{FFEB9C}{\color[HTML]{9C5700} 76.47} \\ \cline{2-10} 
 & KNN & \cellcolor[HTML]{C6EFCE}{\color[HTML]{006100} 95.03} & 94.89 & 95.01 & 94.88 & 95.03 & 94.89 & \cellcolor[HTML]{C6EFCE}{\color[HTML]{006100} 95.01} & 94.86 \\ \cline{2-10} 
 & NB & 43.78 & \cellcolor[HTML]{FFEB9C}{\color[HTML]{9C5700} 58.48} & 57.59 & 63.5 & 43.78 & 58.48 & 38.7 & \cellcolor[HTML]{FFEB9C}{\color[HTML]{9C5700} 56.37} \\ \cline{2-10} 
 & NN & \cellcolor[HTML]{C6EFCE}{\color[HTML]{006100} 94.84} & 94.36 & 94.84 & 94.36 & 94.84 & 94.36 & \cellcolor[HTML]{C6EFCE}{\color[HTML]{006100} 94.84} & 94.36 \\ \cline{2-10} 
 & AdB & \cellcolor[HTML]{C6EFCE}{\color[HTML]{006100} 83.38} & 65.94 & 83.43 & 67.09 & 83.38 & 65.94 & \cellcolor[HTML]{C6EFCE}{\color[HTML]{006100} 83.32} & 65.45 \\ \cline{2-10} 
 & HGBC & 98.23 & \cellcolor[HTML]{FFEB9C}{\color[HTML]{9C5700} 98.46} & 98.24 & 98.46 & 98.23 & 98.46 & 98.23 & \cellcolor[HTML]{FFEB9C}{\color[HTML]{9C5700} 98.46} \\ \cline{2-10} 
\multirow{-10}{*}{3} & GBC & 92.77 & \cellcolor[HTML]{FFEB9C}{\color[HTML]{9C5700} 94.41} & 92.92 & 94.45 & 92.77 & 94.41 & 92.81 & \cellcolor[HTML]{FFEB9C}{\color[HTML]{9C5700} 94.42} \\ \hline
\end{tabular}
\end{table}

\subsubsection*{Performance on ToN-IoT-IoTs}

Table \ref{tab:toniotiots} shows the performance of the HR and FL model for 10 classifiers in terms of accuracy, precision, recall and F1-score at level 2 and 3. 

Level 2:

\begin{itemize}
    \item In terms of accuracy, the HR model demonstrates higher performance for classifiers such as LR(HR:61.79\%, FL: 61.11\%), NB(HR: 57.15\%, FL: 6.93\%), and AdB (HR: 63.23\%, FL: 59.92\%) compared to the FL model. Notably, the NB and AdB classifiers maintain significantly higher accuracy in the HR model compared to the FL model. Conversely, for the remaining classifiers, the FL model slightly outperforms the HR model in terms of accuracy.

    \item When considering the F1-score, the HR model demonstrates slightly better performance for most classifiers, with notable differences observed for NB and AdB classifiers. The HR model exhibits a significant performance gap in favor of higher F1-score for NB and AdB (HR: 54.69\%, FL: 51.1\%) classifiers compared to the FL model.

\end{itemize}

Level 3:

\begin{itemize}
    \item With the exception of NB and AdB classifiers, the FL model exhibits slightly better performance than the HR model for all other classifiers. However, when considering the F1-score metric, the HR model demonstrates higher scores compared to the FL model for most classifiers.
    \item Specifically, RF (HR: 69.81\%, FL: 69.81\%) and ET (HR: 70.16\%, FL: 70.16\%) classifiers show similar F1-score for both models. Overall, while the FL model performs slightly better in terms of overall performance, the HR model showcases higher F1-score for most classifiers, indicating its strength in capturing the balance between precision and recall.
 
\end{itemize}

\begin{table}[!htbp]
\centering
\caption{Performance of Hierarchical and Flat model on ToN-IoT-IoTs}
\label{tab:toniotiots}
\begin{tabular}{|c|c|c|c|c|c|c|c|c|c|}
\hline
Level & Classifier & \begin{tabular}[c]{@{}c@{}}HR.\\  Accuracy\end{tabular} & \begin{tabular}[c]{@{}c@{}}FL. \\      Accuracy\end{tabular} & \begin{tabular}[c]{@{}c@{}}HR.\\      Precision\end{tabular} & \begin{tabular}[c]{@{}c@{}}FL. \\ Precision\end{tabular} & \begin{tabular}[c]{@{}c@{}}HR.\\       Recall\end{tabular} & \begin{tabular}[c]{@{}c@{}}FL. \\      Recall\end{tabular} & \begin{tabular}[c]{@{}c@{}}HR. \\      F1-score\end{tabular} & \begin{tabular}[c]{@{}c@{}}FL. \\      F1-score\end{tabular} \\ \hline
 & DT & 72.62 & \cellcolor[HTML]{FFEB9C}{\color[HTML]{9C5700} 72.69} & 73.14 & 73.29 & 72.62 & 72.69 & 69.65 & \cellcolor[HTML]{FFEB9C}{\color[HTML]{9C5700} 69.71} \\ \cline{2-10} 
 & RF & 73.3 & \cellcolor[HTML]{FFEB9C}{\color[HTML]{9C5700} 73.34} & 74.28 & 74.44 & 73.3 & 73.34 & \cellcolor[HTML]{C6EFCE}{\color[HTML]{006100} 70.38} & 70.37 \\ \cline{2-10} 
 & ET & 73.53 & \cellcolor[HTML]{FFEB9C}{\color[HTML]{9C5700} 73.56} & 74.71 & 74.83 & 73.53 & 73.56 & \cellcolor[HTML]{F2F2F2}{\color[HTML]{FA7D00} \textbf{70.67}} & \cellcolor[HTML]{F2F2F2}{\color[HTML]{FA7D00} \textbf{70.67}} \\ \cline{2-10} 
 & LR & \cellcolor[HTML]{C6EFCE}{\color[HTML]{006100} 61.79} & 61.11 & 50.1 & 48.99 & 61.79 & 61.11 & \cellcolor[HTML]{C6EFCE}{\color[HTML]{006100} 50.24} & 47.63 \\ \cline{2-10} 
 & KNN & 59.59 & \cellcolor[HTML]{FFEB9C}{\color[HTML]{9C5700} 60.92} & 62.08 & 68.12 & 59.59 & 60.92 & 60.26 & \cellcolor[HTML]{FFEB9C}{\color[HTML]{9C5700} 61.92} \\ \cline{2-10} 
 & NB & \cellcolor[HTML]{C6EFCE}{\color[HTML]{006100} 57.15} & 6.93 & 40.6 & 0.71 & 57.15 & 6.93 & \cellcolor[HTML]{C6EFCE}{\color[HTML]{006100} 47.38} & 1.25 \\ \cline{2-10} 
 & NN & 65.13 & \cellcolor[HTML]{FFEB9C}{\color[HTML]{9C5700} 65.31} & 62.66 & 63.51 & 65.13 & 65.31 & 57.5 & \cellcolor[HTML]{FFEB9C}{\color[HTML]{9C5700} 57.06} \\ \cline{2-10} 
 & AdB & \cellcolor[HTML]{C6EFCE}{\color[HTML]{006100} 63.23} & 59.92 & 56.36 & 50.69 & 63.23 & 59.92 & \cellcolor[HTML]{C6EFCE}{\color[HTML]{006100} 54.69} & 51.1 \\ \cline{2-10} 
 & HGBC & 66.76 & \cellcolor[HTML]{FFEB9C}{\color[HTML]{9C5700} 67.15} & 65.84 & 66.91 & 66.76 & 67.15 & 60.02 & \cellcolor[HTML]{FFEB9C}{\color[HTML]{9C5700} 60.25} \\ \cline{2-10} 
\multirow{-10}{*}{2} & GBC & 65.7 & \cellcolor[HTML]{FFEB9C}{\color[HTML]{9C5700} 65.96} & 63.81 & 65.24 & 65.7 & 65.96 & \cellcolor[HTML]{C6EFCE}{\color[HTML]{006100} 58.41} & 58.22 \\ \hline
Level & Classifier & \begin{tabular}[c]{@{}c@{}}HR.  \\  Accuracy\end{tabular} & \begin{tabular}[c]{@{}c@{}}FL.   \\      Accuracy\end{tabular} & \begin{tabular}[c]{@{}c@{}}HR.\\      Precision\end{tabular} & \begin{tabular}[c]{@{}c@{}}FL.  \\  Precision\end{tabular} & \begin{tabular}[c]{@{}c@{}}HR.\\       Recall\end{tabular} & \begin{tabular}[c]{@{}c@{}}FL.   \\      Recall\end{tabular} & \begin{tabular}[c]{@{}c@{}}HR.   \\      F1-score\end{tabular} & \begin{tabular}[c]{@{}c@{}}FL.   \\      F1-score\end{tabular} \\ \hline
 & DT & 72.14 & \cellcolor[HTML]{FFEB9C}{\color[HTML]{9C5700} 72.25} & 72.22 & 72.44 & 72.14 & 72.25 & 69 & \cellcolor[HTML]{FFEB9C}{\color[HTML]{9C5700} 69.12} \\ \cline{2-10} 
 & RF & 72.88 & \cellcolor[HTML]{FFEB9C}{\color[HTML]{9C5700} 72.93} & 73.51 & 73.69 & 72.88 & 72.93 & \cellcolor[HTML]{F2F2F2}{\color[HTML]{FA7D00} \textbf{69.81}} & \cellcolor[HTML]{F2F2F2}{\color[HTML]{FA7D00} \textbf{69.81}} \\ \cline{2-10} 
 & ET & 73.16 & \cellcolor[HTML]{FFEB9C}{\color[HTML]{9C5700} 73.19} & 74.02 & 74.15 & 73.16 & 73.19 & \cellcolor[HTML]{F2F2F2}{\color[HTML]{FA7D00} \textbf{70.16}} & \cellcolor[HTML]{F2F2F2}{\color[HTML]{FA7D00} \textbf{70.16}} \\ \cline{2-10} 
 & LR & 60.91 & \cellcolor[HTML]{FFEB9C}{\color[HTML]{9C5700} 61.13} & 43.43 & 46.05 & 60.91 & 61.13 & \cellcolor[HTML]{C6EFCE}{\color[HTML]{006100} 48.5} & 47.1 \\ \cline{2-10} 
 & KNN & 57.64 & \cellcolor[HTML]{FFEB9C}{\color[HTML]{9C5700} 60.79} & 60.71 & 67.83 & 57.64 & 60.79 & 58.41 & \cellcolor[HTML]{FFEB9C}{\color[HTML]{9C5700} 61.44} \\ \cline{2-10} 
 & NB & \cellcolor[HTML]{C6EFCE}{\color[HTML]{006100} 57.15} & 6.93 & 40.6 & 0.71 & 57.15 & 6.93 & \cellcolor[HTML]{C6EFCE}{\color[HTML]{006100} 47.38} & 1.25 \\ \cline{2-10} 
 & NN & 64.31 & \cellcolor[HTML]{FFEB9C}{\color[HTML]{9C5700} 64.51} & 60.23 & 61.33 & 64.31 & 64.51 & \cellcolor[HTML]{C6EFCE}{\color[HTML]{006100} 56.24} & 55.61 \\ \cline{2-10} 
 & AdB & \cellcolor[HTML]{C6EFCE}{\color[HTML]{006100} 62.09} & 59.25 & 51.55 & 49.56 & 62.09 & 59.25 & \cellcolor[HTML]{C6EFCE}{\color[HTML]{006100} 52.59} & 50.62 \\ \cline{2-10} 
 & HGBC & 66.11 & \cellcolor[HTML]{FFEB9C}{\color[HTML]{9C5700} 66.44} & 64.25 & 66.59 & 66.11 & 66.44 & \cellcolor[HTML]{C6EFCE}{\color[HTML]{006100} 58.98} & 58.41 \\ \cline{2-10} 
\multirow{-10}{*}{3} & GBC & 64.89 & \cellcolor[HTML]{FFEB9C}{\color[HTML]{9C5700} 65.34} & 62.13 & 65.72 & 64.89 & 65.34 & \cellcolor[HTML]{C6EFCE}{\color[HTML]{006100} 57.09} & 56.04 \\ \hline
\end{tabular}
\end{table}

The HR model and FL model on the BoT-IoT dataset performed 100\% accuracy, precision, recall, and F1-score for both models at  level 2 and 3 for most of the classifiers except AdB and NB. In the FL model at level 2, the AdB classifier achieved slightly lower accuracy and precision compared to other classifiers. Similarly, in the HR model at level 3, the LR classifier had a slightly lower accuracy and precision. The NB classifier in both HR and FL models showed slightly lower performance in terms of accuracy, precision, and recall compared to other classifiers. However, its performance still remained quite high, with accuracy and precision above 99\%.

\begin{table}[!htbp]
\centering
\caption{Performance of Hierarchical and Flat model on BoT-IoT}
\label{tab:botiot}
\begin{tabular}{|c|l|l|l|l|l|l|l|l|l|}
\hline
Level & \multicolumn{1}{c|}{Classifier} & \multicolumn{1}{c|}{\begin{tabular}[c]{@{}c@{}}HR.\\  Accuracy\end{tabular}} & \multicolumn{1}{c|}{\begin{tabular}[c]{@{}c@{}}FL. \\      Accuracy\end{tabular}} & \multicolumn{1}{c|}{\begin{tabular}[c]{@{}c@{}}HR.\\      Precision\end{tabular}} & \multicolumn{1}{c|}{\begin{tabular}[c]{@{}c@{}}FL. \\ Precision\end{tabular}} & \multicolumn{1}{c|}{\begin{tabular}[c]{@{}c@{}}HR.\\       Recall\end{tabular}} & \multicolumn{1}{c|}{\begin{tabular}[c]{@{}c@{}}FL. \\      Recall\end{tabular}} & \multicolumn{1}{c|}{\begin{tabular}[c]{@{}c@{}}HR. \\      F1-score\end{tabular}} & \multicolumn{1}{c|}{\begin{tabular}[c]{@{}c@{}}FL. \\      F1-score\end{tabular}} \\ \hline
 & DT & 100 & 100 & 100 & 100 & 100 & 100 & 100 & 100 \\ \cline{2-10} 
 & RF & 100 & 100 & 100 & 100 & 100 & 100 & 100 & 100 \\ \cline{2-10} 
 & ET & 100 & 100 & 100 & 100 & 100 & 100 & 100 & 100 \\ \cline{2-10} 
 & LR & 100 & 100 & 100 & 100 & 100 & 100 & 100 & 100 \\ \cline{2-10} 
 & KNN & 100 & 100 & 100 & 100 & 100 & 100 & 100 & 100 \\ \cline{2-10} 
 & NB & 99.43 & 99.43 & 99.43 & 99.43 & 99.43 & 99.43 & 99.43 & 99.43 \\ \cline{2-10} 
 & NN & 100 & 100 & 100 & 100 & 100 & 100 & 100 & 100 \\ \cline{2-10} 
 & AdB & \cellcolor[HTML]{C6EFCE}{\color[HTML]{006100} 100} & 99.99 & 100 & 99.97 & 100 & 99.99 & \cellcolor[HTML]{C6EFCE}{\color[HTML]{006100} 100} & 99.98 \\ \cline{2-10} 
 & HGBC & \cellcolor[HTML]{C6EFCE}{\color[HTML]{006100} 100} & 99.9 & 100 & 99.97 & 100 & 99.9 & \cellcolor[HTML]{C6EFCE}{\color[HTML]{006100} 100} & 99.94 \\ \cline{2-10} 
\multirow{-10}{*}{2} & GBC & 100 & 100 & 100 & 100 & 100 & 100 & 100 & 100 \\ \hline
Level & \multicolumn{1}{c|}{Classifier} & \multicolumn{1}{c|}{\begin{tabular}[c]{@{}c@{}}HR.  \\  Accuracy\end{tabular}} & \multicolumn{1}{c|}{\begin{tabular}[c]{@{}c@{}}FL.   \\      Accuracy\end{tabular}} & \multicolumn{1}{c|}{\begin{tabular}[c]{@{}c@{}}HR.\\      Precision\end{tabular}} & \multicolumn{1}{c|}{\begin{tabular}[c]{@{}c@{}}FL.  \\  Precision\end{tabular}} & \multicolumn{1}{c|}{\begin{tabular}[c]{@{}c@{}}HR.\\       Recall\end{tabular}} & \multicolumn{1}{c|}{\begin{tabular}[c]{@{}c@{}}FL.   \\      Recall\end{tabular}} & \multicolumn{1}{c|}{\begin{tabular}[c]{@{}c@{}}HR.   \\      F1-score\end{tabular}} & \multicolumn{1}{c|}{\begin{tabular}[c]{@{}c@{}}FL.   \\      F1-score\end{tabular}} \\ \hline
 & DT & 100 & 100 & 100 & 100 & 100 & 100 & 100 & 100 \\ \cline{2-10} 
 & RF & 100 & 100 & 100 & 100 & 100 & 100 & 100 & 100 \\ \cline{2-10} 
 & ET & 100 & 100 & 100 & 100 & 100 & 100 & 100 & 100 \\ \cline{2-10} 
 & LR & 99.99 & 99.99 & 99.99 & 99.99 & 99.99 & 99.99 & 99.99 & 99.99 \\ \cline{2-10} 
 & KNN & 100 & 100 & 100 & 100 & 100 & 100 & 100 & 100 \\ \cline{2-10} 
 & NB & 99.43 & \cellcolor[HTML]{FFEB9C}{\color[HTML]{9C5700} 100} & 99.44 & 100 & 99.43 & 100 & 99.43 & 100 \\ \cline{2-10} 
 & NN & 100 & 100 & 100 & 100 & 100 & 100 & 100 & 100 \\ \cline{2-10} 
 & AdB & \cellcolor[HTML]{C6EFCE}{\color[HTML]{006100} 100} & 79.44 & 100 & 69.99 & 100 & 79.44 & \cellcolor[HTML]{C6EFCE}{\color[HTML]{006100} 100} & 73.2 \\ \cline{2-10} 
 & HGBC & \cellcolor[HTML]{C6EFCE}{\color[HTML]{006100} 100} & 99.65 & 100 & 99.83 & 100 & 99.65 & \cellcolor[HTML]{C6EFCE}{\color[HTML]{006100} 100} & 99.73 \\ \cline{2-10} 
\multirow{-10}{*}{3} & GBC & 100 & 100 & 100 & 100 & 100 & 100 & 100 & 100 \\ \hline
\end{tabular}
\end{table}

Our analysis indicates that the HR model outperforms the FL model in most of the evaluated datasets, such as NSL-KDD, CIC-DoS2017, CIC-DDoS2019, CIC-Darknet2020, ISCXURL2016, and XIIOTID. However, The differences in accuracy and F1-score between the HR and FL models are not statistically significant for most of the classifiers, except for the NB and AdB classifiers. Specifically, for the NSL-KDD dataset, the HR model shows slightly better performance at both Level 2 and Level 3. Moreover, the HR model exhibits improvements over the FL model in terms of root mean square error at the third level, as observed in the NSL-KDD and UNSW-NB15 datasets. Similarly, for the Malmem2022 dataset, although the HR model's performance at the third level may not be as strong, it demonstrates lower RMS error compared to the FL model. The same trend is noticed in the ToN-IoT-IoTs dataset at the second level for RMS. These findings highlight the superiority of the HR model in capturing patterns and reducing variations in these datasets.

\subsubsection*{Performance of HR and FL on attack categories detection}

In this section, we present the analysis of HR and FL model for several datasets(CIC-Darknet2020, CIC-DoS2017, XIITID, Malmem2022, ToN-IoT-Network ) on detecting attack categories at the third level. 

\begin{figure}[!htbp]
   \centering
    \includegraphics[scale = .60]{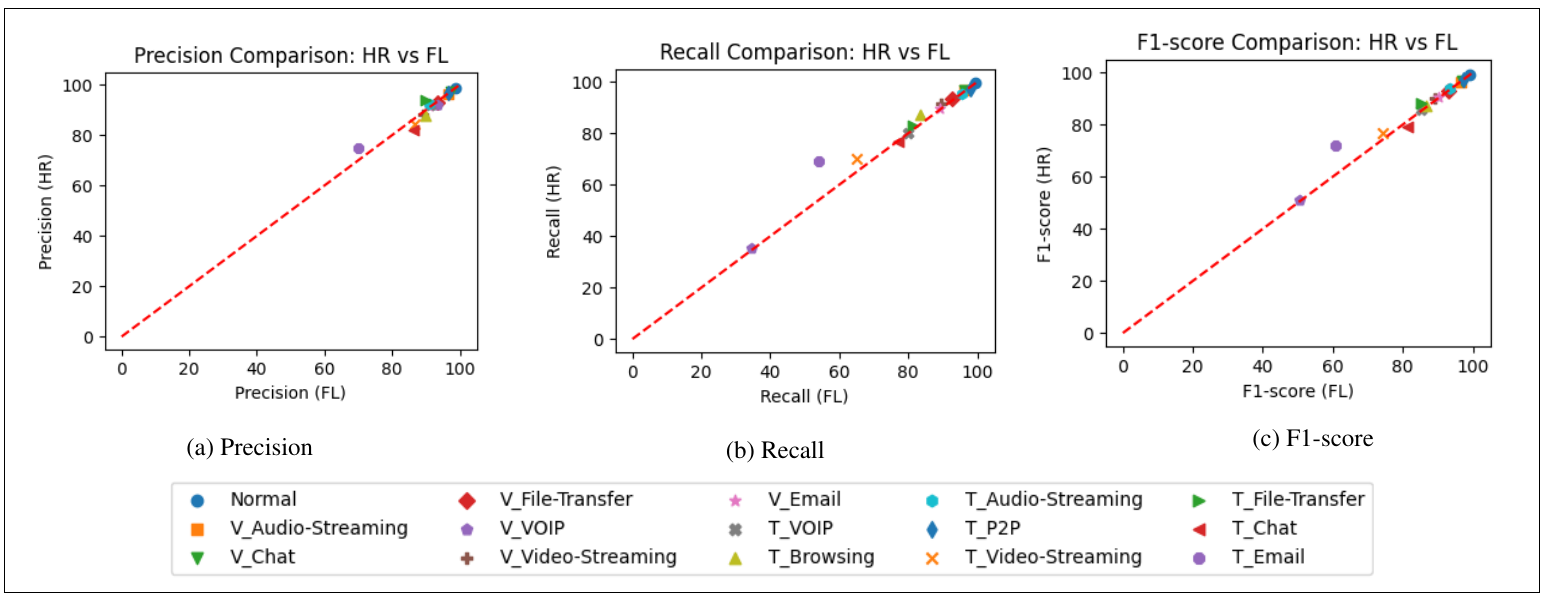}
    \caption{The difference of performance between HR and FL model on CIC Darknet2020's attacks}
    \label{fig:darknetattack}
\end{figure}

CIC-Darknet2020:

Figure \ref{fig:darknetattack} (a), (b) and (c) depicts the performance between the HR (RF) and FL (RF) models for different categories in the CIC Darknet2020 dataset, we can observe the following results:

The categories Normal, V\_Email, T\_VOIP, T\_Audio-Streaming (HR: 92.64\%), T\_P2P, T\_File-Transfer (HR: 93.68\%, FL: 89.69\%), T\_Email (HR: 75\%, FL: 70\%) show that HR model outperforms the FL model in terms of precision for these categories. However, the categories V\_Audio-Streaming, V\_Chat, V\_File-Transfer, V\_VOIP (HR: 92.17\%, FL: 93.5\%), V\_Video-Streaming, T\_Browsing (HR:87.74\%, FL: 89.8\%), T\_Video-Streaming (HR: 88.35\%, FL: 86.75\%), and T\_Chat (HR:81.97\%, FL: 86.21\%), indicate that the FL model performs better in terms of precision for these categories. 

The categories V\_Chat, V\_File-Transfer, V\_VOIP, V\_Video-Streaming, T\_VOIP, T\_Browsing,  T\_Video-Streaming (HR: 70.3\%, FL: 64.85\%,  diff: 5.45), T\_File-Transfer, and T\_Email (HR: 69.23\%, FL:53.85\% , diff: 15.38) show positive differences, suggesting better recall for the HR model in these categories. On the other hand, the categories Normal and V\_Audio-Streaming display slightly higher recall for the FL model in these categories. 

The categories Normal, V\_Chat, T\_VOIP, T\_Audio-Streaming, T\_P2P, T\_File-Transfer (HR: 88.12\%, FL: 85.29\%, diff: 2.83), T\_Video-Streaming (HR: 76.76\%, FL:74.22\%, diff: 2.54), V\_File-Transfer (HR: 88.12\%, FL: 85.29\%, diff: 2.83) and T\_Email (HR: 72\%, FL:60.87\%,  diff: 11.13) show positive differences, suggesting better F1-score for the HR model in these categories. Conversely, the categories V\_Audio-Streaming, T\_P2P and T\_Chat(HR:79.37\% ,FL: 81.3\%, diff: -1.93), display negative differences, indicating better F1-score for the FL model in these categories. 

Overall, the HR and FL models demonstrate variations in their performance for different categories in terms of precision, recall, and F1-score in the CIC Darknet 2020 dataset. The HR model tends to perform better for categories like Normal, T\_VOIP, T\_Audio-Streaming, T\_Video-Stream, T\_File-Transfer, and T\_Email in terms of precision, recall, and F1-score. On the other hand, the FL model performs better for categories like V\_Audio-Streaming, V\_Chat, V\_File-Transfer, and T\_Chat in terms of precision, recall, and F1-score. 

\begin{figure}[!htbp]
   \centering
    \includegraphics[scale = .80]{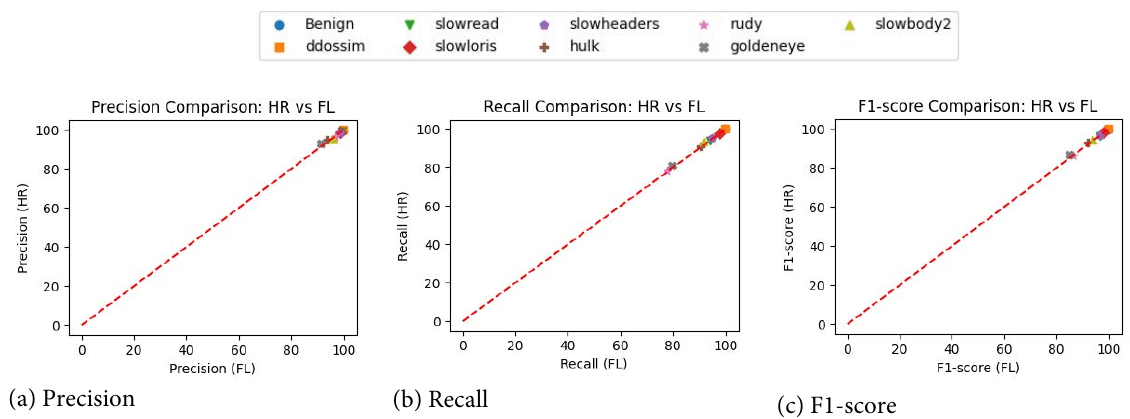}
    \caption{The HR vs FL model on CIC-DoS2017's attacks}
    \label{fig:cicdosattack}
\end{figure}

The results of this analysis suggest that the HR model is generally better than the FL model for the cic-darknet2020 dataset. The performance of HR and FL in detecting attack categories on CIC-DoS2017 is presented in Figure \ref{fig:cicdosattack}.

\begin{figure}[!htbp]
   \centering
    \includegraphics[scale = .60]{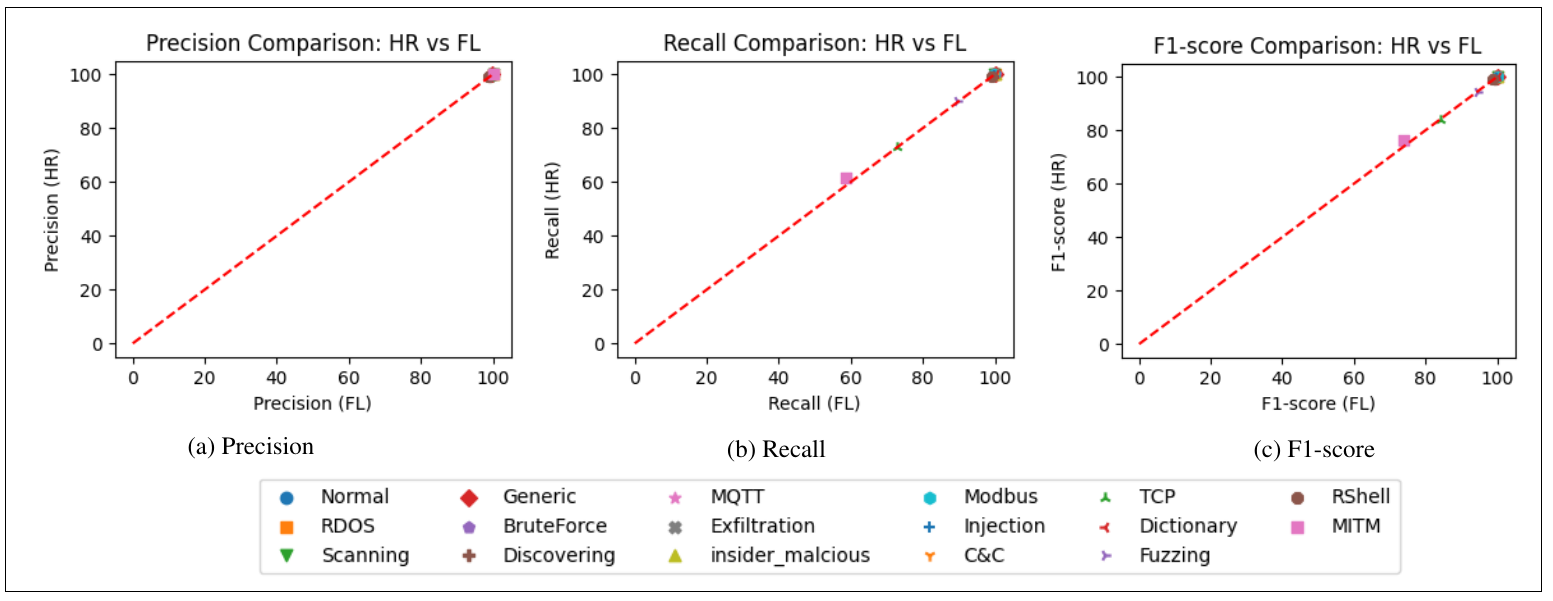}
    \caption{The difference of performance between HR and FL model on XIIOTID's attacks}
    \label{fig:xiiotidtattack}
\end{figure}

XIIOTID:

The comparison of performance between HR model and FL model on XIIOTID dataset is provided in Figure \ref{fig:xiiotidtattack} (a), (b) and (c). The graph in Figure \ref{fig:xiiotidtattack} shows that both the HR and FL models demonstrate similar performance in many categories for the RF classifier, while showing variations in specific categories.

The HR model performs better in terms of precision, recall, and F1-score for the "Fuzzing", RShell, and "MITM" categories. The FL model performs slightly better in terms of precision for the "TCP" category. For other categories, the performance differences are small or zero, indicating similar performance between the models. Therefore, to determine which model is doing well for attack detection in the XIIOTID dataset depends on other factors model's complexity and executing time.

\begin{figure}[!htbp]
   \centering
    \includegraphics[scale = .60]{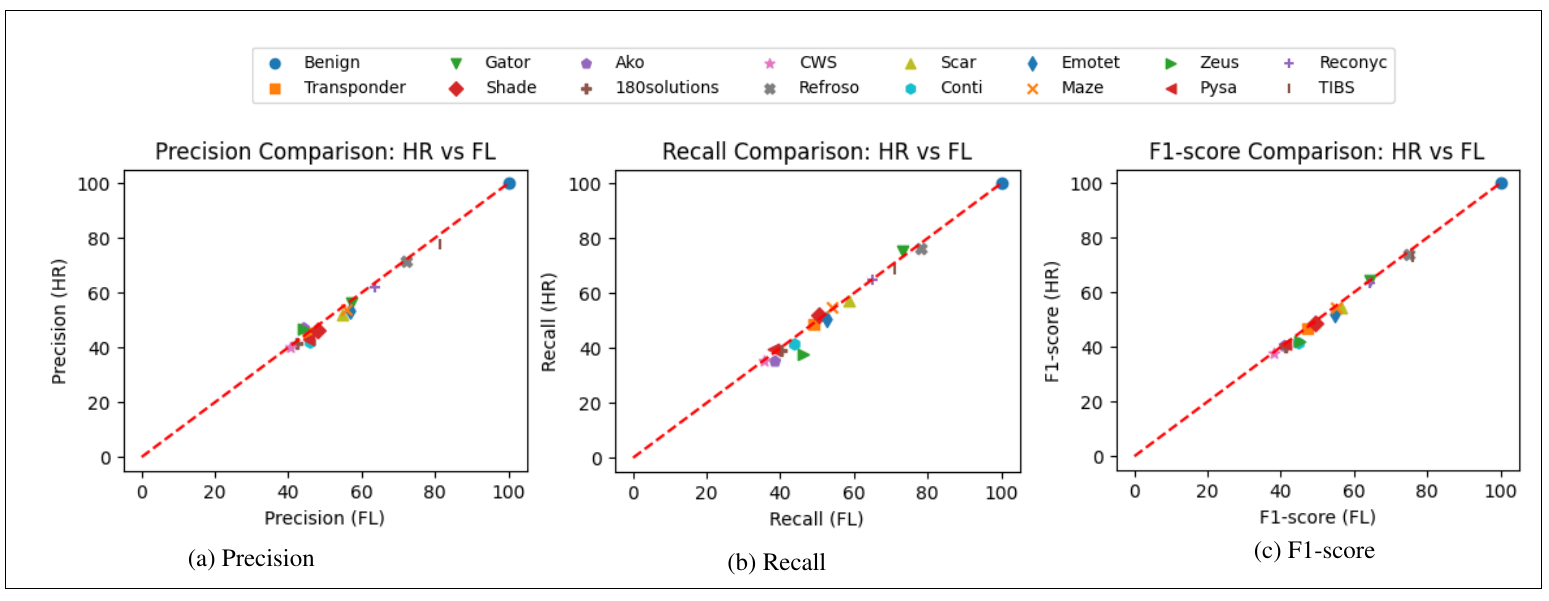}
    \caption{The difference of performance between HR and FL model on CIC-Malmem2022's attacks}
    \label{fig:malmem2022attack}
\end{figure}

CIC-Malmem2022:

Figure \ref{fig:malmem2022attack} depicts the precision, recall, and F1-score of the best classifiers (RF) for the FL model and HR model on the CIC-Malmem 2022 dataset. The results how that the HR model tends to have slightly lower precision, recall, and F1-score compared to the FL model. However, the differences are relatively small.

For most attack classes, the FL model outperforms the HR model in terms of precision, recall, and F1-score. This suggests that the FL model is better at distinguishing between different types of attacks for this dataset. There are a few attack classes where the HR model performs slightly better than the FL model. For example, the HR model achieves higher precision for Ako (HR: 47.17\%, FL: 43.94), Zeus(HR: 46.78\%, FL: 43.96) classes whereas the FL obtains better precison for  Conti (HR:41.85, FL: 46\%), TIBS (HR: 77.66, FL:81.36\%), Pysa(HR:42.97, FL: 45.56), Emotet( HR: 53.6, FL: 56.7), Scar (HR:52.05, FL: 54.7) classes. Likewise, the HR model performs better for Pysa, Gator, Shade in terms of recall and the FL model achieves higher recall for Ako, Zeus, TIBS, Emotet attack categories. However, the FL model outperforms the HR model for most of the classes including  Zeus, TIBS, Emotet, Conti, Scar in terms of F1-score. 

Overall, the FL model shows better performance than the HR model in terms of precision, recall, and F1-score for the majority of attack classes. 

\begin{figure}[!htbp]
   \centering
    \includegraphics[scale = .60]{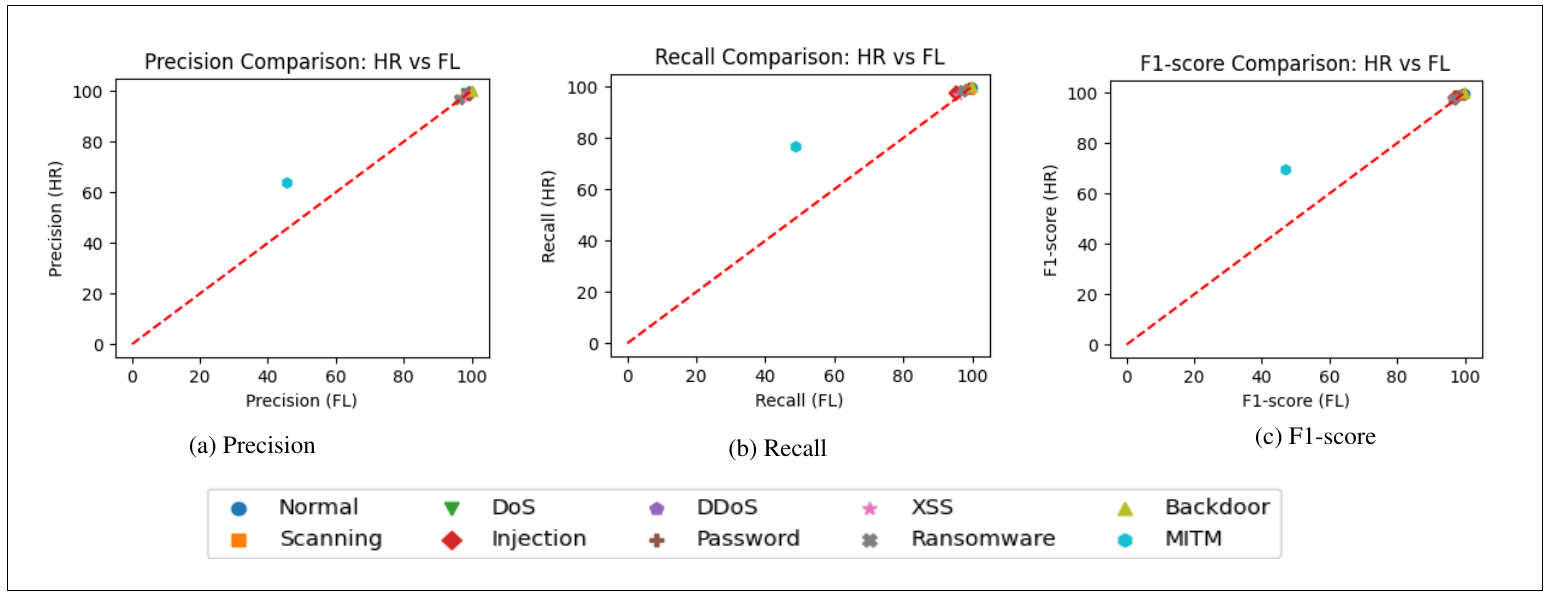}
    \caption{The difference of performance between HR and FL model on ToN-IoT-Network's attacks}
    \label{fig:toniotnetworkattack}
\end{figure}

ToN-IoT-Network:

Figure \ref{fig:toniotnetworkattack} (a), (b) and (c) show the comparison of precision, recall, and F1-score of the HGBC of FL (Flat) model and the HGBC of HR (Hierarchical) model on the ToN IoT Network dataset.

The FL model achieves slightly higher precision, recall, and F1-score compared to the HR model for the most of the attack categories. However, the differences are not noticeable in most cases. However, the HR model achieves higher recall and F1-score for the MITM (Man-in-the-Middle) class compared to the FL model. However, the precision of the HR model is considerably lower for this class. Overall, the FL model generally exhibits superior performance to the HR model in terms of precision, recall, and F1-score for the majority of attack classes in the ToN IoT Network dataset.

\subsubsection*{HR vs FL: False Negative}

\begin{figure}[!htbp]
    \centering
    \subfloat[CIC-DoS2017]{\includegraphics[scale=.60]{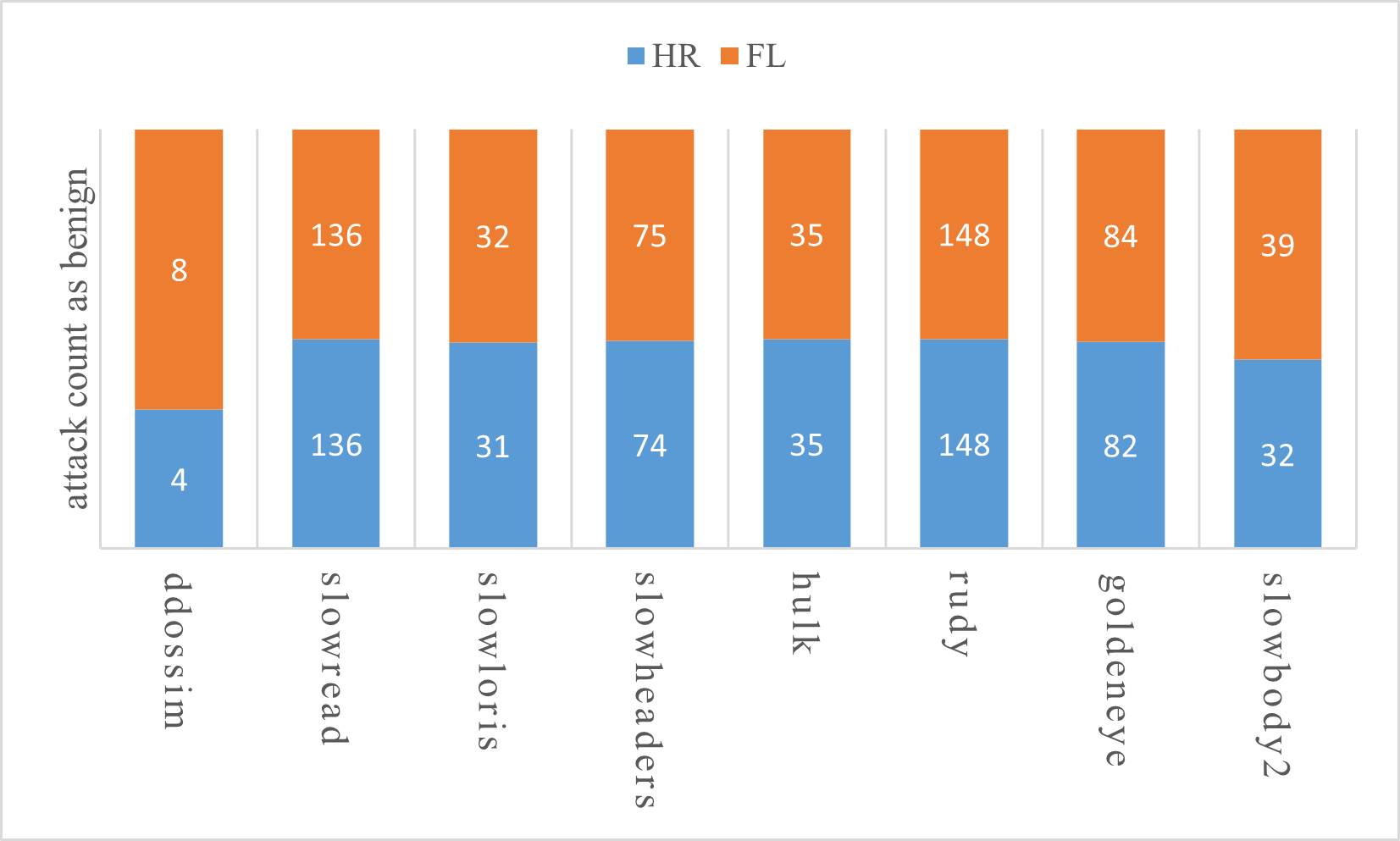}}   
     \subfloat[CIC-Darknet2020]{\includegraphics[scale=.60]{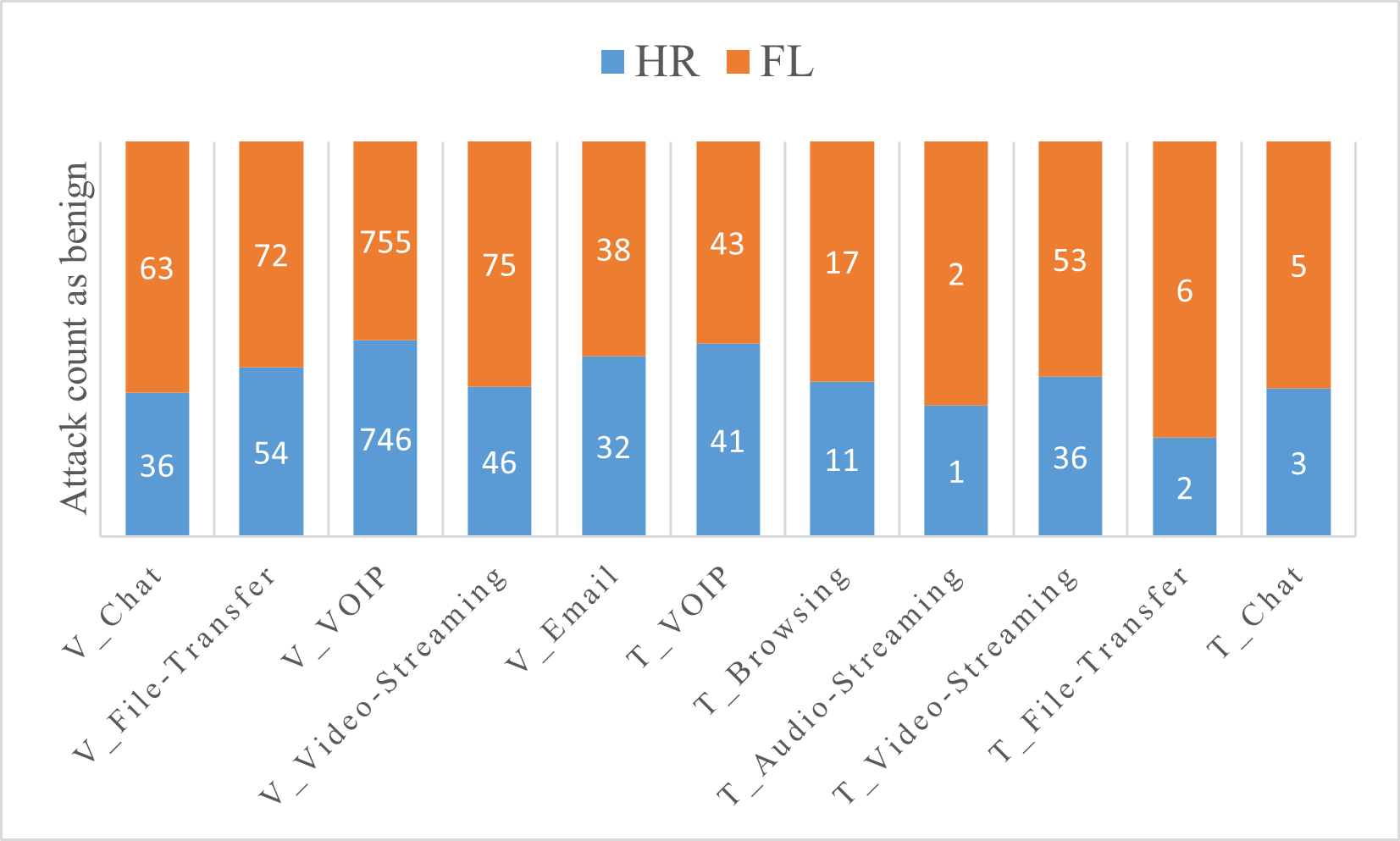}}   
    \caption{The false negative rate of HR and FL model}
    \label{fig:normalattack11}
\end{figure}

\begin{figure}[!htbp]
    \centering
    \subfloat[ToN-IoT-Network]{\includegraphics[scale=.60]{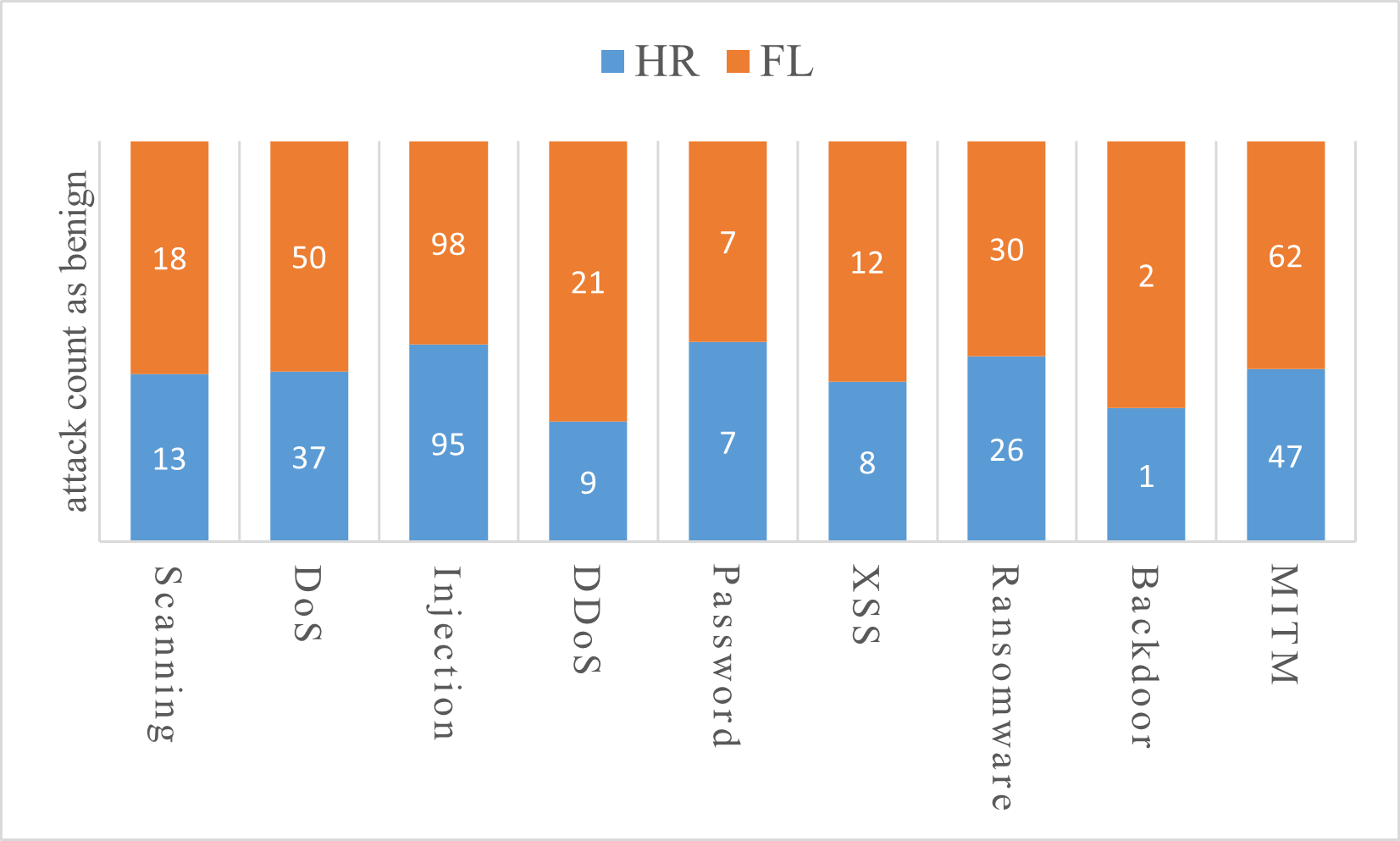}}
    \subfloat[XIIOTID]{\includegraphics[scale=.60]{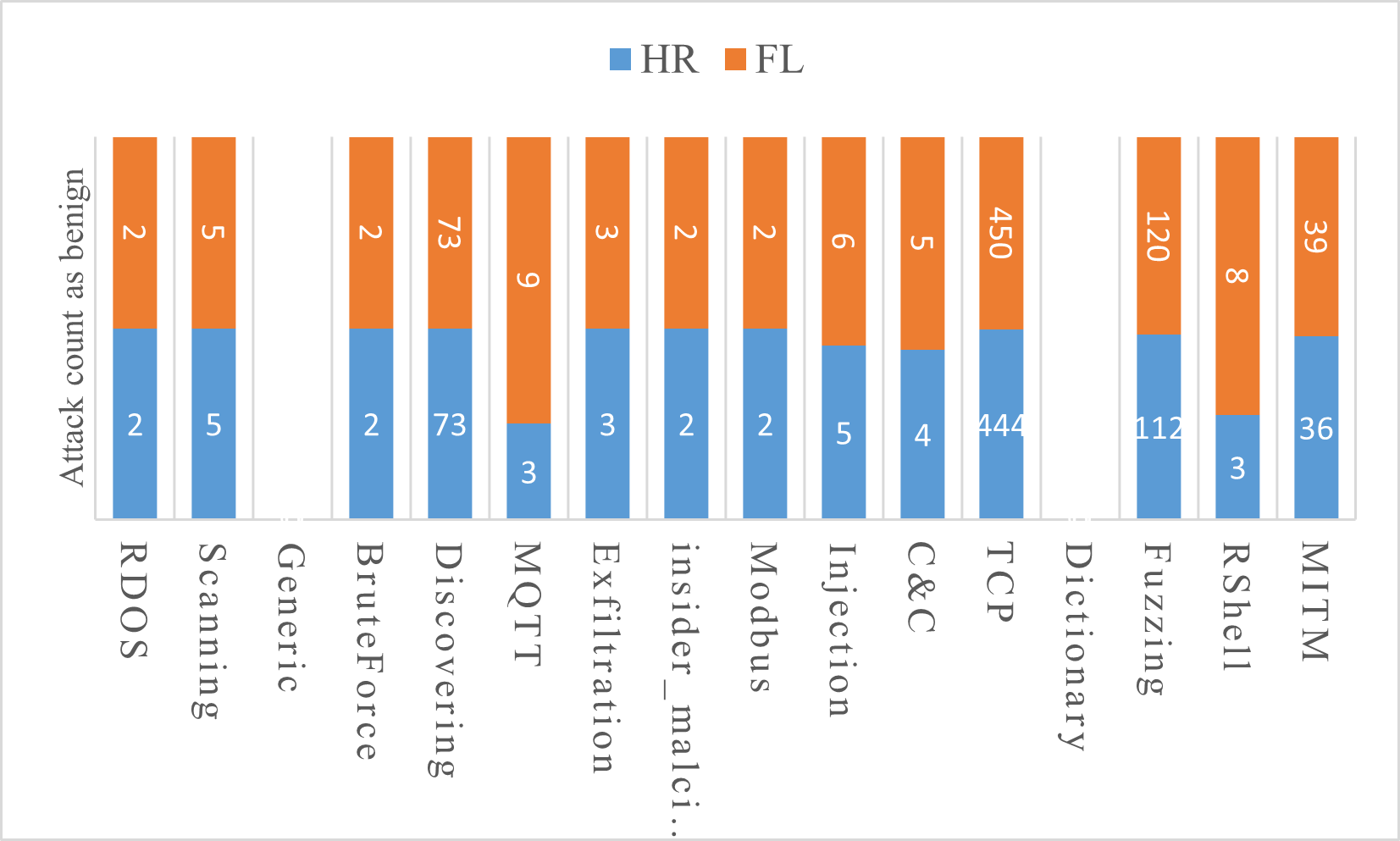}}  
    \caption{The false negative rate of HR and FL model}
    \label{fig:normalattack12}
\end{figure}

\begin{figure}[!htbp]
    \centering
    \subfloat[ISCXURL2016]{\includegraphics[scale=.60]{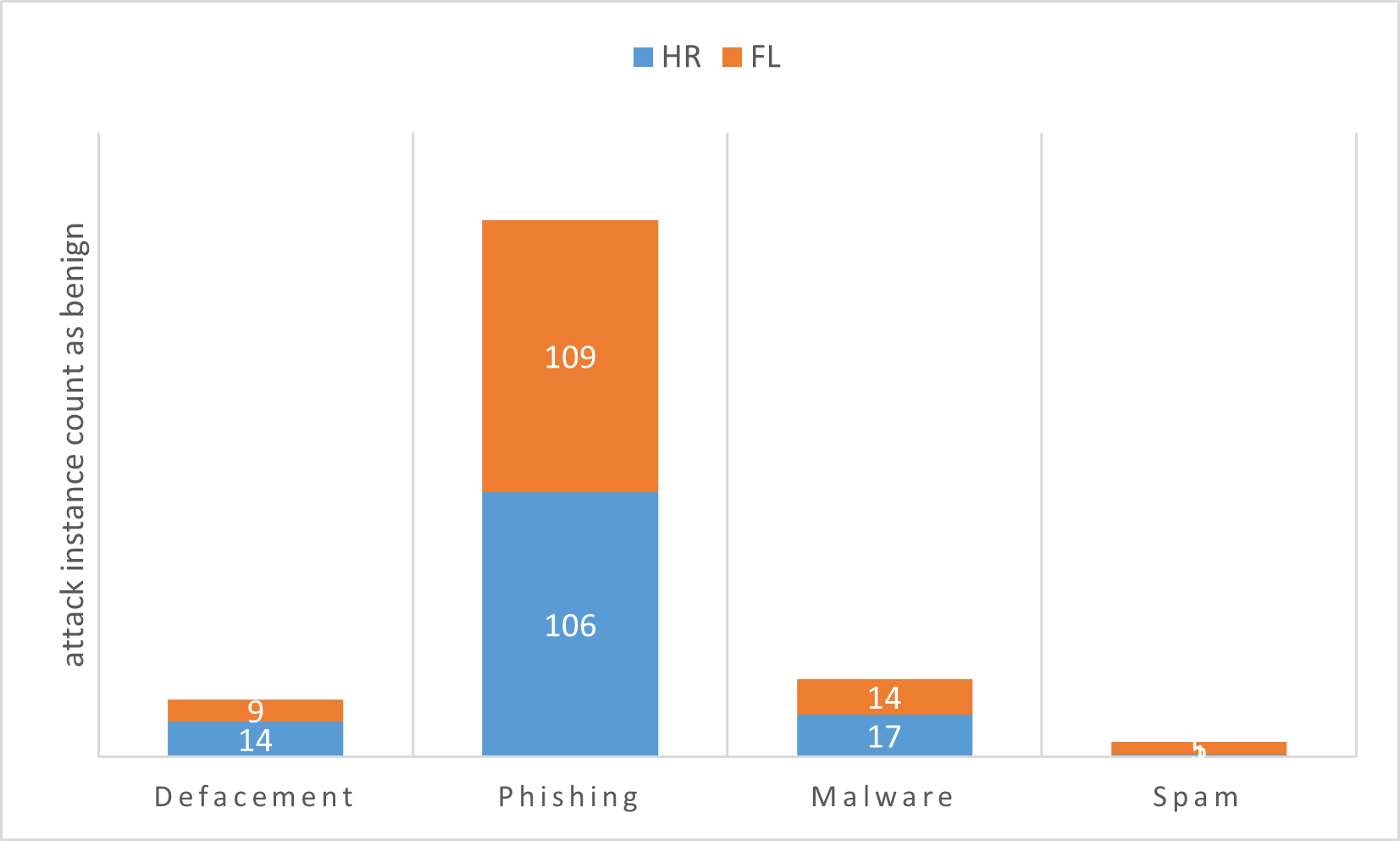}}
    \subfloat[ToN-IoT-IoTs]{\includegraphics[scale=.60]{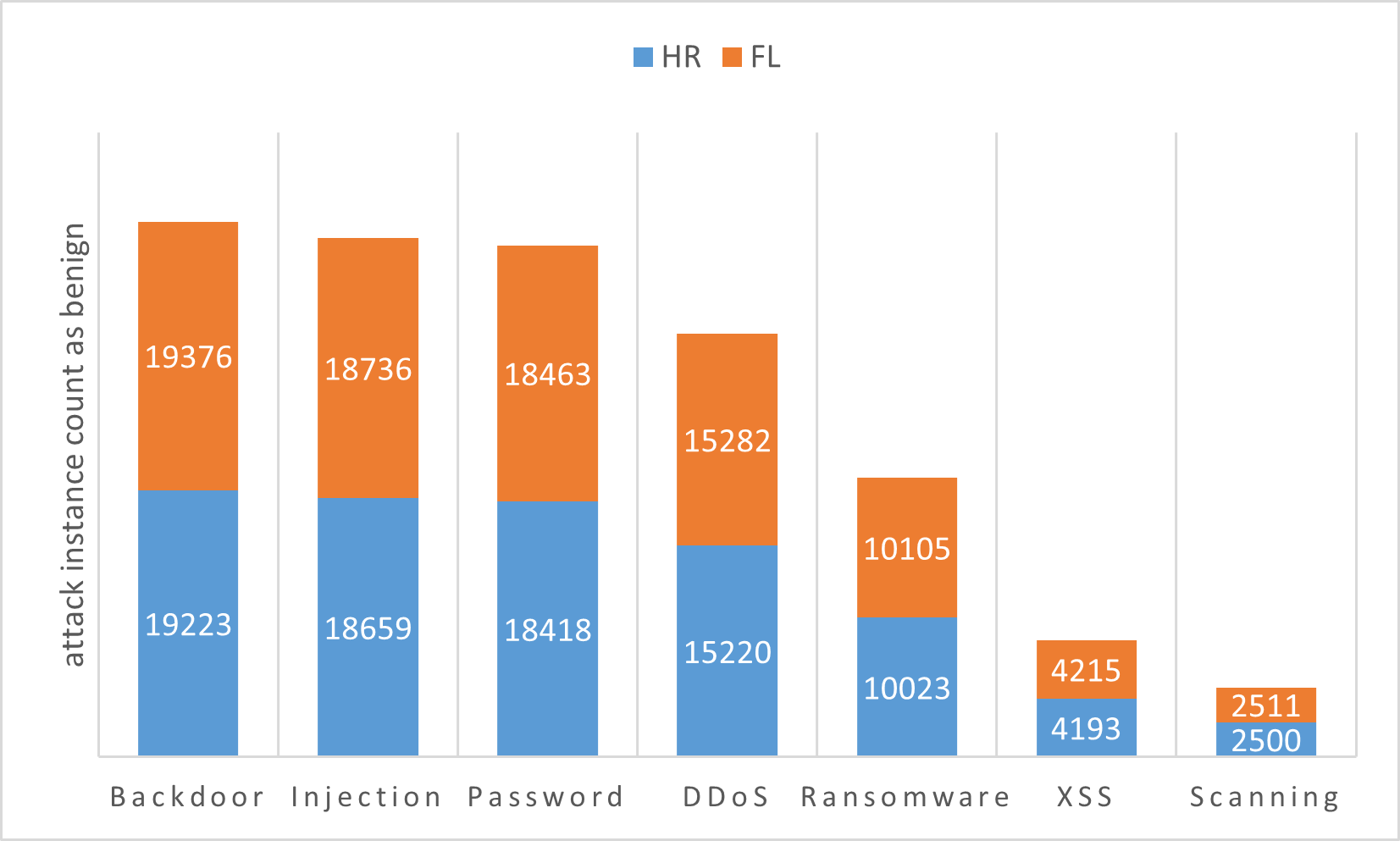}}  
    \caption{The false negative rate of HR and FL model}
    \label{fig:normalattack13}
\end{figure}

In this case, we present the results of CIC-DoS2017, CIC-Darknet2020, ToN-IoT-Network, XIIOTID, ToN-IoT-IoTs, and ISCXURL2016 datasets in Figure \ref{fig:normalattack11}, \ref{fig:normalattack12} and \ref{fig:normalattack13} for the HR and FL model with respect to false negative rate. 

Figure \ref{fig:normalattack11} (a) depicts the false negative rate difference of the HR (RF) and FL (RF) model. The graph shows that there are slight differences observed for attacks such as ddossim (4 vs. 8) , slowloris (31 vs. 32), slowheaders (74 vs. 75),  , goldeneye (82 vs. 84), and slowbody2 (32 vs. 39) in CIC-DoS 2017 dataset. For slowread (136 vs. 136), hulk (35 vs. 35),rudy (148 vs. 148), both models produces the same number of false negatives. 

CIC-Darknet2020 depicted in Figure \ref{fig:normalattack} (b) showed notable differences of the HR (RF) and FL model (RF) for the following attacks: V\_Audio-Streaming (272 vs. 332), V\_Chat (36 vs. 63), V\_File-Transfer (54 vs. 72), V\_Video-Streaming (46 vs. 75), T\_Browsing (11 vs. 17), and T\_Video-Streaming (36 vs. 53).


ToN-IoT-Network and XIIOTID:

Figure \ref{fig:normalattack12} (a) compares the number of attacks identified as benign by the HR (ET) and FL (ET) models for ToN IoT Network dataset. Some notable differences can be observed for certain attack types. Specifically, the HR model identifies fewer attacks as benign for Scanning (13 vs. 18), DoS (37 vs. 50), and Ransomware (26 vs. 30).

Figure \ref{fig:normalattack12} (b) compares the number of attacks identified as benign by the HR (RF) and FL (RF) models for XIIOTID. Both models produce similar results for most attack types, with no significant differences observed. The attack types RDOS, Scanning, Generic, BruteForce, Discovering, insider\_malicious, Modbus, Injection, C\&C, Dictionary, and Exfiltration show identical counts for both models. However, slight differences can be seen for MQTT (3 vs. 9), RShell (3 vs. 8), Fuzzing (112 vs. 120), MITM (36 vs. 39), and TCP (444 vs. 450) attacks, although these differences are relatively small.

\subsubsection*{HR vs FL: False Positive}

\begin{figure}[!htbp]
    \centering
    \subfloat[CIC-DoS2017]{\includegraphics[scale=.60]{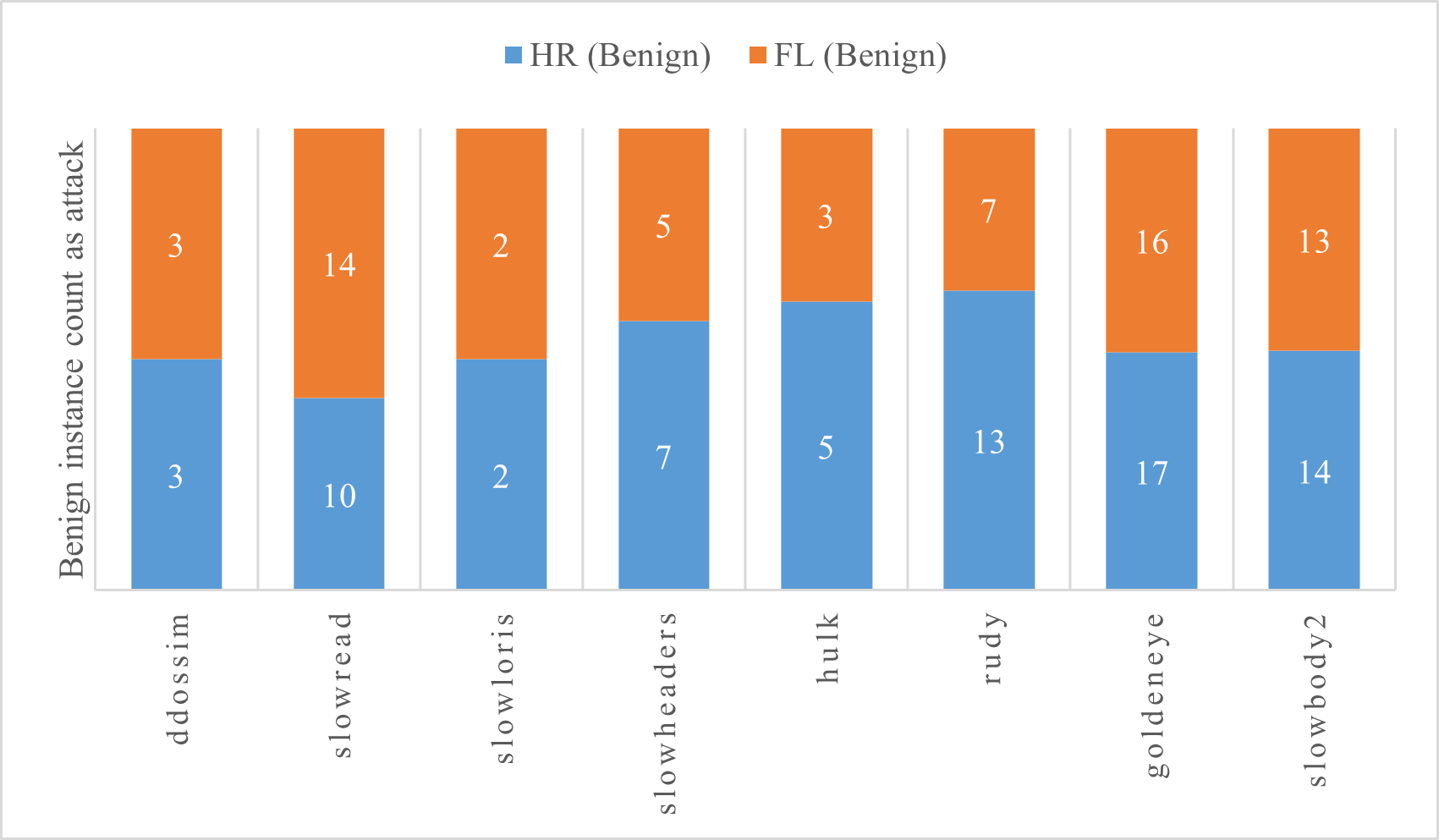}}    
     \subfloat[CIC-Darknet2020]{\includegraphics[scale=.550]{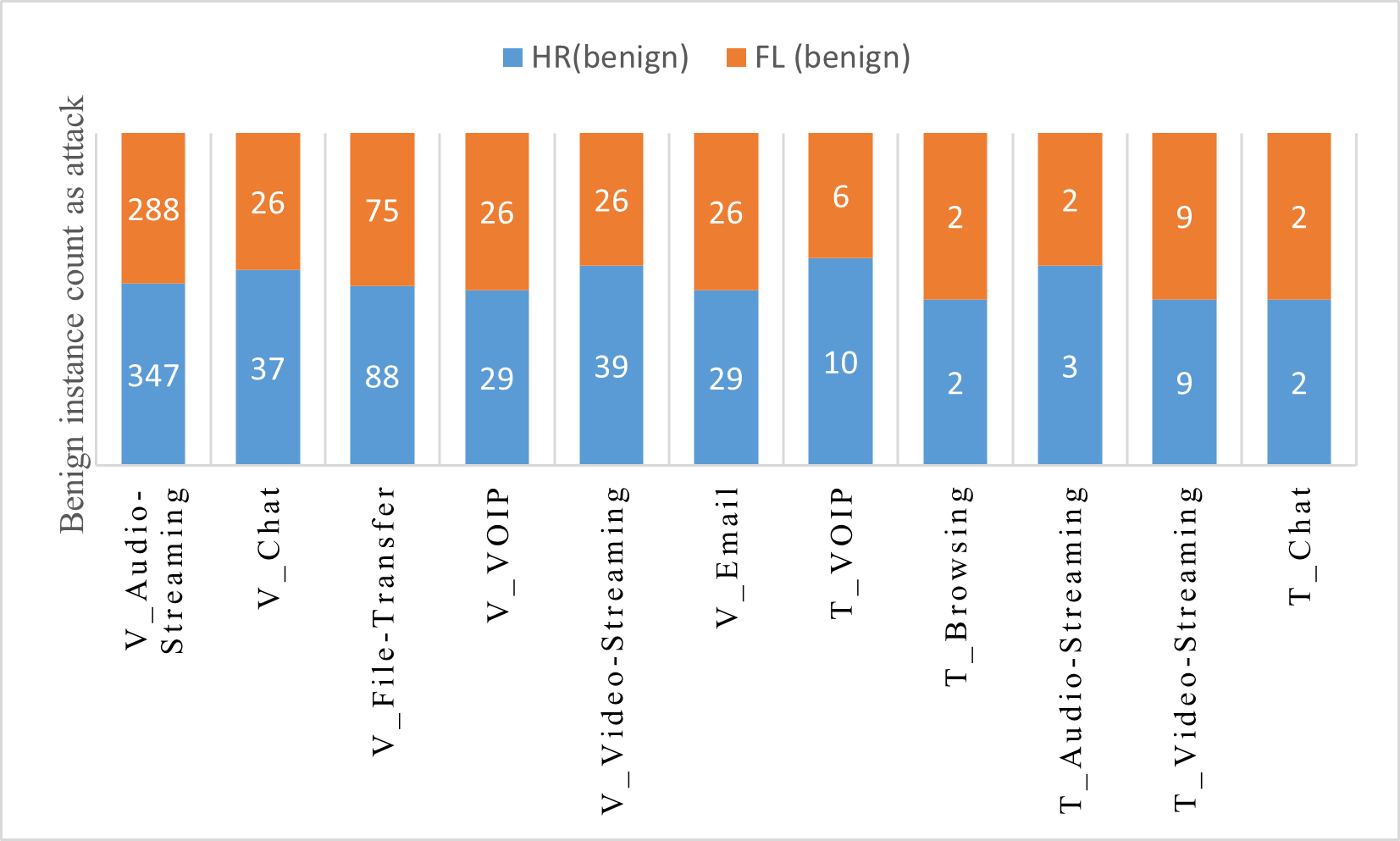}}   
    \caption{The false positive rate of HR and FL model}
    \label{fig:falsepositiverate11}
\end{figure}

\begin{figure}[!htbp]
    \centering
    \subfloat[CIC-Malmem2022]{\includegraphics[scale=.60]{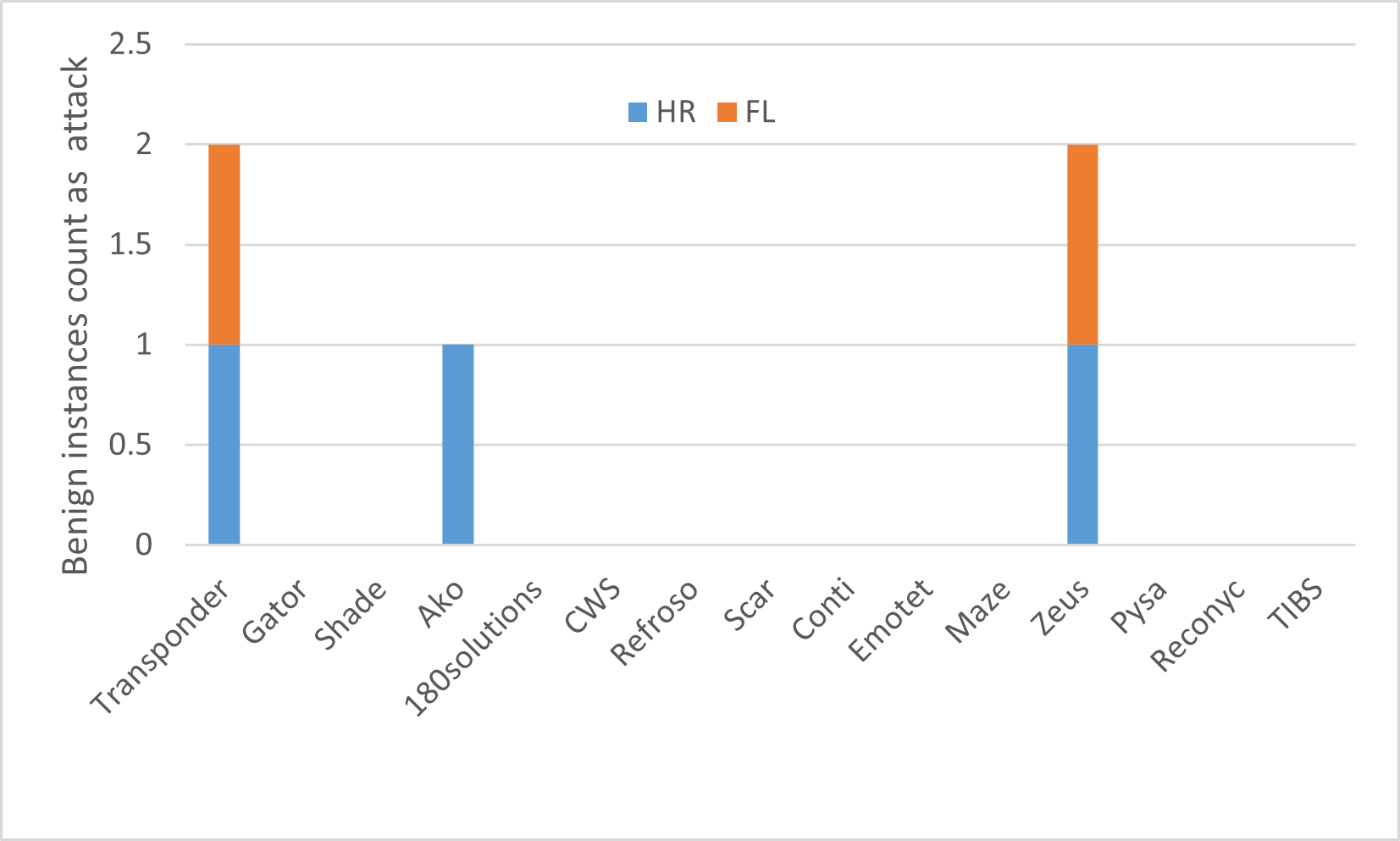}}
    \subfloat[ISCXURL2016]{\includegraphics[scale=.60]{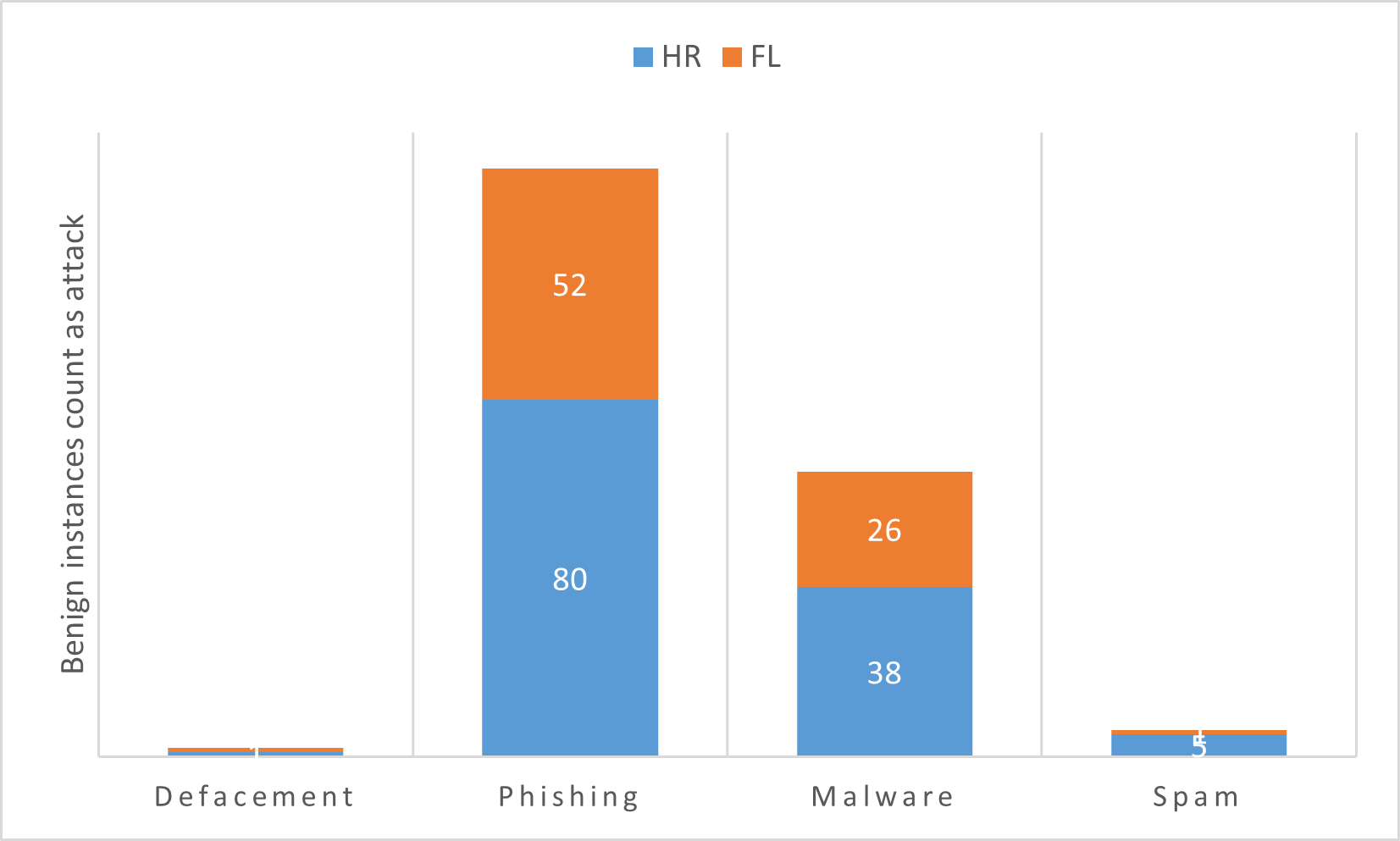}}   \\

     \subfloat[XIIOTID]{\includegraphics[scale=.550]{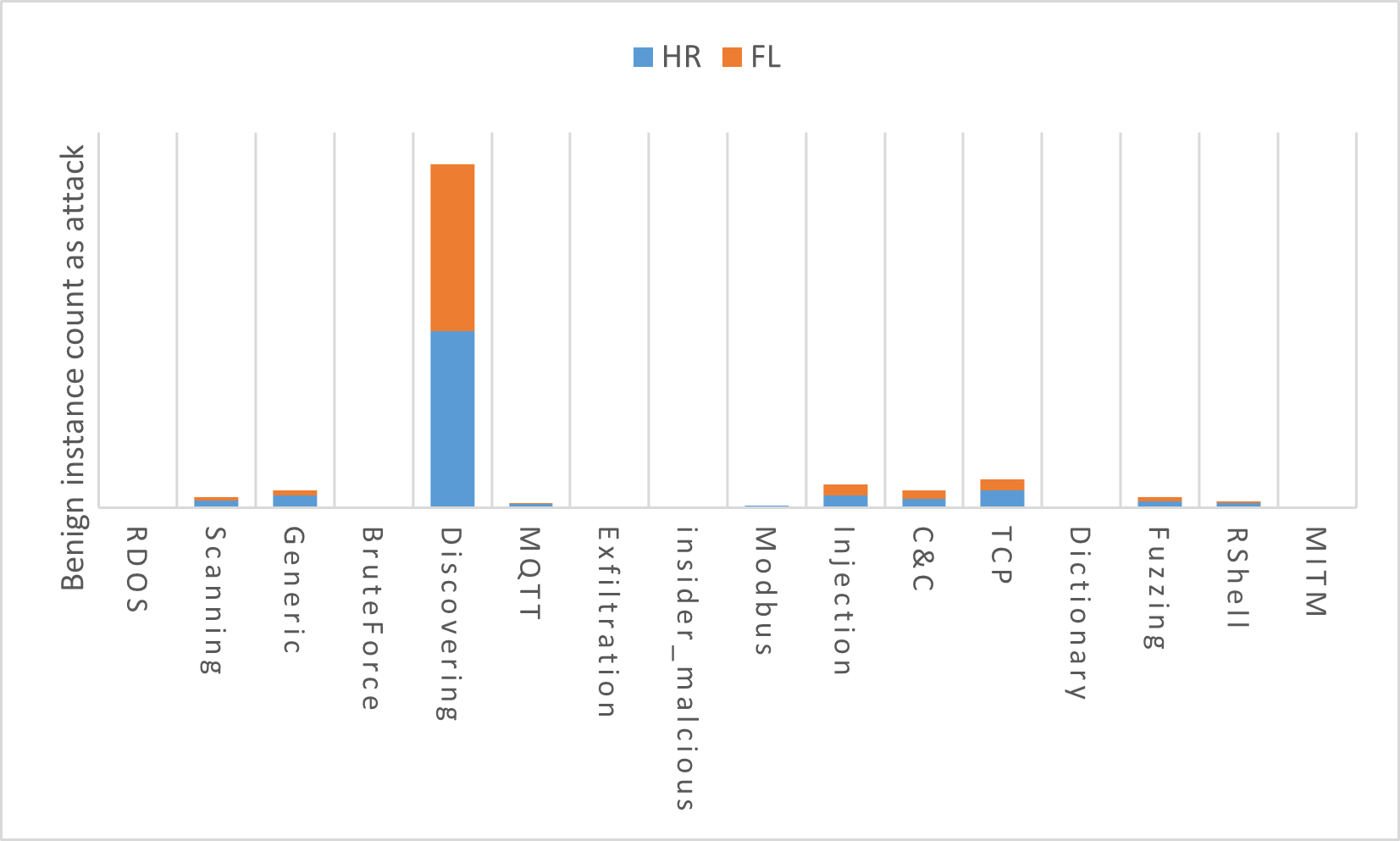}}   
     \subfloat[ToN-IoT-IoTs]{\includegraphics[scale=.550]{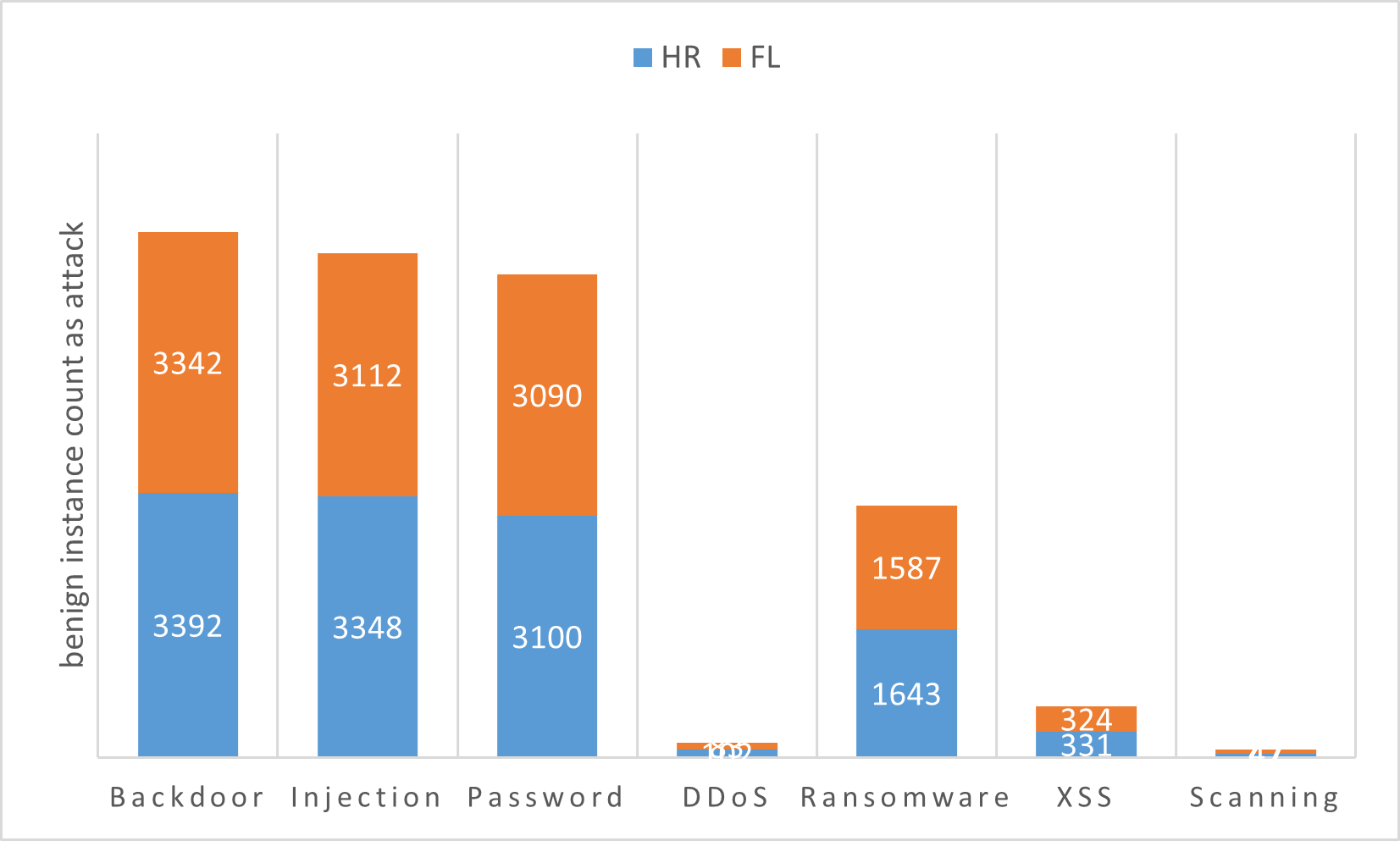}}   
    
    \caption{The false positive rate of HR and FL model}
    \label{fig:falsepositiverate12}
\end{figure}

Figure \ref{fig:falsepositiverate11} and \ref{fig:falsepositiverate12} illustrate the false positive rate of the HR and FL model for CIC-DoS2017, CIC-Darknet2020, CIC-Malmem2022, ISCXURL2016, XIIOTID and ToN-IoT-IoTs datasets. From the results, we can observe that the HR model identified higher number of false positives ( benign instances detected as attacks) than that of the FL model. 

\end{document}